\documentclass[aps,prx,reprint,superscriptaddress,longbibliography,nofootinbib,floatfix]{revtex4-2}
\usepackage[T1]{fontenc}
\usepackage{needspace}
\usepackage{dsfont}
\usepackage{amsmath,amssymb,bm,graphicx,booktabs,longtable}
\usepackage[dvipsnames]{xcolor}
\usepackage{hyperref}
\usepackage{pgfplots}
\usepgfplotslibrary{groupplots}
\pgfplotsset{compat=1.18}
\hypersetup{
  colorlinks=true,
  linkcolor=MidnightBlue,
  citecolor=MidnightBlue,
  urlcolor=MidnightBlue,
  pdftitle={Decoding SO(5) Deconfined Multicriticality: Conformal Bootstrap and Fuzzy-Sphere Spectroscopy},
  pdfauthor={Zhijin Li and Tinghong Shen}
}
\newcommand{\rep}[1]{{\color{black}\mathbf{#1}}}
\newcommand{\SOrephead}{\mbox{$SO(5)$ \textbf{Rep.}}}
\newcommand{\appendixtablestyle}{\small\setlength{\tabcolsep}{6pt}\renewcommand{\arraystretch}{1.2}}
\newcommand{\rowstrut}{\rule[-3pt]{0pt}{13pt}}
\newcommand{\cO}{\mathcal{O}}
\newcommand{\dd}{\mathrm{d}}
\begin{document}%

\title{Decoding \texorpdfstring{$SO(5)$}{SO(5)} Deconfined Multicriticality: Conformal Bootstrap and Fuzzy-Sphere Spectroscopy}
\author{Zhijin Li}
\email{zhijin_li@seu.edu.cn}
\affiliation{School of Physics, Southeast University, 2 Sipailow Street, Nanjing 210096, China}
\affiliation{Shing-Tung Yau Center, Southeast University, 2 Sipailow Street, Nanjing 210096, China}
\author{Tinghong Shen}
\email{thshen@seu.edu.cn}
\affiliation{School of Physics, Southeast University, 2 Sipailow Street, Nanjing 210096, China}
\affiliation{Shing-Tung Yau Center, Southeast University, 2 Sipailow Street, Nanjing 210096, China}
\begin{abstract}
The nature of the deconfined quantum critical point (DQCP) remains unsettled: competing explanations invoke a real multicritical conformal field theory (CFT) or pseudocriticality governed by nearby complex fixed points. We use the conformal bootstrap and fuzzy-sphere regularization to compute $SO(5)$ CFT data and compare our results with large-scale quantum Monte Carlo (QMC) data to investigate these scenarios. Using mixed correlators of the lowest vector $\phi$, singlet $s$, and traceless symmetric scalar $t$, we characterize the bootstrap cone whose tip lies remarkably close to the position indicated by large-scale QMC. We extract an extensive spectrum across $SO(5)$ irreducible representations, together with operator product expansion (OPE) coefficients. We study a fuzzy-sphere realization of the $SO(5)$ nonlinear sigma model with a level-1 Wess--Zumino--Witten term. Using the special conformal generators, we identify 52 conformal primaries in a system with \(N=10\) Landau orbitals, including 31 not previously classified as primaries in the earlier fuzzy-sphere study.
We extend exact diagonalization in the fuzzy sphere computations to $N=12$ using symmetry projection.
The bootstrap spectrum and OPE coefficients show close agreement with fuzzy-sphere data, including 16 of the newly identified primaries in channels accessible to the mixed-correlator bootstrap. The lowest $SO(5)$ singlet scalar is relevant in the bootstrap and QMC results; its fuzzy-sphere data is $\Delta_s=2.816$ at $N=12$ and decreases with system size. Our results suggest that the mixed-correlator bootstrap solution represents a unitary CFT with a relevant $SO(5)$ singlet scalar and is closely connected to the QMC and fuzzy-sphere data for the DQCP. The conformal bootstrap, fuzzy-sphere, and QMC together provide comprehensive evidence for an $SO(5)$ multicritical CFT and critical data to decode the phases in two-dimensional quantum magnets.
\end{abstract}
\maketitle

\section{Introduction}
\label{sec:intro}

The deconfined quantum critical point (DQCP) predicts a direct continuous transition between N\'eel antiferromagnetic order and valence-bond-solid (VBS) order through fractionalized excitations and an emergent gauge field, providing a route beyond the Landau description~\cite{Senthil2004,SenthilLong2004,SenthilReview,CuiYuReview2025}. It offers an organizing principle for square-lattice quantum magnets, where N\'eel, VBS, and possible quantum spin-liquid phases compete~\cite{ReadSachdev1991,LiuQSL2022,LiuQSL2024}. Recently, there has been exciting experimental progress toward realizing the DQCP in real materials. The pressure- and field-tuned studies of SrCu$_2$(BO$_3$)$_2$ connect plaquette-singlet--antiferromagnet transitions to proximate deconfined criticality~\cite{LeeMaterials2019,CuiMaterials2023}, and recent NMR work suggests a phase diagram related to $SO(5)$ DQCP physics~\cite{CuiMaterials2025}. The universality classes and asymptotic transition behavior remain under investigation~\cite{GuoMaterials2025}.

Despite extensive numerical evidence for emergent $SO(5)$ symmetry and approximate scaling, the continuum origin of the N\'eel--VBS transition remains unsettled: does the observed regime reflect proximity to a real multicritical conformal field theory (CFT), or walking associated with complex fixed points? We address this question by testing the spectrum and interactions of a candidate selected by a conic feature of the bootstrap bounds against independent fuzzy-sphere and QMC data.

\subsection{Deconfined transitions and emergent symmetries}
N\'eel and VBS orders break different microscopic symmetries, so a direct continuous boundary is not the generic outcome of Landau theory with independent order parameters. In the proposed noncompact $\mathrm{CP}^{1}$ description, the N\'eel vector is a bilinear of bosonic spinons coupled to an emergent gauge field, while monopoles carry VBS quantum numbers. Square-lattice Berry phases allow a quadrupled monopole perturbation; its proposed dangerous irrelevance permits an emergent continuous VBS angle at criticality while selecting discrete directions in the ordered phase~\cite{SenthilLong2004,Motrunich2004}. This mechanism requires an infrared fixed point stable against the remaining perturbations allowed by the microscopic symmetries.

The quantum Monte Carlo (QMC) made this question accessible in the lattice simulations. In particular the $J$--$Q$ model exhibits a direct N\'eel--VBS boundary with large anomalous dimensions and emergent VBS angular symmetry~\cite{Sandvik2007}. Subsequent simulations found unconventional scaling together with drifting exponents and pronounced finite-size corrections~\cite{MelkoKaul2008,Lou2009,Sandvik2010}. Explanations ranged from logarithmic corrections and dangerously irrelevant variables~\cite{Sandvik2010,Shao2016} to first-order flow~\cite{Jiang2008,Kuklov2008}. Studies of $SU(N)$ generalizations and gauge formulations reinforced the difficulty of determining the asymptotic behavior~\cite{Kaul2011,ChenFlow2013}. Large classical-loop simulations found strong scaling violations without conventional first-order signatures over the accessible range~\cite{NahumScaling}. Together these studies exposed a persistent tension between approximate scaling and its fixed-point interpretation.

Emergent $SO(5)$ symmetry further constrains the continuum description~\cite{NahumSO5}. The three N\'eel and two VBS components form a vector,
\begin{equation}
\phi_i\sim(N_x,N_y,N_z,V_x,V_y),\qquad i=1,\ldots,5.
\label{eq:vector}
\end{equation}
An $SO(5)$ nonlinear sigma model with a level-one Wess--Zumino--Witten term expresses this unification and connects it to gauge-theory dualities~\cite{TanakaHu2005,SenthilFisher2006,WangDualities2017}. The topological term distinguishes this proposal from the ordinary $O(5)$ Wilson--Fisher theory. The N\'eel--VBS tuning perturbation belongs to a traceless symmetric representation; a stable one-parameter transition additionally requires the absence of a relevant $SO(5)$ singlet.

Related questions arise for the easy-plane DQCP and its conjectured dual, two-flavor fermionic QED$_3$ with enhanced $O(4)$ symmetry~\cite{WangDualities2017}. Monopole bootstrap constraints~\cite{LiNfTwo2022} and recent monopole-condensation scenarios  \cite{ChesterKomargodski2025,DumitrescuNiroThorngren2024} emphasize that anomaly and duality constraints alone do not establish the dynamical existence of an interacting CFT.

\subsection{Bootstrap constraints and fuzzy-sphere studies}
The conformal bootstrap tests stability through crossing symmetry and reflection positivity~\cite{BootstrapReview}. Bounds excluded the numerical dimensions under the assumption of a unitary $SO(5)$ CFT with an irrelevant leading singlet~\cite{LiQED}; complementary analyses constrained monopole dimensions needed for emergent continuous symmetry~\cite{NakayamaOhtsuki}. These results leave open both a theory with a relevant singlet and scaling that does not approach a real conformal fixed point.

A prominent realization of the latter possibility is walking near complex-conjugate fixed points~\cite{WangDualities2017,Gorbenko2018,GorbenkoPotts2018}: a physical renormalization-group (RG) trajectory evolves slowly before reaching a weakly first-order transition. Nonlinear sigma models with topological terms provide a framework for this deconfined pseudocriticality scenario~\cite{MaWang2020,Nahum2020}. Near a simple fixed-point collision, the merger direction is nearly marginal, making the singlet spectrum a useful diagnostic of this mechanism.

Fuzzy-sphere regularization brought direct spectral information to this debate~\cite{HeZhuFuzzyReview2026}. Building on Landau-level formulations and sign-free $SO(5)$ simulations~\cite{Ippoliti2018,WangSO52021}, it uses the state--operator correspondence to extract dimensions and three-point data, as demonstrated for the Ising CFT~\cite{ZhuFuzzy2023,HuOPE2023}. For the $SO(5)$ model, Zhou \emph{et al.} obtained a broad spectrum and approximate conformal towers, interpreting the singlet's size-dependent drift as evidence for pseudocriticality~\cite{Zhou2024}. Chen \emph{et al.} varied independent N\'eel and VBS couplings, reducing $SO(5)$ to $SO(3)\times SO(2)$, and reported a disordered region with boundaries meeting at a multicritical point~\cite{Chen2024}. Their subsequent spectroscopy varied an $SO(5)$-preserving interaction range to reduce finite-size effects~\cite{ChenConformal}. Liao \emph{et al.} extended continuum QMC to larger spheres and tori, studying the symmetric point and the anisotropic phase boundaries~\cite{Liao2026}. These differing interpretations make the evolution toward consistent continuum operator data a central issue.


%
\subsection{Multicriticality and the present work}
Large-scale QMC on extended $J$--$Q$ models has observed tiny discontinuities and proposed that the weakly first-order trajectories pass near a real $SO(5)$ multicritical point~\cite{Takahashi2024}. An additional relevant singlet then controls departure from the scaling regime. Weak first order is compatible with either this mechanism or complex-fixed-point walking; the eventual transition alone cannot distinguish them. The QMC dimensions characterize intermediate scales and provide benchmarks for identifying the nearby theory. Studies of the $SU(N)$ $X$--$Q$ model found weakly first-order behavior for $N=2$ and strongly first-order behavior for $N>2$, emphasizing the role of microscopic interactions in sustaining emergent VBS angular fluctuations~\cite{FlynnSandvik2026}.

Entanglement measurements supply complementary probes of the putative continuum theory. Square-lattice studies report anomalous logarithmic terms, including for smooth boundaries, and behavior consistent with weak first order~\cite{ZhaoEntanglement2022,SongSubleadingEntanglement2024,DengEntanglement2024}. Similar anomalies occur in small-$N$ $SU(N)$ models~\cite{SongSUNEntanglement2025}, whereas suitably chosen bipartitions in the $SU(2)$ model yield corner coefficients compatible with an $SO(5)$ CFT~\cite{DEmidioEntanglement2024}. Recent honeycomb-lattice results likewise find CFT-compatible smooth-cut and corner scaling, alongside geometry-dependent period-three finite-size effects~\cite{KumarEntanglement2026}. Additional edge contributions for nonordinary cuts further underscore the importance of boundary effects~\cite{ZhuEntanglement2026}. These diagnostics constrain symmetry and scaling, but their finite-size and boundary dependence leaves the distinction between a real multicritical CFT and pseudocriticality open.

The mixed vector--singlet--tensor correlator has been bootstrapped by allowing  a relevant singlet \cite{ChesterSu2024}. They selected an extremal solution by fixing the vector dimension to a large-$N$ monopole estimate and maximizing the tensor dimension. Our subsequent vector-correlator study identified a sharp bootstrap cone whose tip lies close to large-scale QMC estimates~\cite{LiShen2026}. The cone selects a distinct numerical solution: although some leading dimensions are nearby, its operator product expansion (OPE) data differ from those of the earlier extremal solution in \cite{ChesterSu2024}. The cone was not identified in Ref.~\cite{ChesterSu2024}. Its physical identification requires tests extending beyond the few dimensions used to locate it.

Here we characterize this cone using mixed correlator bootstrap and independent fuzzy-sphere calculations. Mixing leaves its location approximately unchanged while supplying dimensions and OPE coefficients across many more representation and spin sectors. On the fuzzy sphere, special conformal generators resolve additional primaries among previously ambiguous higher levels, providing independent tests of the bootstrap spectrum. Symmetry projection extends the calculation to $N=12$ Landau orbitals and tracks the lowest singlet to larger sizes. Agreement across many primary dimensions and OPE coefficients, together with the low-lying QMC comparisons, supports the physical identification of the cone. The fuzzy-sphere singlet dimension decreases steadily with system size, suggesting a relevant singlet.

The combined evidence strongly supports a unitary $SO(5)$ multicritical CFT with a relevant singlet. The singlet supplies an additional symmetry-preserving unstable direction alongside the tensor deformation selecting N\'eel or VBS order, explaining why a one-parameter lattice scan can miss the fixed point while displaying its scaling regime. OPE comparisons test operator interactions beyond agreement of exponents, while crossing and positivity in conformal bootstrap constrain their consistency. These data characterize a candidate local organizing theory for quantum magnets and supply inputs for studying the observed crossover; its connections to ordered and spin-liquid phases require further analysis of the relevant RG flows.

Section~\ref{sec:overview} presents the competing scenarios, the complementary methods, and the physical implications of the data. Sections~\ref{sec:bootstrap} and \ref{sec:fuzzy} give the bootstrap and fuzzy-sphere analyses. Section~\ref{sec:discussion} discusses lattice crossover, alternative explanations, and applications to quantum magnets and spin liquids.

\section{Overview of results and physical implications}
\label{sec:overview}

\input{fig_overview}%

In this work we compute a large set of CFT data of DQCP using mixed correlator conformal bootstrap and fuzzy sphere regularization. 
Our principal result is an extensive agreement between the conformal data obtained from the mixed-correlator bootstrap and from the fuzzy sphere, together with close agreement with the low-lying dimensions accessible to large-scale QMC. The comparison includes both a substantially enlarged spectrum of conformal primaries and their OPE coefficients. It supports a unitary $SO(5)$ multicritical CFT and gives a quantitative description of the theory proposed to organize the observed N\'eel--VBS scaling regime. We first explain the competing RG interpretations and the complementary information provided by the three methods, then present the content and physical implications of the combined evidence. Figure~\ref{fig:overview} summarizes their complementary roles and the two RG interpretations.

\subsection{Multicriticality and complex-fixed-point pseudocriticality}
The observation of a weakly first-order N\'eel--VBS transition leaves an important question unresolved: what accounts for its broad regime of approximate scaling and emergent $SO(5)$ symmetry? The two principal explanations assign this regime to different structures in the space of theories. In the multicritical scenario, there is a real, unitary $SO(5)$ fixed point with an additional relevant symmetry-preserving perturbation. A microscopic trajectory can approach this fixed point, display its scaling behavior over a substantial range, and subsequently depart along a direction that the chosen lattice scan does not tune away~\cite{ChesterSu2024,Takahashi2024,LiShen2026}. A first-order transition on that trajectory is therefore compatible with a nearby real CFT.

In the complex-fixed-point scenario, a pair of fixed points has moved away from real couplings. Their proximity produces slow RG evolution along a physical trajectory, generating approximate scale invariance without an accessible real fixed point controlling the asymptotic transition~\cite{Gorbenko2018,MaWang2020,Nahum2020}. This provides a mechanism for pseudocritical behavior and has motivated an interpretation of earlier fuzzy-sphere results~\cite{Zhou2024}. We refer to this possibility specifically as \emph{complex-fixed-point pseudocriticality}: a long transient scaling regime can also arise near a real multicritical point, so pseudocritical behavior alone does not distinguish the two mechanisms.

The leading scalars that we use to investigate this question are
\begin{equation}
\phi_i\in\rep{5},\qquad s\in\rep{1},\qquad
t_{ij}=t_{ji},\quad t_{ii}=0,\quad t\in\rep{14}.
\label{eq:external}
\end{equation}
The singlet scalar $s$ is especially informative. Its relevance, $\Delta_s<3$, supplies a symmetry-preserving unstable direction at a real fixed point. Its distance from marginality also tests the simplest fixed-point-collision description, for which the collision direction becomes marginal at the merger. A singlet substantially below three is difficult to identify with that nearly marginal direction. 

A second distinct diagnostic is whether the spectrum and interactions admit a description compatible with unitarity and conformal crossing symmetry conditions. A Hermitian finite-size Hamiltonian has real energy levels in either scenario, and approximate conformal towers alone do not decide this question. The combined spectrum and OPE comparison is more restrictive: bootstrap data compatible with crossing and positivity are compared with the independently extracted many-body energies and matrix elements. We will focus on the two diagnostics: the singlet relevance and a consistent CFT data compatible with unitarity.

\subsection{Complementary roles of QMC, the fuzzy sphere, and bootstrap}
Large-scale QMC probes the nature of the N\'eel--VBS transition directly in microscopic square-lattice spin models. It resolves the N\'eel--VBS transition, the emergence of approximate $SO(5)$, and the low-lying scaling behavior, including its eventual departure toward weak first order~\cite{Takahashi2024}. The extracted dimensions identify the physically relevant neighborhood of the bootstrap landscape, where our previous vector-correlator study found a sharp cone~\cite{LiShen2026}. QMC of the Landau-level-regularized $SO(5)$ WZW model supplies complementary evidence for the proposed multicritical phase structure~\cite{Liao2026}.

The fuzzy sphere gives access to rich CFT data. Through the state--operator correspondence, angular momentum and $SO(5)$ representations organize many-body energies into conformal families, while matrix elements yield three-point coefficients~\cite{ZhuFuzzy2023,HuOPE2023,Zhou2024,ChenConformal}. This makes it possible to compare far more than the few dimensions available in lattice studies. Recent advances in fuzzy-sphere regularization include the use of special conformal generators $K_\mu$ to distinguish primaries from descendants~\cite{FanGenerators2024,FardelliGenerators2024,FardelliOPE2026} and symmetry projection to enable calculations at larger system sizes~\cite{BaranExactL2026,ZhouHeSO3ED2026}.

The bootstrap constrains the same dimensions and OPE coefficients through crossing symmetry and reflection positivity~\cite{BootstrapReview}. Mixing $\phi$, $s$, and $t$ requires operators observed in different channels to participate in one consistent conformal algebra, and thereby tests the joint interpretation of the order parameter, the symmetry-preserving deformation, and the N\'eel--VBS tuning operator. Its particular contribution is a unitarity test that spectra or emergent symmetry alone cannot supply.

The three methods are therefore complementary: QMC locates the candidate, the fuzzy sphere identifies its many-body realization, and the bootstrap checks the compatibility of its operator data with a unitary CFT. Higher primaries and OPE coefficients that were not used to locate the solution provide the strongest tests of the multicritical interpretation.
\subsection{A comprehensive spectrum and set of OPE coefficients}
The mixed $\phi$--$s$--$t$ system characterizes the cone far more completely than the original $\langle \phi \phi \phi \phi \rangle$ analysis. An operator exchanged between two vectors must also have compatible couplings in the singlet and tensor correlators; for example, the tensor in the vector OPE must be consistent with its interactions with $s$ and $t$. This interlocking structure is the main physical gain from mixed correlators~\cite{MixedIsing2014,ChesterSu2024}.

The bootstrap data come from the $\Lambda=23$ solution at the lower tip of the cone, $(\Delta_\phi,\Delta_s,\Delta_t)=(0.614,2.258,1.464)$, with next-scalar gaps $\Delta_{\phi'}\geq3$, $\Delta_{s'}\geq4$, and $\Delta_{t'}\geq4$. The extremal functional at this point gives the higher spectrum and its couplings; the stress tensor is resolved as an isolated exchange whose couplings are fixed by its Ward identity (Appendix~\ref{app:EFMinterpretation}). The fuzzy-sphere dimensions are $\Delta=2(E-E_0)/(E_J-E_0)$, which fixes $\Delta_J=2$, at the coupling $g_*(N)$ where $\Delta_T=3$; at $N=12$, $g_*=1.0092383$.

\begin{table}[tbp]
\caption{Representative primary dimensions. Bootstrap: the $\Lambda=23$ mixed-correlator solution, whose $\phi,s,t$ entries are the optimized external dimensions. Fuzzy sphere: strict primaries at $N=12$ and $g_*(12)$; $\Delta_J=2$ and $\Delta_T=3$ fix the calibration. QMC: $J$--$Q$ estimates of Ref.~\cite{Takahashi2024}. $S^-$ is the lowest parity-odd singlet; $\rep{35}=[1,2]$ and $\rep{35}^{\prime}=[0,4]$. A bootstrap dash marks a channel outside the mixed system; a QMC dash marks a missing estimate.}
\label{tab:spectrum}
\begingroup\small\setlength{\tabcolsep}{3pt}
\begin{ruledtabular}\begin{tabular}{lccrrr}
Op. & \SOrephead & $\ell^{\mathcal P}$ & Bootstrap & $N=12$ & QMC~\cite{Takahashi2024}\\\hline
\rowstrut $S^-$ & $\rep{1}$ & $0^{-}$ & --- & 5.303 & --- \\
\rowstrut $s$ & $\rep{1}$ & $0^{+}$ & 2.258 & 2.816 & 2.273(4) \\
\rowstrut $T_{\mu\nu}$ & $\rep{1}$ & $2^{+}$ & 3.000 & 3.000 & --- \\
\rowstrut $\phi$ & $\rep{5}$ & $0^{-}$ & 0.614 & 0.590 & 0.607(4) \\
\rowstrut  & $\rep{5}$ & $0^{-}$ & --- & 4.308 & --- \\
\rowstrut $J_\mu$ & $\rep{10}$ & $1^{+}$ & 2.000 & 2.000 & --- \\
\rowstrut  & $\rep{10}$ & $1^{+}$ & --- & 3.159 & --- \\
\rowstrut $t$ & $\rep{14}$ & $0^{+}$ & 1.464 & 1.469 & 1.417(7) \\
\rowstrut  & $\rep{14}$ & $0^{+}$ & --- & 4.339 & --- \\
\rowstrut $\mathcal M_{6\pi}$ & $\rep{30}$ & $0^{-}$ & 2.493 & 2.592 & --- \\
\rowstrut  & $\rep{35}$ & $1^{-}$ & --- & 3.040 & --- \\
\rowstrut  & $\rep{35}^{\prime}$ & $0^{+}$ & --- & 4.863 & --- \\
\rowstrut $\mathcal M_{8\pi}$ & $\rep{55}$ & $0^{+}$ & 3.692 & 3.926 & 3.723(11) \\
\end{tabular}\end{ruledtabular}\endgroup
\end{table}

\begin{figure*}[t]
\centering\includegraphics[width=\textwidth]{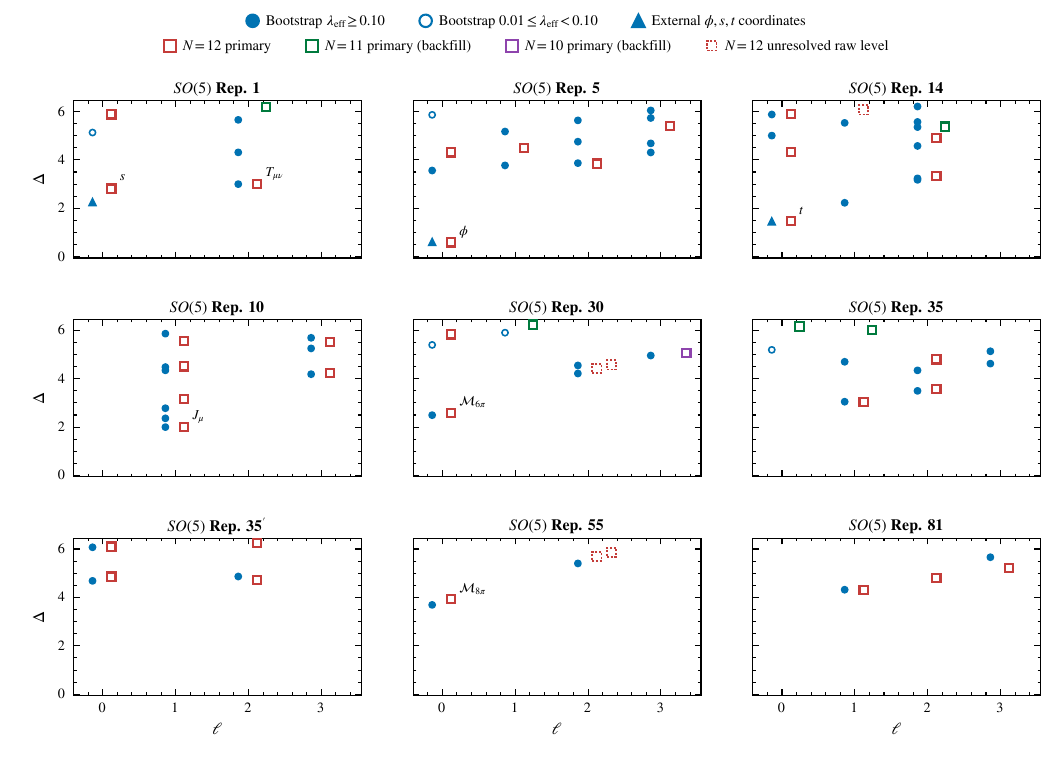}
\caption{Spectrum of the mixed $\phi$--$s$--$t$ bootstrap and the fuzzy sphere in the common $SO(5)$ channels, for $\ell\leq3$ and $\Delta\leq6.25$. Blue circles: exchanges of the $\Lambda=23$ bootstrap solution, filled for combined OPE-coefficient magnitude $\lambda_{\rm eff}\geq0.10$ (49 roots) and open for $0.01\leq\lambda_{\rm eff}<0.10$ (5 roots), where $\lambda_{\rm eff}$ combines the OPE coefficients for all allowed external pairs in the crossing normalization [Eq.~\eqref{eq:efmdiagnostics}]; blue triangles: optimized external dimensions. Solid squares: fuzzy-sphere primaries at $N=12$ (red, 32 levels), supplemented by $N=11$ (green, 5) and $N=10$ (purple, 1) where the $N=12$ continuation is not resolved. Dashed red squares: five unresolved levelts at $N=12$ (notes to Table~\ref{tab:unresolvedlevelweights}). 
Horizontal offsets separate the two methods.}
\label{fig:mixedspectrum}
\end{figure*}

Table~\ref{tab:spectrum} summarizes representative dimensions. The lowest rank-three and rank-four symmetric traceless scalars, in $\rep{30}$ and $\rep{55}$, contain the monopole components $\mathcal M_{6\pi}$ and $\mathcal M_{8\pi}$, where the subscripts label monopole flux. At $N=12$, the vector, tensor, and the two monopole multiplets differ from the bootstrap values by about $0.4\%$ to $6\%$, and the low-lying dimensions agree closely with the large-scale QMC estimates. The finite-size singlet remains above the bootstrap and QMC values and decreases with system size (Sec.~\ref{sec:fuzzyspectrum}). Figure~\ref{fig:mixedspectrum} extends the comparison to all common representation and spin channels, including higher primary families beyond the leading scalars.

On the fuzzy-sphere side, analysis using the special conformal generators $K_\mu$ identifies 52 nonvacuum primaries at $N=10$ and 44 at $N=12$ among the levels with $\ell\leq3$ and $\Delta\leq6.25$ (Appendix~\ref{app:reviewedN12}). The published $N=10$ spectrum of Ref.~\cite{Zhou2024} lists the levels with $\Delta\leq5.5$; among them we identify 18 primaries not assigned there, and 13 further primaries lie between $\Delta=5.5$ and $6.25$. Table~\ref{tab:newprimarymain} lists all 31, together with the nearest bootstrap exchanges in the 16 cases probed by the mixed system. Section~\ref{sec:fuzzyspectrum} describes the identification.

\begin{table*}[t]
\caption{Primaries at $N=10$ not assigned in Table~VI of Ref.~\cite{Zhou2024}, at $g=0.9437$. Upper rows: levels of that table ($\Delta\leq5.5$), with the assignment of Ref.~\cite{Zhou2024} $\to$ the present $K$-generator analysis (P primary, D descendant, U unassigned). Rows below $5.5<\Delta\leq6.25$: primaries above the published window. The last two rows are published primaries whose assignment is unresolved here. $\Delta_{\rm FS}=2(E-E_0)/(E_J-E_0)$; $\Delta_{\rm bootstrap}$ is the nearest $\Lambda=23$ root in the same channel with combined OPE-coefficient magnitude $\lambda_{\rm eff}\geq0.01$ [Eq.~\eqref{eq:efmdiagnostics}]. A dash marks a channel outside the mixed system.}
\label{tab:newprimarymain}
\centering\begingroup\small\setlength{\tabcolsep}{3.5pt}\renewcommand{\arraystretch}{1.06}
\begin{tabular}{@{}ccccc@{\hspace{14pt}}ccccc@{}}
\toprule
\multicolumn{5}{c}{Channels probed by the mixed bootstrap} &
\multicolumn{5}{c}{Channels outside the mixed system} \\
\cmidrule(r){1-5}\cmidrule(l){6-10}
\SOrephead & $\ell^{\mathcal P}$ & Assignment & $\Delta_{\mathrm{FS}}$ & $\Delta_{\mathrm{bootstrap}}$ &
\SOrephead & $\ell^{\mathcal P}$ & Assignment & $\Delta_{\mathrm{FS}}$ & $\Delta_{\mathrm{bootstrap}}$ \\
\midrule
$\rep{5}$ & $3^{-}$ & $\mathrm{U}\to\mathrm{P}$ & 5.382 & 5.732 & $\rep{10}$ & $1^{-}$ & $\mathrm{U}\to\mathrm{P}$ & 5.144 & --- \\
$\rep{10}$ & $1^{+}$ & $\mathrm{U}\to\mathrm{P}$ & 4.515 & 4.478 & $\rep{10}$ & $3^{-}$ & $\mathrm{U}\to\mathrm{P}$ & 4.418 & --- \\
$\rep{10}$ & $3^{+}$ & $\mathrm{U}\to\mathrm{P}$ & 5.497 & 5.688 & $\rep{14}$ & $2^{-}$ & $\mathrm{U}\to\mathrm{P}$ & 5.299 & --- \\
$\rep{14}$ & $2^{+}$ & $\mathrm{U}\to\mathrm{P}$ & 4.887 & 4.577 & $\rep{14}$ & $3^{-}$ & $\mathrm{U}\to\mathrm{P}$ & 5.500 & --- \\
$\rep{14}$ & $2^{+}$ & $\mathrm{U}\to\mathrm{P}$ & 5.376 & 5.356 & $\rep{35}$ & $1^{+}$ & $\mathrm{U}\to\mathrm{P}$ & 4.839 & --- \\
$\rep{30}$ & $3^{-}$ & $\mathrm{D}\to\mathrm{P}$ & 5.063 & 4.956 & $\rep{35}$ & $2^{+}$ & $\mathrm{U}\to\mathrm{P}$ & 4.805 & --- \\
$\rep{35}$ & $2^{-}$ & $\mathrm{U}\to\mathrm{P}$ & 4.787 & 4.342 & $\rep{35}$ & $3^{+}$ & $\mathrm{D}\to\mathrm{P}$ & 5.336 & --- \\
$\rep{35}^{\prime}$ & $0^{+}$ & $\mathrm{U}\to\mathrm{P}$ & 4.885 & 4.685 & $\rep{35}^{\prime}$ & $2^{-}$ & $\mathrm{D}\to\mathrm{P}$ & 5.169 & --- \\
$\rep{81}$ & $3^{+}$ & $\mathrm{U}\to\mathrm{P}$ & 5.185 & 5.660 & $\rep{81}$ & $2^{+}$ & $\mathrm{U}\to\mathrm{P}$ & 4.792 & --- \\
\multicolumn{5}{c}{\textit{$5.5<\Delta\leq6.25$}} & \multicolumn{5}{c}{\textit{$5.5<\Delta\leq6.25$}} \\
$\rep{1}$ & $0^{+}$ & $\mathrm{P}$ & 5.904 & 6.255 & $\rep{5}$ & $2^{+}$ & $\mathrm{P}$ & 5.606 & --- \\
$\rep{14}$ & $0^{+}$ & $\mathrm{P}$ & 5.923 & 5.876 & $\rep{10}$ & $3^{-}$ & $\mathrm{P}$ & 5.607 & --- \\
$\rep{30}$ & $0^{-}$ & $\mathrm{P}$ & 5.826 & 5.395 & $\rep{35}$ & $3^{+}$ & $\mathrm{P}$ & 5.712 & --- \\
$\rep{30}$ & $1^{-}$ & $\mathrm{P}$ & 6.247 & 5.900 & $\rep{105}$ & $2^{-}$ & $\mathrm{P}$ & 6.105 & --- \\
$\rep{35}$ & $0^{-}$ & $\mathrm{P}$ & 6.174 & 6.771 & $\rep{154}$ & $1^{-}$ & $\mathrm{P}$ & 5.760 & --- \\
$\rep{35}$ & $1^{-}$ & $\mathrm{P}$ & 6.012 & 6.856 & $\rep{154}$ & $2^{-}$ & $\mathrm{P}$ & 6.227 & --- \\
$\rep{35}^{\prime}$ & $0^{+}$ & $\mathrm{P}$ & 6.136 & 6.071 &  &  &  &  &  \\
\multicolumn{5}{c}{\textit{Published primaries now unresolved}} &  &  &  &  &  \\
$\rep{30}$ & $2^{-}$ & $\mathrm{P}\to\mathrm{U}$ & 4.400 & 4.546 &  &  &  &  &  \\
$\rep{35}$ & $3^{-}$ & $\mathrm{P}\to\mathrm{U}$ & 4.622 & 4.621 &  &  &  &  &  \\
\bottomrule
\end{tabular}\endgroup
\end{table*}

The OPE coefficients test the couplings between the identified operators. For external scalars $A,B$ and an exchanged primary $\mathcal O$ in representation $R$, matching the internal tensor normalizations gives the mixed-crossing coefficient
\begin{equation}
\beta_{AB\mathcal O}=\left(\frac{d_R}{d_A d_B}\right)^{1/4}
\lambda_{\rm iso}(A\otimes B\to\mathcal O),
\label{eq:lambdaconversionmain}
\end{equation}
where $d_A,d_B,d_R$ are representation dimensions and $\lambda_{\rm iso}$ uses isometric Clebsch--Gordan tensors. We quote all OPE coefficients in the $v$-singlet convention $\lambda$ of Ref.~\cite{LiShen2026}, extended from the vector channels to every channel by the same normalization rule: $\lambda=\lambda_{\rm iso}$ in general, $\lambda=\beta$ when a singlet takes part, and $\lambda=\lambda_{\rm iso}/\sqrt2$ for an exchanged $\rep{10}$. For example, $\lambda_{\phi\phi s}=\beta_{\phi\phi s}$, $\lambda_{\phi\phi t}=(25/14)^{1/4}\beta_{\phi\phi t}$, $\lambda_{\phi\phi J}=(5/8)^{1/4}\beta_{\phi\phi J}$, and $\lambda_{ttt}=14^{1/4}\beta_{ttt}$. Appendix~\ref{app:opeunit} lists the factor $\lambda/\beta$ for every channel.

\begin{table}[tbp]
\caption{OPE coefficients in all mixed correlator channels as specified in Appendix~\ref{app:opeunit}. Bootstrap: the $\Lambda=23$ solution with an isolated stress tensor. Fuzzy sphere: $\phi\phi s$, $stt$, and $sss$ use $s$ insertion, $\phi\phi t$ and $ttt$ use $t$ insertion, and $J$ and $T$ use $\phi$ insertion. The column $1/N\to0$ gives the intercept of a linear fit $a+b/N$ over $N=8$--12; a dash marks a quantity that is not extrapolated.}
\label{tab:ope}
\begingroup\small\setlength{\tabcolsep}{3pt}
\begin{ruledtabular}\begin{tabular}{lrrrrr}
Quantity & Bootstrap & $1/N\to0$ & $N=12$ & $11$ & $10$\\\hline
\rowstrut $\lambda_{\phi\phi s}/\lambda_{\phi\phi t}$ & 0.255 & --- & 0.203 & 0.204 & 0.204 \\
\rowstrut $\lambda_{ttt}/\lambda_{\phi\phi t}$ & 2.111 & --- & 2.118 & 2.121 & 2.125 \\
\rowstrut $\lambda_{stt}/\lambda_{\phi\phi t}$ & 0.594 & --- & 0.549 & 0.548 & 0.548 \\
\rowstrut $\lambda_{sss}/\lambda_{\phi\phi t}$ & 0.231 & --- & 0.256 & 0.218 & 0.171 \\
\rowstrut $|\lambda_{\phi\phi J}|$ & 0.768 & 0.781 & 0.762 & 0.760 & 0.758 \\
\rowstrut $|\lambda_{\phi\phi T}|$ & 0.350 & 0.363 & 0.273 & 0.264 & 0.254 \\
\end{tabular}\end{ruledtabular}\endgroup
\end{table}

Table~\ref{tab:ope} shows that the bootstrap and fuzzy-sphere OPE data are closely consistent. The $ttt$ and $stt$ ratios agree to $0.3\%$ and $8\%$, and $|\lambda_{\phi\phi J}|$ agrees to $0.7\%$; fixing $\Delta_J=2$ for the energy calibration does not impose this agreement. The ratio $\lambda_{\phi\phi s}/\lambda_{\phi\phi t}$ is $0.255$ in the bootstrap and $0.203$ at $N=12$. The stress-tensor coupling grows with $N$, from $0.254$ at $N=10$ to $0.273$ at $N=12$, and a linear $1/N$ fit gives $0.363$, close to the bootstrap value $0.350$; the current coupling extrapolates to $0.781$. 

An analytic benchmark comes from the large-flavor expansion of the QED$_3$--Gross--Neveu--Yukawa model~\cite{GuoLiCentral2026}. Extrapolating to two two-component fermion flavors gives $c_J\simeq1.78$ and $c_T\simeq8.29$, where $c_{J,T}=(4\pi)^2C_{J,T}$. In our conventions, the Ward identities give $|\lambda_{\phi\phi J}|=c_J^{-1/2}$ and $|\lambda_{\phi\phi T}|=3\Delta_\phi/(2\sqrt{c_T})$, yielding $0.749$ and $0.320$, respectively, with the latter evaluated at our bootstrap $\Delta_\phi=0.614$. These values are reasonably close to the bootstrap and extrapolated fuzzy-sphere results in Table~\ref{tab:ope}, although the large-flavor expansion is not controlled at two flavors.\par These data characterize the candidate multicritical theory beyond its critical exponents. The dimensions determine the growth or decay of perturbations under coarse graining, while $\lambda_{\phi\phi s}$, $\lambda_{\phi\phi t}$, $\lambda_{sss}$, and $\lambda_{stt}$ connect the order parameter to the singlet and tensor deformations and constrain their interactions. They provide the ingredients for the local response to symmetry-preserving and N\'eel--VBS deformations, and for connecting the CFT to the crossover seen in QMC (Sec.~\ref{sec:discussion}).

The singlet requires a separate convergence analysis because it is central to the RG interpretation. Its effective dimension decreases from $2.998$ at $N=5$ to $2.816$ at $N=12$ along the stress-matched sequence, and fixed-coupling scans show the same trend (Sec.~\ref{sec:fuzzy}). The data thus support a relevant singlet, while its limiting dimension remains less precisely determined than the bootstrap value $2.258$ or the QMC data $2.273$.
\subsection{The proposed \texorpdfstring{$SO(5)$}{SO(5)} multicritical CFT}
The combined results support a coherent interpretation of the three numerical settings. The bootstrap cone identifies a candidate conformal theory in the neighborhood selected by QMC. The agreement of its spectrum and OPE coefficients with the fuzzy-sphere results supports the proposed $SO(5)$ multicritical CFT. The relevant singlet then supplies the additional unstable direction required by multicriticality, while the mixed-bootstrap determination is compatible with the implemented unitary crossing conditions. The interpretation rests on this conjunction of the three approaches.

The tuning structure can be expressed explicitly. For the embedding of the N\'eel and VBS orders in Eq.~\eqref{eq:vector}, the tensor component that distinguishes the two orders is
\begin{equation}
\mathcal T=\sqrt{\frac65}\left(
\frac13\sum_{a=1}^{3}t_{aa}
-\frac12\sum_{a=4}^{5}t_{aa}\right),
\label{eq:anisotropy}
\end{equation}
where the positive factor gives unit two-point normalization in Eq.~\eqref{eq:norm}; the unnormalized traceless weight matrix has squared norm $5/6$.  The leading deformations are
\begin{equation}
S=S_*+u_s\int\dd^3x\,s+u_t\int\dd^3x\,\mathcal T+\cdots,
\label{eq:deformations}
\end{equation}
with RG eigenvalues $y_s=3-\Delta_s$ and $y_t=3-\Delta_t$. The tensor perturbation distinguishes N\'eel from VBS order; the singlet preserves $SO(5)$ while moving the system away from the fixed point. When both are relevant, a microscopic family allowing both generally requires at least two independent tunings to reach the point. Within an exactly $SO(5)$-symmetric family the tensor perturbation is forbidden, so the same theory can be reached with fewer tunings. This is how a critical point in the symmetric subspace acquires a multicritical role in the larger space of quantum-magnet Hamiltonians. The distinction underlies the phase diagrams explored by Chen \emph{et al.} and Liao \emph{et al.} through independent N\'eel and VBS couplings~\cite{Chen2024,Liao2026}.

A lattice trajectory that misses the singlet tuning can nevertheless pass close to the fixed point and share a substantial scaling regime with the fuzzy-sphere theory. Departure from this neighborhood can then lead to the weakly first-order behavior observed in QMC. The relevant singlet permits such a departure; it does not alone determine its eventual phase or force a first-order endpoint. Establishing the global arrangement of N\'eel, VBS, and possible spin-liquid phases requires the allowed additional perturbations and their flows. The comprehensive CFT data specify the local organizing theory and its interactions, providing a basis for that investigation.

These data strongly support a real, unitary multicritical CFT as the origin of the observed spectrum, operator interactions, and lattice scaling regime. Sections~\ref{sec:bootstrap} and \ref{sec:fuzzy} give the numerical analyses; Sec.~\ref{sec:discussion} develops the consequences for crossover and the phase diagram.
\section{Mixed-correlator conformal bootstrap}
\label{sec:bootstrap}
In our bootstrap studies, we first place the bootstrap cone in the global landscape of vector-correlator bounds, then introduce the mixed $\phi$--$s$--$t$ system and present the conformal data of the mixed-correlator solution.

\subsection{The single-correlator landscape and the bootstrap cone}
\begin{figure*}[tbp]
\centering
\begin{minipage}[t]{.495\textwidth}\centering
\includegraphics[width=\linewidth]{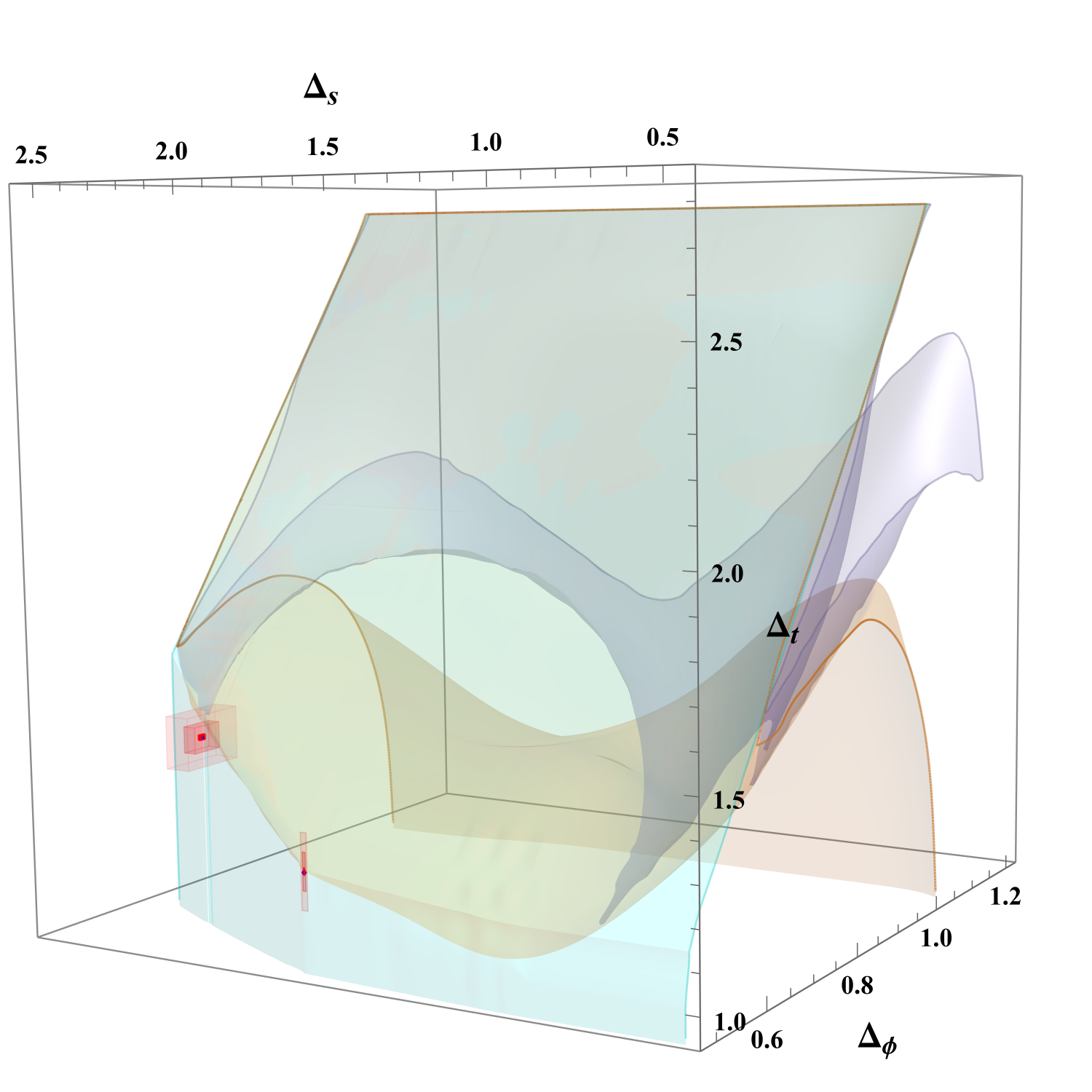}\\[-2pt]
\small (a) Global vector-correlator landscape
\end{minipage}\hfill
\begin{minipage}[t]{.495\textwidth}\centering
\includegraphics[width=\linewidth]{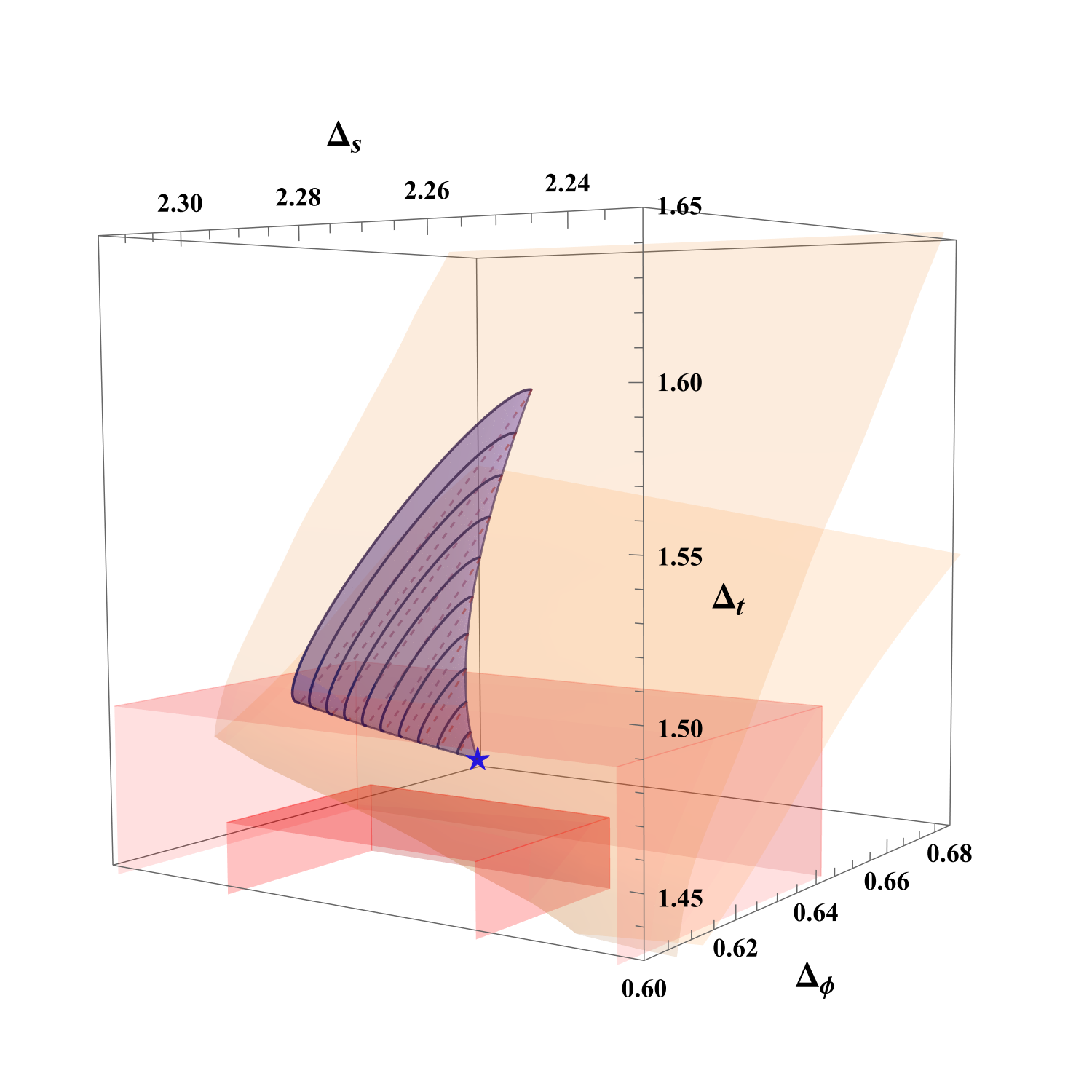}\\[-2pt]
\small (b) Zoom near the conic tip
\end{minipage}
\caption{Vector-correlator bootstrap regions at $\Lambda=27$ in $(\Delta_\phi,\Delta_s,\Delta_t)$. (a) Translucent cyan: no additional scalar gap; pale-orange: $\Delta_{s'},\Delta_{t'}\geq3$; purple: $\Delta_{s'},\Delta_{t'}\geq4$. Red nested rectangular error boxes indicate the quoted QMC estimates expanded by factors $1,5,10$ of their reported errors. The upper-left group has $(\Delta_\phi,\Delta_s,\Delta_t)=(0.607(4),2.273(4),1.417(7))$, the $J$--$Q$ estimates of Ref.~\cite{Takahashi2024}; the lower group represents the O(5) vector-model estimates of Refs.~\cite{HasenbuschPelissettoVicari2005,Hasenbusch2022}. The boxes expand the quoted coordinate errors separately along each axis. (b) Local view of the pale-orange region and the purple conic structure, with the cone computed at the same $\Lambda=27$ and gaps $(4,4)$. The larger and smaller red boxes show the $10\sigma$ and $5\sigma$ expansions of the $J$--$Q$ estimates. The dark-blue star marks the navigator candidate with minimal $\Delta_s$ in the singlet channel of $\langle\phi\phi\phi\phi\rangle$. }
\label{fig:landscape}
\end{figure*}

The vector four-point function probes the representations in
\begin{equation}
\rep{5}\otimes\rep{5}=\rep{1}\oplus\rep{14}\oplus\rep{10}.
\end{equation}
Bose symmetry permits even spacetime spin in the singlet and traceless symmetric channels and odd spin in the antisymmetric channel. We resolve the leading scalars $s$ and $t$ and impose gaps on the next scalar primaries,
\begin{equation}
\Delta_{s'}\geq g_s,\qquad \Delta_{t'}\geq g_t.
\label{eq:gaps}
\end{equation}
Figure~\ref{fig:landscape} shows how these assumptions organize the allowed region in $(\Delta_\phi,\Delta_s,\Delta_t)$ at $\Lambda=27$. The translucent cyan region has no additional scalar gaps; the pale-orange and purple regions have $(g_s,g_t)=(3,3)$ and $(4,4)$, respectively.

With the $(3,3)$ gaps, the critical $O(5)$ vector model is associated with a prominent kink in the global boundary. Its position agrees with the vector-model estimates of Refs.~\cite{HasenbuschPelissettoVicari2005,Hasenbusch2022}. This familiar feature provides a useful reference for interpreting the rest of the landscape: crossing and positivity distinguish a particular interacting theory through a sharp boundary structure. Raising both gaps to four removes the vector-model region and carves out the conic structure near the DQCP estimates. The cone is a prominent feature of the full landscape under these stronger assumptions, rather than a feature visible only after selecting a narrow slice. It therefore plays a role analogous to the vector-model kink, but for a candidate with larger gaps above its leading singlet and tensor scalars.

More remarkably, the tip of this cone lies close to the large-scale $J$--$Q$ estimates~\cite{Takahashi2024},
\begin{equation}
(\Delta_\phi,\Delta_s,\Delta_t)_{\rm QMC}
=(0.607(4),\,2.273(4),\,1.417(7)).
\end{equation}
The local view in Fig.~\ref{fig:landscape}(b) makes this proximity explicit. The red boxes show the quoted QMC errors and their enlargements. The blue star minimizes $\Delta_s$ in the vector-correlator bootstrap and marks a representative solution at the conic tip, which we follow with mixed correlators.

Our earlier work identified the cone and compared its vector-channel data with QMC and the fuzzy sphere~\cite{LiShen2026}. The sharp feature and its proximity to QMC locate the candidate; the mixed system below supplies the additional spectrum and OPE coefficients that test it.
\subsection{Mixed correlators, symmetry sectors, and the OPE scan}
We take real scalar primaries $\phi_i$, $s$, and $t_{ij}$ in the $SO(5)$ representations $\rep5$, $\rep1$, and $\rep{14}$, with
\begin{align}
\langle\phi_i(x)\phi_j(0)\rangle&=\frac{\delta_{ij}}{|x|^{2\Delta_\phi}},
\qquad \langle s(x)s(0)\rangle=\frac{1}{|x|^{2\Delta_s}},\nonumber\\
\langle t_{ij}(x)t_{kl}(0)\rangle&=\frac{\Pi_{ij,kl}}{|x|^{2\Delta_t}},\nonumber\\
\Pi_{ij,kl}&=\frac{\delta_{ik}\delta_{jl}+\delta_{il}\delta_{jk}}{2}
-\frac{\delta_{ij}\delta_{kl}}5.
\label{eq:norm}
\end{align}
The coupled four-point functions have the external-field multisets
\begin{equation}
\begin{gathered}
\phi\phi\phi\phi,\ ssss,\ tttt,\ \phi\phi ss,\ \phi\phi tt,\\
sstt,\ \phi\phi st,\ sttt,
\end{gathered}
\end{equation}
with the independent orderings included in crossing.

Table~\ref{tab:mixedselection} gives the exchanged primary sectors for every external pair. We use the physical parity convention of the fuzzy-sphere study~\cite{Zhou2024}, $\mathcal P_\phi=-1$ and $\mathcal P_s=\mathcal P_t=+1$. Since the three-point tensor of two scalars and a spin-$\ell$ primary contains no Levi-Civita tensor, a nonzero coupling requires
\begin{equation}
\mathcal P_{\mathcal O}=\mathcal P_X\mathcal P_Y
\quad\hbox{in }X\times Y,
\label{eq:mixedparity}
\end{equation}
independently of the even/odd-spin restriction for identical external operators.

\begin{table}[tb]
\caption{Primary sectors accessible to the $\phi$--$s$--$t$ system. $\mathcal P$ is intrinsic spacetime parity in the convention of Ref.~\cite{Zhou2024}; it is distinct from exchange symmetry. ``All'' means all nonnegative integer spins. The rows for each pair specify its complete representation, spin, and parity selection rules.}
\label{tab:mixedselection}
\begin{ruledtabular}
\begin{tabular}{lccc}
Pair & $SO(5)$ representations & $\mathcal P$ & $\ell$ \\
\hline
$\phi\phi$ & $\rep1,\rep{14}$ & $+$ & even \\
$\phi\phi$ & $\rep{10}$ & $+$ & odd \\
$ss$ & $\rep1$ & $+$ & even \\
$tt$ & $\rep1,\rep{14},\rep{35}',\rep{55}$ & $+$ & even \\
$tt$ & $\rep{10},\rep{81}$ & $+$ & odd \\
$s\phi$ & $\rep5$ & $-$ & all \\
$st$ & $\rep{14}$ & $+$ & all \\
$t\phi$ & $\rep5,\rep{30},\rep{35}$ & $-$ & all \\
\end{tabular}
\end{ruledtabular}
\end{table}

The new products include
\begin{align}
\rep5\otimes\rep{14}&=\rep5\oplus\rep{30}\oplus\rep{35},\nonumber\\
\operatorname{Sym}^2(\rep{14})&=\rep1\oplus\rep{14}\oplus\rep{35}'\oplus\rep{55},\nonumber\\
\wedge^2(\rep{14})&=\rep{10}\oplus\rep{81}.
\label{eq:mixedproducts}
\end{align}
Mixing thus accesses nine irreducible representations. Relative to the vector-only problem, it adds $\rep5$, $\rep{30}$, $\rep{35}$, $\rep{35}'$, $\rep{55}$, and $\rep{81}$, as well as odd-spin $\rep{14}^{+}$ through $st$; the two dimension-35 representations have $B_2$ labels $[1,2]$ and $[0,4]$. In particular, $t\phi$ and $tt$ probe the rank-three and rank-four symmetric traceless multiplets containing $\mathcal M_{6\pi}$ and $\mathcal M_{8\pi}$. Primaries in other sectors, such as the parity-odd singlet and the rank-five $\rep{91}$ scalar of Ref.~\cite{Zhou2024}, do not appear in this system.

The crossing equations have the matrix form
\begin{equation}
\vec V_{\mathds1}+\sum_{\mathcal R,\ell,\mathcal O}
\bm\beta_{\mathcal O}^{T}\vec V^{\mathcal R}_{\Delta,\ell}
\bm\beta_{\mathcal O}=0,
\label{eq:crossing}
\end{equation}
where the identity is excluded from the sum and $\bm\beta_{\mathcal O}$ collects the mixed-crossing OPE coefficients of Eq.~\eqref{eq:lambdaconversionmain}. A singlet exchange carries $(\beta_{\phi\phi\mathcal O},\beta_{ss\mathcal O},\beta_{tt\mathcal O})$, an even-spin $\rep{14}$ carries $(\beta_{\phi\phi\mathcal O},\beta_{st\mathcal O},\beta_{tt\mathcal O})$, and a $\rep5$ exchange couples to both $s\phi$ and $t\phi$. Mixing therefore constrains the relative couplings of the same operator across different correlators. Reflection positivity requires positive semidefinite OPE matrices, and the dimensions obey the three-dimensional unitarity bounds.

The five independent couplings among the external scalars,
\begin{equation}
\bm\beta_{\rm ext}=
(\beta_{\phi\phi s},\beta_{\phi\phi t},
\beta_{sss},\beta_{stt},\beta_{ttt})^{T},
\label{eq:externalopevector}
\end{equation}
appear in several channels tied by permutation symmetry; for instance, $\beta_{\phi\phi s}$ enters $s$ exchange in $\phi\phi$ and $\phi$ exchange in $s\phi$. We scan the direction of this vector~\cite{ChesterSu2024}, defining
\begin{gather}
r_{XYZ}=\beta_{XYZ}/\beta_{\phi\phi t},\nonumber\\
\bm r=(r_{\phi\phi s},1,r_{sss},r_{stt},r_{ttt})^{T},\qquad
\bm\beta_{\rm ext}=\beta_{\phi\phi t}\bm r.
\label{eq:operatioscan}
\end{gather}
At fixed external dimensions and ratios, the resolved $\phi,s,t$ exchanges contribute
\begin{equation}
\beta_{\phi\phi t}^{\,2}\,
\bm r^{T}\vec V_{\rm ext}(\Delta_\phi,\Delta_s,\Delta_t)\bm r,
\label{eq:externalscanblock}
\end{equation}
and a semidefinite program tests each trial point~\cite{SDPB}. We vary the three dimensions and four ratios jointly to minimize $\Delta_s$ near the cone, using Skydiving at $\Lambda=19$ and $23$~\cite{Skydiving2023}. The extremal functional at the optimized point then yields the higher spectrum and its OPE coefficients, including those of the current and stress tensor.
\subsection{The mixed-correlator candidate and its conformal data}
The mixed search follows the conic region using the principal gaps
\begin{equation}
\Delta_{\phi'}\geq3,\qquad \Delta_{s'}\geq4,\qquad \Delta_{t'}\geq4,
\label{eq:mixedprincipal}
\end{equation}
with the remaining channel assumptions listed in Table~\ref{tab:channels}. Minimizing $\Delta_s$ while optimizing the four OPE ratios selects the lower tip of the cone. At $\Lambda=23$,
\begin{equation}
(\Delta_\phi,\Delta_s,\Delta_t)
=(0.613735,\,2.258388,\,1.464017).
\label{eq:mixedcandidate}
\end{equation}
Including the mixed correlators leaves the candidate in the conic region found with the vector correlator, while attaching much richer conformal data to it. The external dimensions and OPE ratios are determined together, so the solution fixes the interactions among the leading deformation operators as well as their dimensions; $r_{\phi\phi s}$ couples the order parameter to the singlet direction, while $r_{sss}$ and $r_{stt}$ probe interactions that a vector-only correlator cannot determine.

\begin{table}[tb]
\caption{Mixed-bootstrap candidate coordinates and external OPE ratios. The principal gaps are $g_\phi=3$ and $g_s=4$ in all columns. All three candidates use Skydiving. The two $\Lambda=19$ runs differ in arithmetic precision. Entries are rounded numerical estimates, not uncertainty intervals.}
\label{tab:mixedstability}
\begin{ruledtabular}
\begin{tabular}{lrrr} 
& \shortstack{$\Lambda=19$\\$g_t=4$} & \shortstack{$\Lambda=23$\\$g_t=4$} & \shortstack{$\Lambda=19$\\$g_t=3.95$} \\
\hline
$\Delta_\phi$ & 0.607 & 0.614 & 0.599 \\
$\Delta_s$ & 2.240 & 2.258 & 2.199 \\
$\Delta_t$ & 1.446 & 1.464 & 1.422 \\
$r_{\phi\phi s}$ & 0.294 & 0.294 & 0.296 \\
$r_{ttt}$ & 1.260 & 1.262 & 1.254 \\
$r_{stt}$ & 0.685 & 0.687 & 0.679 \\
$r_{sss}$ & 0.268 & 0.268 & 0.257 \\
\end{tabular}
\end{ruledtabular}
\end{table}

Table~\ref{tab:mixedstability} shows that the solution is stable. Between $\Lambda=19$ and $23$, the external dimensions change by about $1\%$ and the four OPE ratios by less than $0.35\%$. Lowering the tensor-scalar gap to $g_t=3.95$ at $\Lambda=19$ shifts the dimensions by $1.3$--$1.8\%$, and the singlet remains clearly relevant, with $\Delta_s\simeq2.20$--$2.26$ across all candidates. Solver settings and the $\Lambda=15$ record are given in Appendix~\ref{app:bootstrapdetails}.

The extremal functional at this point supplies the higher dimensions and coupling vectors in all sectors of Table~\ref{tab:mixedselection}. To determine the stress-tensor coupling, we solve the crossing problem at the same external point with $T_{\mu\nu}$ as an isolated exchange at $\Delta=3$. Its couplings to $\phi\phi$, $ss$, and $tt$ are fixed by the Ward identity to be proportional to $(\Delta_\phi,\Delta_s,\Delta_t)$, and the remaining spin-two singlet spectrum starts above $3$. This gives $|\lambda_{\phi\phi T}|=0.350$ and leaves the rest of the low-lying spectrum unchanged at the displayed precision. We use this solution as the bootstrap reference throughout.

The rank-three and rank-four scalar multiplets have $\Delta_{\mathcal M_{6\pi}}=2.493$ and $\Delta_{\mathcal M_{8\pi}}=3.692$, with couplings $|\lambda_{t\phi\mathcal M_{6\pi}}|=1.426$ and $|\lambda_{tt\mathcal M_{8\pi}}|=1.783$. Further spinning exchanges occur in the new $\rep{35}$, $\rep{35}'$, and $\rep{81}$ channels, and even the original vector channels acquire couplings to $ss$ and $tt$. This enlarged operator algebra is the principal bootstrap result of the present work. Figure~\ref{fig:mixedspectrum} and Tables~\ref{tab:spectrum} and \ref{tab:ope} compare it with the fuzzy sphere and QMC; the complete root weights are listed in Table~\ref{tab:efmfull}.

The cone solution is distinct from the extremal candidate studied in \cite{ChesterSu2024}, which used the same correlator system with $\Delta_\phi=0.630$ fixed from a large-$N$ monopole estimate and maximized $\Delta_t$ at $\Lambda=19$, obtaining $\Delta_s=2.359$ and $\Delta_t=1.519$. Here the cone of Ref.~\cite{LiShen2026} and its proximity to QMC determine the search region, and all three external dimensions and four ratios are optimized jointly. The two solutions have different external OPE ratios, and the earlier candidate does not reproduce the fuzzy-sphere couplings of Table~\ref{tab:ope}.
\section{Fuzzy-sphere calculation}
\label{sec:fuzzy}
\begingroup
\makeatletter
\setlength{\@fptop}{0pt}\setlength{\@fpsep}{14pt}\setlength{\@fpbot}{0pt plus 1fil}
\makeatother
\renewcommand{\topfraction}{0.98}\renewcommand{\dbltopfraction}{0.98}
\renewcommand{\floatpagefraction}{0.70}\renewcommand{\dblfloatpagefraction}{0.70}
\renewcommand{\textfraction}{0.02}\setcounter{topnumber}{4}\setcounter{dbltopnumber}{4}\setcounter{totalnumber}{6}

In our fuzzy sphere studies, we adopt the fuzzy sphere implementation for DQCP proposed in Ref.~\cite{Zhou2024} with two additions: symmetry projection, which reaches $N=12$ orbitals, and the use of special conformal generators $K_\mu$ to separate primaries from descendants. We describe the method, the conformal checks, the spectrum and OPE coefficients used in Sec.~\ref{sec:overview}. 

\subsection{Symmetry projection and primary identification}
\label{sec:fuzzymethods}
We use the SO(5)-symmetric Hamiltonian of Ref.~\cite{Zhou2024}, with four fermion flavors in the half-filled lowest Landau level on a sphere ($N_e=2N$ for $N$ orbitals). Its long-wavelength description is the SO(5) nonlinear sigma model with a level-1 WZW term. The density pseudopotential sets the energy unit, and $g=V/U$ is the ratio of pairing to density couplings (Appendices~\ref{app:fuzzysetup} and \ref{app:fuzzyWZW}).

We diagonalize the Hamiltonian in blocks of spatial spin $\ell$ and internal subgroup spins $(s_A,s_B)$ of $\mathrm{SU}(2)_A\times\mathrm{SU}(2)_B\subset\mathrm{Spin}(5)$, resolving the exchange eigenvalue $X$ when $s_A=s_B$~\cite{BaranExactL2026,ZhouHeSO3ED2026}. The SO(5) Casimir distinguishes representations, and the particle--hole transformation gives the physical parity $\mathcal P$. This exact block reduction extends the calculation to $N=12$ (Appendix~\ref{app:symmetryprojectors}).

Energies and quantum numbers alone do not distinguish primaries from descendants. In radial quantization, a primary satisfies $K_\mu|\cO\rangle=0$ for every special conformal generator. Following Refs.~\cite{FanGenerators2024,FardelliGenerators2024,FardelliOPE2026}, we construct the generators from angular moments of the energy density and form
\begin{equation}
Q_K=\sum_\mu K_\mu^\dagger K_\mu,\qquad
\langle\psi|Q_K|\psi\rangle=\sum_\mu\|K_\mu|\psi\rangle\|^2 .
\label{eq:Knorm}
\end{equation}
The small-$Q_K$ subspace defines the primary space in each symmetry sector. An energy eigenstate is classified as a primary when its squared projection $w_P$ onto this space is at least $0.90$, and as a descendant when its projection onto the descendants of a single parent family reaches the same value. Strict primaries pass this test at three values of the classification threshold $\kappa_c$ and for reduced energy windows (Appendix~\ref{app:Kprojectors}). 

For the spectral comparison, we normalize the SO(5) current to $\Delta_J=2$, giving $\Delta_{\cO}^{\rm eff}=2(E_{\cO}-E_0)/(E_J-E_0)$, and tune $V/U$ at each size so that the stress tensor has $\Delta_T=3$, following Ref.~\cite{Zhou2024}. This defines the calibration sequence $g_*(N)$; the remaining energies and matrix elements are predictions.
\begin{figure}[!t]
\centering\includegraphics[width=\linewidth]{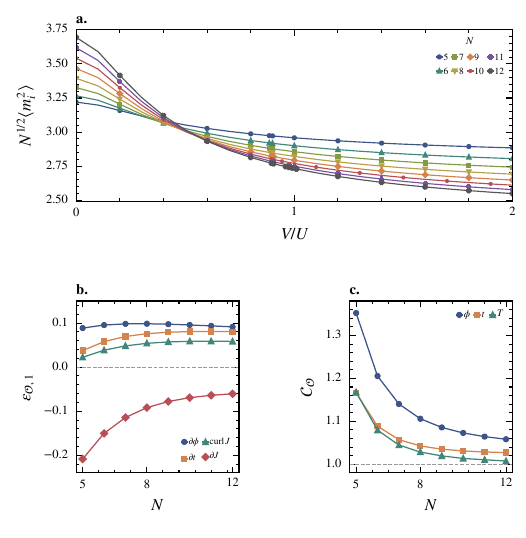}
\caption{Conformal diagnostics through $N=12$. (a) Single-component order parameter $N^{1/2}\langle m_i^2\rangle$, with $m_i=N^{-1}\sum_m c_m^\dagger\Gamma_i c_m$ for a fixed SO(5) component $i$, scanned in $g$ for $N=5$--12.  (b) At $g_*(N)$, first-descendant spacing residuals $\epsilon_{\cO,1}=\Delta_{\rm desc}^{\rm eff}-\Delta_{\cO}^{\rm eff}-1$, relative to the dashed zero line. The branches are $\partial\phi:(\rep5,1^-)$, $\partial t:(\rep{14},1^+)$, $\mathrm{curl}\,J:(\rep{10},1^-)$, and the spin-two derivative of $J:(\rep{10},2^+)$. (c) Raising-norm ratios $\mathcal C_{\cO}=\sum_\mu\|P_\mu|\cO\rangle\|^2/(6\Delta_{\cO})$ for $\phi,t,T_{\mu\nu}$ at the matching points; the dashed line is 1.} 
\label{fig:fuzzyvalidation}
\end{figure}
\subsection{Finite-size tests of conformal symmetry}
\label{sec:fuzzyvalidation}
Following Ref.~\cite{Zhou2024}, we locate the spectroscopy regime from the size dependence of the SO(5) vector order parameter $m_i=N^{-1}\sum_m c_m^\dagger\Gamma_i c_m$ for a fixed component $i$. The reversal of the size ordering of $N^{1/2}\langle m_i^2\rangle$ across the coupling scan places $g_*(N)$ in the critical regime of the accessible spheres [Fig.~\ref{fig:fuzzyvalidation}(a)].

At the stress-matched couplings, the first-descendant gaps of $\phi$, $t$, and $J$ deviate from unity by less than $0.1$ at $N=12$ [Fig.~\ref{fig:fuzzyvalidation}(b)], testing energies beyond the current and stress anchors. Generator matrix elements give a complementary test~\cite{FanGenerators2024,FardelliGenerators2024}: for a normalized primary in three dimensions,
\begin{equation}
\mathcal C_{\cO}=\frac{\sum_\mu\|P_\mu|\cO\rangle\|^2}{6\Delta_{\cO}}=1.
\label{eq:raisingnormtest}
\end{equation}
The ratios for $\phi$, $t$, and $T_{\mu\nu}$ approach unity with increasing size and reach $1.058$, $1.027$, and $1.008$ at $N=12$ [Fig.~\ref{fig:fuzzyvalidation}(c)]; the singlet fixes the generator normalization. These checks establish approximate conformal organization of the spectrum used below.
\subsection{Operator spectrum and finite-size evolution}
\label{sec:fuzzyspectrum}
Among the levels with $\ell\leq3$ and $\Delta\leq6.25$, analysis using the special conformal generators $K_\mu$ identifies 52 nonvacuum primaries at $N=10$, 50 at $N=11$, and 44 at $N=12$. At $N=10$ and $g=0.9437$ we can compare with Table~VI of Ref.~\cite{Zhou2024}, which lists the 129 levels with $\Delta\leq5.5$ and assigns 23 of them as primaries using quantum numbers and approximately integer level spacings. In this window we identify 39 primaries: 21 of the published ones, 15 previously unassigned levels, and three levels previously classified as descendants; two published primaries remain unresolved.\footnote{We count the P labels in Table~VI of Ref.~\cite{Zhou2024}, which gives 23 primaries and 76 descendants, consistent with Sec.~III~C of that work.} A further 13 primaries lie between $\Delta=5.5$ and $6.25$. Both analyses give the same dimensions at the displayed precision. The complete $N=12$ assignments are listed in Table~\ref{tab:N12fullspectrum}.

Sixteen of the 31 additional primaries in Table~\ref{tab:newprimarymain} lie in channels probed by the mixed bootstrap; the others have parity or spin outside those channels. For example, the $(\rep{14},2^+)$ primary at $5.376$ and the $(\rep{10},1^+)$ primary at $4.515$ lie within $0.4\%$ and $0.8\%$ of bootstrap exchanges, and the $(\rep{14},0^+)$ and $(\rep{35}',0^+)$ primaries above the published window lie within about $1\%$. Twelve of the 18 primaries within the published window have $N=12$ continuations, whose dimensions change by less than $0.7\%$ from $N=10$ in the six families probed by the bootstrap (Appendix~\ref{app:publishedcomparison}).

\paragraph{The lowest singlet.}
The relevance of $s$ determines whether the fixed point has an additional SO(5)-preserving unstable direction. Along the stress-matched sequence in Table~\ref{tab:singletcomparison}, $\Delta_s^{\rm eff}$ decreases from $2.998$ at $N=5$ to $2.816$ at $N=12$. At fixed coupling, where only the current fixes the energy scale, the singlet dimension also decreases monotonically with $N$ (Fig.~\ref{fig:fixedgsinglet}). A relevant singlet, together with the relevant N\'eel--VBS tensor deformation, supplies the two unstable directions of the proposed multicritical point. The downward drift alone does not distinguish this interpretation from complex-fixed-point walking~\cite{Zhou2024}; its significance comes from the accompanying agreement of the primary spectrum and OPE data with the unitary mixed-bootstrap solution.

\begin{table}[tbp]
\caption{The tuned finite-size singlet sequence, bootstrap candidate, and QMC estimate. The fuzzy-sphere rows give finite-size measurements at $g_*(N)$; the bootstrap row gives the separately optimized mixed $\phi$--$s$--$t$, $\Lambda=23$ candidate. Values are rounded to the displayed precision; the QMC uncertainty is given in parentheses.}
\label{tab:singletcomparison}
\begingroup\small\setlength{\tabcolsep}{7pt}
\begin{ruledtabular}\begin{tabular}{lrl}
$N$ & $g_*(N)$ & $\Delta_s$\\\hline
\rowstrut 5 & 0.8643 & 2.998 \\
\rowstrut 6 & 0.8614 & 2.936 \\
\rowstrut 7 & 0.8716 & 2.894 \\
\rowstrut 8 & 0.8904 & 2.865 \\
\rowstrut 9 & 0.9150 & 2.845 \\
\rowstrut 10 & 0.9437 & 2.831 \\
\rowstrut 11 & 0.9753 & 2.822 \\
\rowstrut 12 & 1.0092 & 2.816 \\
\hline
\multicolumn{2}{l}{$v$-$s$-$t$ bootstrap, $\Lambda=23$} & 2.258 \\
\multicolumn{2}{l}{QMC ($J$--$Q$)~\cite{Takahashi2024}} & 2.273(4) \\
\end{tabular}\end{ruledtabular}\endgroup
\end{table}

\begin{figure}[tbp]
\centering\includegraphics[width=\columnwidth]{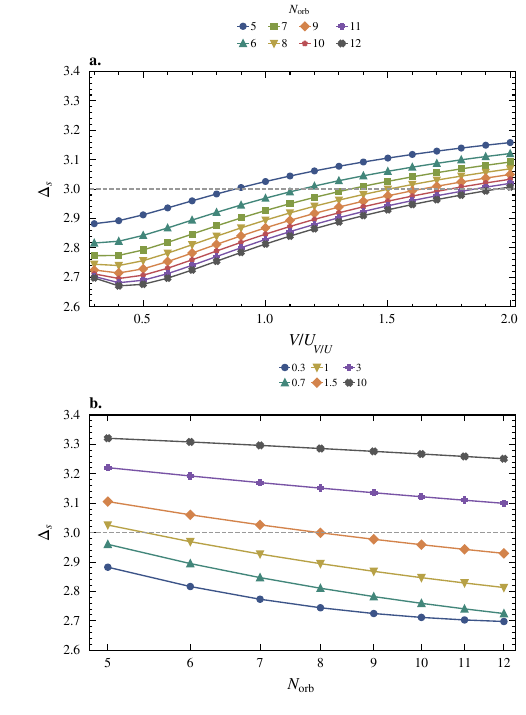}
\caption{Fixed-coupling singlet dimensions for $N=5$--12. At each $(N,g)$, the lowest nonvacuum scalar with $C_2=0$, $\mathcal P=+$, and $X=+$ defines $E_s$, with $\Delta_s^{\rm eff}=2(E_s-E_0)/(E_J-E_0)$ using that point\textquotesingle s current gap. (a) Coupling scans, with color and shape identifying $N$. (b) Size sequences at fixed $g$, with a logarithmic $N$ axis and color identifying $g$. Dashed lines mark marginality, $\Delta_s=3$; solid lines connect measurements.}
\label{fig:fixedgsinglet}
\end{figure}

\subsection{OPE coefficients from matrix elements}
\label{sec:fuzzyope}
The cylinder matrix element of a microscopic insertion between two primary states estimates the corresponding three-point coefficient. For input state $X$, microscopic insertion $Y$ of angular rank $r$, and output state $Z$,
\begin{equation}
\lambda_{\rm iso}(X\otimes Y\to Z)=
\sqrt{\frac{(2r+1)!!}{r!}}\frac{m_{ZX}^{(Y)}}{Z_Y C_{\rm iso}},
\label{eq:extractionmain}
\end{equation}
where $m_{ZX}^{(Y)}$ is the normalized microscopic matrix element, $Z_Y$ the vacuum-to-state overlap of the insertion, and $C_{\rm iso}$ the internal isometric Clebsch--Gordan component. The singlet contraction subtracts the disconnected term $\langle0|O_s|0\rangle\langle Z|X\rangle$. The resulting $\lambda_{\rm iso}$ is converted to the convention $\lambda$ of Table~\ref{tab:ope} by the same channel factors as the bootstrap data (Appendix~\ref{app:opeunit}). Different insertions estimate the same coefficient with different finite-size corrections; Table~\ref{tab:ope} specifies the insertion used for each entry, and Appendix~\ref{app:opeextractiondetails} compares the alternatives.
\section{Discussion and outlook}
\label{sec:discussion}

\subsection{Multicritical interpretation and remaining questions}
The combined spectrum and OPE data strongly support an $SO(5)$ multicritical interpretation of the DQCP scaling regime. The mixed $\phi$--$s$--$t$ bootstrap approximately preserves the cone identified in the vector correlator~\cite{LiShen2026}, while resolving substantially more of its operator content. Its low-lying dimensions are consistent with the large-scale $J$--$Q$ estimates~\cite{Takahashi2024}, and many higher dimensions and couplings agree with the fuzzy-sphere results. This extended comparison tests the operator interactions and conformal families associated with the cone, beyond the leading exponents that initially connected it to quantum magnets.

The independent identification of additional primaries is particularly significant. Applying the special conformal generators~\cite{FanGenerators2024,FardelliGenerators2024,FardelliOPE2026} resolves 52 primaries at $N=10$ and 44 at $N=12$ in the same spectral window, including levels whose primary character was not established in the earlier fuzzy-sphere study~\cite{Zhou2024}. Their classification does not use proximity to bootstrap dimensions, so agreement in the shared symmetry channels provides an independent test of the cone's physical interpretation. Together with the OPE comparisons, these results support identifying the two numerical descriptions with the same CFT: the dimensions locate operator families, and their couplings to different external fields test the identification. An accidental proximity of energy levels would not explain consistent three-point data in several channels.

Symmetry projection extends the calculation to $N=12$ Landau orbitals. The lowest $SO(5)$ singlet dimension decreases with larger system size, supporting its relevance. Its finite-size values remain above the strongly relevant bootstrap singlet, and determining the continuum limit from larger systems is an important quantitative test.

This distinction also matters for the comparison with pseudocriticality. In the fixed-point collision mechanism, the direction responsible for walking becomes nearly marginal near the collision~\cite{Kaplan2009,Gorbenko2018,MaWang2020,Nahum2020}. The strongly relevant bootstrap singlet is difficult to identify with that direction. More general complex CFTs can contain relevant singlets, and a Hermitian finite-size Hamiltonian has real energies even in a walking regime, so neither approximate conformal towers nor a decreasing singlet dimension alone excludes such explanations. A complex-fixed-point interpretation would, however, need to account quantitatively for the enlarged spectrum and OPE comparison, together with the unitary mixed-crossing solution and its stability under the numerical checks in Sec.~\ref{sec:bootstrap}.

The resulting evidence favors a real, unitary multicritical theory. Independently resolved primary families and three-point coefficients, including the current and stress-tensor couplings, now provide a broad basis for testing the proposed theory well beyond a comparison of a few critical exponents.
\subsection{Consequences for QMC and the phase diagram}
A real multicritical point can organize the scaling regime of a lattice transition that is ultimately weakly first order. The two leading deformations in Eq.~\eqref{eq:deformations} have distinct roles: the singlet preserves $SO(5)$, whereas the tensor component distinguishes N\'eel and VBS order. A one-parameter lattice trajectory can tune the latter while retaining a singlet perturbation, approach the fixed point, and subsequently leave its scaling regime. Within an exactly $SO(5)$-symmetric Hamiltonian the tensor direction is forbidden; if no other relevant perturbation is allowed, one tuning can reach the same point in that subspace. The anisotropic phase diagram and spectroscopy of Chen \emph{et al.}, and the larger-system continuum QMC study of Liao \emph{et al.}, provide complementary evidence for this tuning structure~\cite{Chen2024,ChenConformal,Liao2026}.

The CFT data make this interpretation quantitative near the fixed point. For singlet and tensor scaling fields $u_s,u_t$, the leading finite-size variables are $u_sL^{y_s}$ and $u_tL^{y_t}$, with $y_s=3-\Delta_s$ and $y_t=3-\Delta_t$. If the singlet controls departure from the scaling regime, the local crossover length obeys
\begin{equation}
\xi_{\mathrm{cross}}\sim |u_s|^{-1/y_s}.
\label{eq:crossoverscale}
\end{equation}
Here lengths are measured in a fixed microscopic unit; the mapping from lattice interactions to $u_s$ and the proportionality factor are nonuniversal. This relation describes how proximity to the multicritical surface can extend the apparent critical regime.  

Connecting this local description to QMC requires identifying which combinations of microscopic interactions couple to the singlet and tensor. Dangerously irrelevant anisotropies can introduce additional scales~\cite{Shao2016,Takahashi2024}. The enlarged operator data provide a basis for separating these effects and testing whether different microscopic trajectories can be described by deformations of the same multicritical theory. A useful test is to vary an additional symmetry-preserving lattice interaction while retuning the N\'eel--VBS transition. If the trajectories pass near the proposed fixed point, changing their singlet overlap should alter the extent of the scaling window and the leading responses of observables. Comparing several trajectories would then test the deformation picture more directly than fitting each drifting exponent separately. 

The OPE coefficients supply information beyond these linearized exponents. For example, $\lambda_{\phi\phi s}$ and $\lambda_{\phi\phi t}$ enter the order parameter's response to the two deformations, while $\lambda_{sss}$ and $\lambda_{stt}$ constrain their interactions. With microscopic matching and a specified renormalization prescription, conformal perturbation theory can use these data to study responses and crossover functions near the fixed point. 

The possible connection to quantum spin liquids is especially interesting. Extended square-lattice models exhibit proposed spin-liquid regimes adjoining deconfined transitions~\cite{LiuQSL2022,LiuQSL2024}, and fuzzy-sphere and continuum QMC studies report neighboring disordered regimes~\cite{Chen2024,Liao2026}. The relevant singlet supplies a direction in which to explore such phases, but does not specify their infrared character. Establishing that connection requires following the flow and testing fractionalization, topological order, or gapless excitations. The Wess--Zumino--Witten structure and microscopic embedding remain essential to the deconfined interpretation~\cite{TanakaHu2005,SenthilFisher2006,WangDualities2017}; the local CFT data constrain how this structure is realized near the proposed multicritical point.
\subsection{Outlook}
The immediate numerical priorities follow directly from the two fuzzy-sphere improvements developed here. Larger symmetry-resolved calculations can test whether the singlet's downward trend persists and determine its continuum dimension. Applying the special conformal generators to additional levels, and measuring their matrix elements, can sharpen the identification of bootstrap-predicted families and test more of the operator algebra. On the bootstrap side, higher resolution and further checks of the spectral assumptions can assess the stability of the candidate and resolve uncertain reconstructed amplitudes. These tests would strengthen the comparison beyond the leading spectrum and external-scalar OPE ratios.

We also plan to include the conserved $SO(5)$ currents $J_\mu^{[ab]}$ and stress tensor $T_{\mu\nu}$ as external bootstrap operators. Their mixed correlators with $\phi,s,t$ access further OPE structures, while conservation and Ward identities relate the data to $C_J$, $C_T$, and conserved-operator three-point functions. Relevant precedents are the three-dimensional Ising scalar--stress-tensor bootstrap~\cite{ChangIsingStress2025}, the $O(2)$ scalar--current and mixed current--stress-tensor studies~\cite{ReehorstScalarCurrent2020,DommesCurrentStress2026}, and the four-dimensional Abelian-current bootstrap~\cite{KarateevCurrents4D2026}. The three-dimensional non-Abelian current bootstrap is particularly relevant to an $SO(5)$ extension~\cite{HeNonAbelianCurrents2024}. Reliable conserved-sector amplitudes would supply additional fuzzy-sphere comparisons and connect to large-flavor estimates in related gauge theories~\cite{GuoLiCentral2026}.


The central application is to use the enlarged spectrum and OPE data to explain the QMC scaling regime and investigate neighboring phases through controlled deformations. The independently identified higher primaries provide new evidence for the cone's interpretation, while the larger-system singlet sequence defines a particularly important remaining test. Together these advances make the proposed $SO(5)$ multicritical CFT a quantitatively testable description of deconfined magnetism.
\begin{acknowledgments}
We thank Denis Bernard, Wenan Guo, Yifei He, Zhengxin Liu, Adam Nahum, Hui Shao, Kay Wiese, Rong Yu, Weiqiang Yu and Wei Zhu for helpful discussions and communications. ZL also thanks the Laboratoire de Physique de l'\'Ecole normale sup\'erieure
(LPENS), Paris, for its hospitality and support during the final stage of this work. This research was supported by Startup Funding No.~4007022314 from Southeast University and by the National Natural Science Foundation of China under Grant No.~12375061. We employed the Codex and Claude in the fuzzy sphere computations. Codex assisted with manuscript organization and language. The authors verified the text and take full responsibility for the manuscript.
\end{acknowledgments}
\section*{Data availability}
The numerical data supporting this work, including the bootstrap and fuzzy-sphere spectra, OPE data, and supporting diagnostics, are openly available on Zenodo~\cite{ShenLiData2026}. Version 1.0.0, which corresponds to this article, has \href{https://doi.org/10.5281/zenodo.24073249}{doi:10.5281/zenodo.24073249}, and the concept \href{https://doi.org/10.5281/zenodo.24073248}{doi:10.5281/zenodo.24073248} resolves to the latest version; the maintained repository is \url{https://github.com/dantengs/dqcp-vst-mixing-data}. The code for the fuzzy-sphere spectra, conformal-generator analysis, and OPE extraction is available at \url{https://github.com/dantengs/fuzzysphere-dqcp}. The data are released under the CC BY 4.0 license and the code under the MIT license. The archive documents the OPE convention of Appendix~\ref{app:opeunit} and gives the bootstrap reference records and the fuzzy-sphere OPE coefficients in that convention.
\appendix
\onecolumngrid\setlength{\LTcapwidth}{\textwidth}
\section{Mixed-bootstrap assumptions and reconstruction}
\label{app:bootstrapdetails}
This appendix specifies the channel assumptions and numerical settings underlying Sec.~\ref{sec:bootstrap}, and explains how the reconstructed exchanges are selected for Fig.~\ref{fig:mixedspectrum}. The complete crossing equations, configuration records, and extremal spectra are retained in the numerical data accompanying this work.
\subsection{Crossing channels and spectral assumptions}
The mixed crossing system and four-ratio OPE scan are defined in Sec.~\ref{sec:bootstrap}. Table~\ref{tab:channels} lists the additional scalar gap assumptions; the complete channel content and spin restrictions are given in Table~\ref{tab:mixedselection}. The external $\phi,s,t$ contributions are included separately, with their repeated couplings tied by permutation symmetry. These gaps are assumptions on the spectrum; the external dimensions and OPE ratios are coordinates optimized within the resulting allowed region.
\begingroup
\appendixtablestyle\begin{longtable}{lll}
\caption{Additional scalar gap assumptions. The resolved external $\phi,s,t$ exchanges are supplied separately. The complete channel content and spin restrictions are given in Table~\ref{tab:mixedselection}. A prime distinguishes the second dimension-35 irrep.}\label{tab:channels}\\
\toprule \SOrephead & external pairs & scalar gap\\\midrule
\endfirsthead
\toprule \SOrephead & external pairs & scalar gap\\\midrule
\endhead\bottomrule\endfoot

\rowstrut $\rep{1}$ & $(\phi \phi ,ss,tt)$ & $4$ \\
\rowstrut $\rep{14}$ & $(\phi \phi ,st,tt)$ & $4$ \\
\rowstrut $\rep{5}$ & $(s\phi ,t\phi )$ & $3$ \\
\rowstrut $\rep{30}$ & $(t\phi )$ & $2$ \\
\rowstrut $\rep{35}$ & $(t\phi )$ & $3$ \\
\rowstrut $\rep{35}^{\prime}$ & $(tt)$ & $3$ \\
\rowstrut $\rep{55}$ & $(tt)$ & $3$ \\
\end{longtable}\endgroup

We use the $B_2$ Dynkin convention
\[
\begin{array}{c|rrrrrrrrr}
\mathcal R&\rep1&\rep5&\rep{10}&\rep{14}&\rep{30}&\rep{35}&\rep{35}'&\rep{55}&\rep{81}\\
{[a,b]}&[0,0]&[1,0]&[0,2]&[2,0]&[3,0]&[1,2]&[0,4]&[4,0]&[2,2].
\end{array}
\]
For these correlators, the internal invariant tensors can be built from Kronecker deltas. A single $SO(5)$ epsilon tensor would require at least two indices from one external factor; only the symmetric tensor $t_{ij}$ carries multiple indices, and that contraction vanishes. Pairs of epsilon tensors reduce to delta contractions. This property does not assume an internal $O(5)$ extension of the full theory. Intrinsic spacetime parity and the separate identical-pair spin restrictions are given in Eq.~\eqref{eq:mixedparity} and Table~\ref{tab:mixedselection}.

For the vector-model comparison in Fig.~\ref{fig:landscape}, the conventions of Refs.~\cite{HasenbuschPelissettoVicari2005,Hasenbusch2022} give $\Delta_\phi=(1+\eta)/2$, $\Delta_s=3-1/\nu$, and $\Delta_t=3-y_{2,2}$, where $y_{2,2}$ is the rank-two quadratic crossover eigenvalue.
\subsection{Numerical resolution and stability}
The candidate search varies three external dimensions and four OPE ratios jointly, minimizing $\Delta_s$ as described in Sec.~\ref{sec:bootstrap}. The principal $\Lambda=15,19,23$ candidates use Skydiving. A separate $\Lambda=23$ Navigator calculation checks the singlet spin-four gap entry~\cite{Navigator2021}. Table~\ref{tab:solverrecords} summarizes the arithmetic and stopping conditions. At the main $\Lambda=23$ point, SDPB reports primal--dual optimality with duality gap $1.87\times10^{-48}$, primal error $4.94\times10^{-16}$ and dual error $5.27\times10^{-67}$. The isolated-stress solve at the same point (Appendix~\ref{app:EFMinterpretation}) reaches duality gap $6.58\times10^{-31}$, primal error $3.78\times10^{-257}$, and dual error $1.88\times10^{-138}$.
\begingroup
\appendixtablestyle\begin{longtable}{lrrrrr}
\caption{SDPB arithmetic and stopping conditions for the mixed $\phi$--$s$--$t$ calculations. Columns report the configured arithmetic precision and stopping tolerances. The main $\Lambda=23$ result uses the Skydiving record. The Navigator row is for the spin-four reconstruction check in Appendix~\ref{app:EFMinterpretation}.}\label{tab:solverrecords}\\
\toprule $\Lambda$, setting & precision (bits) & gap tol. & primal tol. & dual tol. & max. iter.\\\midrule
\endfirsthead
\toprule $\Lambda$, setting & precision (bits) & gap tol. & primal tol. & dual tol. & max. iter.\\\midrule
\endhead\bottomrule\endfoot

\rowstrut 15, $g_t=4$ & 785 & $1.00\times10^{-30}$ & $1.00\times10^{-15}$ & $1.00\times10^{-15}$ & 1000 \\
\rowstrut 19, $g_t=4$ & 785 & $1.00\times10^{-25}$ & $1.00\times10^{-15}$ & $1.00\times10^{-15}$ & 1000 \\
\rowstrut 19, $g_t=3.95$ & 1024 & $1.00\times10^{-25}$ & $1.00\times10^{-15}$ & $1.00\times10^{-15}$ & 1000 \\
\rowstrut 23, Skydiving & 1024 & $1.00\times10^{-25}$ & $1.00\times10^{-15}$ & $1.00\times10^{-15}$ & 1000 \\
\rowstrut 23, Navigator check & 1024 & $1.00\times10^{-15}$ & $1.00\times10^{-60}$ & $1.00\times10^{-60}$ & 2000 \\
\end{longtable}\endgroup

Table~\ref{tab:mixedstability} already compares the principal $\Lambda=19,23$ candidates and the $g_t=3.95$ test. The additional $\Lambda=15$ Skydiving record gives $(\Delta_\phi,\Delta_s,\Delta_t)=(0.595,2.201,1.413)$ at the principal gaps $(g_\phi,g_s,g_t)=(3,4,4)$. Its external ratios, in the order $(r_{\phi\phi s},r_{ttt},r_{stt},r_{sss})$, are $(0.293,1.257,0.682,0.280)$. These rounded values extend the resolution comparison without duplicating the main-text table.

The principal $\Lambda=19$ and $23$ runs use different arithmetic precision, as do the two $\Lambda=19$ gap tests. Full-precision coordinates and configuration records are retained with the numerical data.

\appendixtablestyle\begin{longtable}{lrrr}
\caption{The $\min\Delta_s$ from $\phi$--$s$--$t$  bootstrap  at three derivative orders. All three points use Skydiving. All columns use the principal gaps $\Delta_{\phi'}\geq3$, $\Delta_{s'}\geq4$, $\Delta_{t'}\geq4$. The table reports optimization coordinates rounded to three decimals.}\label{tab:vstcutoffs}\\
\toprule 
Quantity & $\Lambda=15$ & $19$ & $23$\\\midrule
\endfirsthead
\toprule 
Quantity & $\Lambda=15$ & $19$ & $23$\\\midrule
\endhead\bottomrule\endfoot

\rowstrut $\Delta_\phi$ & 0.595 & 0.607 & 0.614 \\
\rowstrut $\Delta_s$ & 2.201 & 2.240 & 2.258 \\
\rowstrut $\Delta_t$ & 1.413 & 1.446 & 1.464 \\
\rowstrut $\beta_{\phi \phi s}/\beta_{\phi \phi t}$ & 0.293 & 0.294 & 0.294 \\
\rowstrut $\beta_{ttt}/\beta_{\phi \phi t}$ & 1.257 & 1.260 & 1.262 \\
\rowstrut $\beta_{stt}/\beta_{\phi \phi t}$ & 0.682 & 0.685 & 0.687 \\
\rowstrut $\beta_{sss}/\beta_{\phi \phi t}$ & 0.280 & 0.268 & 0.268 \\
\end{longtable}

\begingroup
\appendixtablestyle\begin{longtable}{lrrr}
\caption{The $\min\Delta_s$ sensitivity to the next tensor-scalar gap at $\Lambda=19$. Columns compare the two saved Skydiving runs. Their arithmetic settings also differ (Table~\ref{tab:solverrecords}); the two columns jointly change the next tensor gap from $4$ to $3.95$ and the arithmetic precision from $785$ to $1024$ bits.}\label{tab:gapsensitivity}\\
\toprule quantity & $g_t=4$ & $g_t=3.95$ & relative change (\%)\\\midrule
\endfirsthead
\toprule quantity & $g_t=4$ & $g_t=3.95$ & relative change (\%)\\\midrule
\endhead\bottomrule\endfoot

\rowstrut $\Delta_\phi$ & 0.607 & 0.599 & -1.280 \\
\rowstrut $\Delta_s$ & 2.240 & 2.199 & -1.800 \\
\rowstrut $\Delta_t$ & 1.446 & 1.422 & -1.695 \\
\rowstrut $\beta_{\phi \phi s}/\beta_{\phi \phi t}$ & 0.294 & 0.296 & +0.775 \\
\rowstrut $\beta_{ttt}/\beta_{\phi \phi t}$ & 1.260 & 1.254 & -0.509 \\
\rowstrut $\beta_{stt}/\beta_{\phi \phi t}$ & 0.685 & 0.679 & -0.816 \\
\rowstrut $\beta_{sss}/\beta_{\phi \phi t}$ & 0.268 & 0.257 & -4.004 \\
\end{longtable}\endgroup

\subsection{Extremal reconstruction and the comparison spectrum}
\label{app:EFMinterpretation}
The extremal-functional reconstruction associates scaling dimensions with functional roots and determines their external-pair matrices of OPE-coefficient products. At each root we retain the complete matrix, including all significant positive components. Removing the fixed block prefactor gives the mixed-crossing matrix
\begin{equation}
\Gamma_{ij}=\sum_a\beta_{i,a}\beta_{j,a},\qquad \lambda_{\rm eff}^{\,2}=\operatorname{tr}\Gamma=\sum_{i,a}\left|\frac{\lambda_{i,a}}{\kappa_i}\right|^2,
\label{eq:efmdiagnostics}
\end{equation}
where $i$ labels the allowed external pairs, $a$ labels the reconstructed components at a root, and $\lambda_{i,a}=\kappa_i\beta_{i,a}$ with the channel-dependent conversion factors in Table~\ref{tab:conversion}. The multicomponent OPE vectors use $(\phi\phi,ss,tt)$ for $\rep1$, $(\phi\phi,st,tt)$ for even-spin $\rep{14}$, $(\phi\phi,tt)$ for $\rep{10}$, and $(s\phi,t\phi)$ for $\rep5$. The combined OPE-coefficient magnitude $\lambda_{\rm eff}$ measures the total contribution in the crossing normalization; it is not an individual OPE coefficient. Figure~\ref{fig:mixedspectrum} displays 49 roots with $\lambda_{\rm eff}\geq0.10$ and five with $0.01\leq\lambda_{\rm eff}<0.10$; roots pinned to imposed gaps record assumptions rather than spectral predictions and are not shown. Table~\ref{tab:efmfull} lists the complete diagonal weights in the reported convention, $\sum_a\lambda_{i,a}^2=\kappa_i^2\Gamma_{ii}$.

\paragraph{Isolated stress tensor.}
Solving the crossing problem at the optimized point with a continuous spin-two singlet spectrum starting at $\Delta=3$ yields a single root at $\Delta=3$ whose matrix has three significant components, with secondary-to-leading eigenvalue ratio $0.081$. Its diagonal magnitudes in the order $(\phi\phi,ss,tt)$ are $(0.349994,2.410297,0.839764)$. A single stress tensor contributes a rank-one matrix along the Ward direction $w=(\Delta_\phi,\Delta_s,\Delta_t)$. We therefore solve the problem again at the same external point, with $T_{\mu\nu}$ placed in a one-dimensional block at $\Delta=3$ along $w$ and the remaining spin-two singlet spectrum starting at $3+10^{-6}$. The isolated block gives $c=0.5699043873$ and
\begin{equation}
c\,w=(0.3497700271,\,1.2870653837,\,0.8343499238)
\label{eq:isolatedstresscoefficients}
\end{equation}
for $(\beta_{\phi\phi T},\beta_{ssT},\beta_{ttT})$. By rule (a) of Appendix~\ref{app:opeunit}, $|\lambda_{\phi\phi T}|=0.349770$, displayed as $0.350$ in Table~\ref{tab:ope}. This solve reaches duality gap $6.58\times10^{-31}$ with navigator objective $+1.008\times10^{-17}$, and the external coordinates are held at their optimized values.

The original matrix is reproduced by the isolated Ward contribution and the threshold continuum root,
\begin{equation}
\Gamma_{\rm original}\simeq c^2ww^T+\Gamma_{\rm cont}.
\label{eq:stressmatrixdecomposition}
\end{equation}
The largest absolute entry of the difference, divided by the largest entry of $\Gamma_{\rm original}$, is $7.83\times10^{-7}$; normalizing each difference by its own entry gives at most $3.66\times10^{-6}$. The continuum root carries $0.13\%$, $71.5\%$, and $1.29\%$ of the original diagonal weights in the $\phi\phi$, $ss$, and $tt$ directions. The $\phi\phi$ stress coupling is therefore essentially the full root weight, whereas most of the $ss$ weight belongs to the near-threshold spin-two singlet continuum. Positivity of $\Gamma_{\rm original}-c^2ww^T$ gives the upper limit $|\lambda_{\phi\phi T}|\leq0.349770$, which the isolated solution saturates, and the independent vector-correlator estimate is $0.349456$.

All 53 other bootstrap points of Fig.~\ref{fig:mixedspectrum} coincide in the two solutions within $1.93\times10^{-5}$ in dimension and $0.022\%$ in $\lambda_{\rm eff}$, and all 225 roots with $\Delta\leq10$ outside the stress channel agree to three decimals. We use the isolated-stress solution as the bootstrap reference throughout.

\paragraph{Free-theory normalization check.}
At a near-free $\Lambda=15$ point with $(\Delta_\phi,\Delta_s,\Delta_t)=(0.5001,1.001753662,1.0002)$, the same conversions give
\[
\begin{array}{c|cccc}
\text{coefficient}&\lambda_{\phi\phi s}&\lambda_{\phi\phi t}&|\lambda_{\phi\phi J}|&|\lambda_{\phi\phi T}|\\\hline
\text{free value}&\sqrt{2/5}&\sqrt2&1/\sqrt2&\sqrt{3/40}\\
\text{numerical/free}&0.998122&0.999326&0.999862&1.000323
\end{array}
\]
For stress exchange the complete matrix gives $0.273950$, whereas the leading component alone gives $0.129569$; the complete matrix is therefore used whenever a root has several components.

\paragraph{Resolution dependence of higher couplings.}
Higher reconstructed couplings can be more sensitive to resolution than the scanned external ratios. The lowest odd-spin $\rep{14}^{+}$ exchange, accessible through $st$, illustrates this:
\[
\begin{array}{c|c|c|c}
\Lambda & g_t & \Delta & |\lambda_{st\mathcal O}|\\\hline
19 &4&2.197412&0.248825\\
19 &3.95&2.202876&0.237211\\
23 &4&2.228650&0.192230
\end{array}
\]
Between the two $g_t=4$ records its dimension changes by $1.42\%$ and its amplitude by $22.7\%$.

Table~\ref{tab:efmfull} lists the complete diagonal weights of the mixed-crossing root matrices by external pair $AB\to C$, with the external-pair ordering defined above.
\begingroup
\fontsize{8.3}{10.4}\selectfont
\setlength{\tabcolsep}{2.4pt}
\begin{table}[p]
\caption{Complete OPE weights $\sum_a\lambda_{AB,a}^2$ of the $\Lambda=23$ bootstrap roots by external pair $AB$, in the convention of Table~\ref{tab:ope} (Appendix~\ref{app:opeunit}). The exchanged root is specified by its representation $\mathcal R_C$, spin $\ell_C$, and dimension $\Delta_C$. For a root with a single component this is the squared OPE coefficient. The $(\rep1,2)$ root at $\Delta=3$ is the isolated stress tensor. $\dagger$ marks a root at an imposed gap or at the threshold $\ell+1+10^{-6}$ of a spinning continuum, and $\ddagger$ a root with several components.}\label{tab:efmfull}
\centering
\fontsize{8.3}{10.4}\selectfont\renewcommand{\arraystretch}{1.0}
\begin{minipage}[t]{.244\textwidth}\vspace{0pt}\centering
\begin{tabular}{@{}crr@{}}\toprule
\multicolumn{3}{c}{$AB=\phi \phi $} \\[2pt]
$(\mathcal R_C,\ell_C)$ & $\Delta_C$ & $\lambda^2$-weight\\\midrule
\rule[-2pt]{0pt}{9.7pt}$(\rep{1},0)$ & 4.000$^{\dagger}$ & $0.00147$ \\
\rule[-2pt]{0pt}{9.7pt}$(\rep{1},0)$ & 5.130 & $2.48\times10^{-7}$ \\
\rule[-2pt]{0pt}{9.7pt}$(\rep{1},2)$ & 3.000 & $0.122$ \\
\rule[-2pt]{0pt}{9.7pt}$(\rep{1},2)$ & 3.000$^{\dagger}$ & ${1.57\times10^{-4}}^{\ddagger}$ \\
\rule[-2pt]{0pt}{9.7pt}$(\rep{1},2)$ & 4.311 & $0.00828$ \\
\rule[-2pt]{0pt}{9.7pt}$(\rep{1},2)$ & 5.654 & $5.03\times10^{-7}$ \\
\rule[-2pt]{0pt}{9.7pt}$(\rep{1},4)$ & 5.000$^{\dagger}$ & ${0.00346}^{\ddagger}$ \\
\rule[-2pt]{0pt}{9.7pt}$(\rep{1},4)$ & 5.170 & $0.0209$ \\
\rule[-2pt]{0pt}{9.7pt}$(\rep{10},1)$ & 2.000 & $0.590$ \\
\rule[-2pt]{0pt}{9.7pt}$(\rep{10},1)$ & 2.362 & $9.12\times10^{-4}$ \\
\rule[-2pt]{0pt}{9.7pt}$(\rep{10},1)$ & 2.779 & $0.103$ \\
\rule[-2pt]{0pt}{9.7pt}$(\rep{10},1)$ & 4.345 & $0.00123$ \\
\rule[-2pt]{0pt}{9.7pt}$(\rep{10},1)$ & 4.478 & $0.00383$ \\
\rule[-2pt]{0pt}{9.7pt}$(\rep{10},1)$ & 5.859 & $4.16\times10^{-6}$ \\
\rule[-2pt]{0pt}{9.7pt}$(\rep{10},3)$ & 4.000$^{\dagger}$ & $4.27\times10^{-11}$ \\
\rule[-2pt]{0pt}{9.7pt}$(\rep{10},3)$ & 4.185 & $0.122$ \\
\rule[-2pt]{0pt}{9.7pt}$(\rep{10},3)$ & 5.250 & $0.00490$ \\
\rule[-2pt]{0pt}{9.7pt}$(\rep{10},3)$ & 5.688 & $3.42\times10^{-6}$ \\
\rule[-2pt]{0pt}{9.7pt}$(\rep{14},0)$ & 4.000$^{\dagger}$ & ${0.0102}^{\ddagger}$ \\
\rule[-2pt]{0pt}{9.7pt}$(\rep{14},0)$ & 5.003 & $2.50\times10^{-7}$ \\
\rule[-2pt]{0pt}{9.7pt}$(\rep{14},0)$ & 5.875 & $2.01\times10^{-6}$ \\
\rule[-2pt]{0pt}{9.7pt}$(\rep{14},2)$ & 3.000$^{\dagger}$ & $2.88\times10^{-5}$ \\
\rule[-2pt]{0pt}{9.7pt}$(\rep{14},2)$ & 3.176 & $0.0429$ \\
\rule[-2pt]{0pt}{9.7pt}$(\rep{14},2)$ & 3.238 & $0.502$ \\
\rule[-2pt]{0pt}{9.7pt}$(\rep{14},2)$ & 4.577 & $0.00461$ \\
\rule[-2pt]{0pt}{9.7pt}$(\rep{14},2)$ & 5.356 & $1.57\times10^{-5}$ \\
\rule[-2pt]{0pt}{9.7pt}$(\rep{14},2)$ & 5.568 & $0.00335$ \\
\rule[-2pt]{0pt}{9.7pt}$(\rep{14},2)$ & 6.206 & $3.65\times10^{-11}$ \\
\rule[-2pt]{0pt}{9.7pt}$(\rep{14},4)$ & 5.000$^{\dagger}$ & ${0.00151}^{\ddagger}$ \\
\rule[-2pt]{0pt}{9.7pt}$(\rep{14},4)$ & 5.028 & $8.25\times10^{-5}$ \\
\rule[-2pt]{0pt}{9.7pt}$(\rep{14},4)$ & 5.238 & $0.116$ \\
\addlinespace[4pt]
\multicolumn{3}{c}{$AB=ss$} \\[2pt]
$(\mathcal R_C,\ell_C)$ & $\Delta_C$ & $\lambda^2$-weight\\\midrule
\rule[-2pt]{0pt}{9.7pt}$(\rep{1},0)$ & 4.000$^{\dagger}$ & $1.49$ \\
\rule[-2pt]{0pt}{9.7pt}$(\rep{1},0)$ & 5.130 & $0.00617$ \\
\rule[-2pt]{0pt}{9.7pt}$(\rep{1},2)$ & 3.000 & $1.66$ \\
\rule[-2pt]{0pt}{9.7pt}$(\rep{1},2)$ & 3.000$^{\dagger}$ & ${4.15}^{\ddagger}$ \\
\rule[-2pt]{0pt}{9.7pt}$(\rep{1},2)$ & 4.311 & $0.415$ \\
\rule[-2pt]{0pt}{9.7pt}$(\rep{1},2)$ & 5.654 & $1.96$ \\
\rule[-2pt]{0pt}{9.7pt}$(\rep{1},4)$ & 5.000$^{\dagger}$ & ${1.68}^{\ddagger}$ \\
\rule[-2pt]{0pt}{9.7pt}$(\rep{1},4)$ & 5.170 & $0.931$ \\
\bottomrule\end{tabular}\end{minipage}
\hfill%
\begin{minipage}[t]{.244\textwidth}\vspace{0pt}\centering
\begin{tabular}{@{}crr@{}}\toprule
\multicolumn{3}{c}{$AB=tt$} \\[2pt]
$(\mathcal R_C,\ell_C)$ & $\Delta_C$ & $\lambda^2$-weight\\\midrule
\rule[-2pt]{0pt}{9.7pt}$(\rep{1},0)$ & 4.000$^{\dagger}$ & $0.0747$ \\
\rule[-2pt]{0pt}{9.7pt}$(\rep{1},0)$ & 5.130 & $0.00373$ \\
\rule[-2pt]{0pt}{9.7pt}$(\rep{1},2)$ & 3.000 & $0.696$ \\
\rule[-2pt]{0pt}{9.7pt}$(\rep{1},2)$ & 3.000$^{\dagger}$ & ${0.00907}^{\ddagger}$ \\
\rule[-2pt]{0pt}{9.7pt}$(\rep{1},2)$ & 4.311 & $0.178$ \\
\rule[-2pt]{0pt}{9.7pt}$(\rep{1},2)$ & 5.654 & $0.0426$ \\
\rule[-2pt]{0pt}{9.7pt}$(\rep{1},4)$ & 5.000$^{\dagger}$ & ${0.0825}^{\ddagger}$ \\
\rule[-2pt]{0pt}{9.7pt}$(\rep{1},4)$ & 5.170 & $0.150$ \\
\rule[-2pt]{0pt}{9.7pt}$(\rep{10},1)$ & 2.000 & $4.03$ \\
\rule[-2pt]{0pt}{9.7pt}$(\rep{10},1)$ & 2.362 & $0.113$ \\
\rule[-2pt]{0pt}{9.7pt}$(\rep{10},1)$ & 2.779 & $1.44$ \\
\rule[-2pt]{0pt}{9.7pt}$(\rep{10},1)$ & 4.345 & $0.0412$ \\
\rule[-2pt]{0pt}{9.7pt}$(\rep{10},1)$ & 4.478 & $0.652$ \\
\rule[-2pt]{0pt}{9.7pt}$(\rep{10},1)$ & 5.859 & $0.332$ \\
\rule[-2pt]{0pt}{9.7pt}$(\rep{10},3)$ & 4.000$^{\dagger}$ & $0.0107$ \\
\rule[-2pt]{0pt}{9.7pt}$(\rep{10},3)$ & 4.185 & $1.75$ \\
\rule[-2pt]{0pt}{9.7pt}$(\rep{10},3)$ & 5.250 & $0.107$ \\
\rule[-2pt]{0pt}{9.7pt}$(\rep{10},3)$ & 5.688 & $0.631$ \\
\rule[-2pt]{0pt}{9.7pt}$(\rep{14},0)$ & 4.000$^{\dagger}$ & ${1.19}^{\ddagger}$ \\
\rule[-2pt]{0pt}{9.7pt}$(\rep{14},0)$ & 5.003 & $0.235$ \\
\rule[-2pt]{0pt}{9.7pt}$(\rep{14},0)$ & 5.875 & $0.0926$ \\
\rule[-2pt]{0pt}{9.7pt}$(\rep{14},2)$ & 3.000$^{\dagger}$ & $0.156$ \\
\rule[-2pt]{0pt}{9.7pt}$(\rep{14},2)$ & 3.176 & $0.127$ \\
\rule[-2pt]{0pt}{9.7pt}$(\rep{14},2)$ & 3.238 & $5.21$ \\
\rule[-2pt]{0pt}{9.7pt}$(\rep{14},2)$ & 4.577 & $0.0249$ \\
\rule[-2pt]{0pt}{9.7pt}$(\rep{14},2)$ & 5.356 & $1.53$ \\
\rule[-2pt]{0pt}{9.7pt}$(\rep{14},2)$ & 5.568 & $0.430$ \\
\rule[-2pt]{0pt}{9.7pt}$(\rep{14},2)$ & 6.206 & $5.86\times10^{-4}$ \\
\rule[-2pt]{0pt}{9.7pt}$(\rep{14},4)$ & 5.000$^{\dagger}$ & ${0.0233}^{\ddagger}$ \\
\rule[-2pt]{0pt}{9.7pt}$(\rep{14},4)$ & 5.028 & $0.0146$ \\
\rule[-2pt]{0pt}{9.7pt}$(\rep{14},4)$ & 5.238 & $1.52$ \\
\rule[-2pt]{0pt}{9.7pt}$(\rep{35}^{\prime},0)$ & 4.685 & $0.460$ \\
\rule[-2pt]{0pt}{9.7pt}$(\rep{35}^{\prime},0)$ & 6.071 & $0.231$ \\
\rule[-2pt]{0pt}{9.7pt}$(\rep{35}^{\prime},2)$ & 4.865 & $2.37$ \\
\rule[-2pt]{0pt}{9.7pt}$(\rep{35}^{\prime},4)$ & 5.000$^{\dagger}$ & $0.0559$ \\
\rule[-2pt]{0pt}{9.7pt}$(\rep{55},0)$ & 3.692 & $3.18$ \\
\rule[-2pt]{0pt}{9.7pt}$(\rep{55},2)$ & 3.000$^{\dagger}$ & $0.00991$ \\
\rule[-2pt]{0pt}{9.7pt}$(\rep{55},2)$ & 5.405 & $2.34$ \\
\rule[-2pt]{0pt}{9.7pt}$(\rep{55},4)$ & 5.000$^{\dagger}$ & $0.0251$ \\
\rule[-2pt]{0pt}{9.7pt}$(\rep{81},1)$ & 2.000$^{\dagger}$ & $0.0371$ \\
\rule[-2pt]{0pt}{9.7pt}$(\rep{81},1)$ & 4.319 & $2.34$ \\
\rule[-2pt]{0pt}{9.7pt}$(\rep{81},3)$ & 4.000$^{\dagger}$ & $0.0260$ \\
\rule[-2pt]{0pt}{9.7pt}$(\rep{81},3)$ & 5.660 & $1.24$ \\
\bottomrule\end{tabular}\end{minipage}
\hfill%
\begin{minipage}[t]{.244\textwidth}\vspace{0pt}\centering
\begin{tabular}{@{}crr@{}}\toprule
\multicolumn{3}{c}{$AB=s\phi $} \\[2pt]
$(\mathcal R_C,\ell_C)$ & $\Delta_C$ & $\lambda^2$-weight\\\midrule
\rule[-2pt]{0pt}{9.7pt}$(\rep{5},0)$ & 3.000$^{\dagger}$ & $0.00782$ \\
\rule[-2pt]{0pt}{9.7pt}$(\rep{5},0)$ & 3.561 & $0.587$ \\
\rule[-2pt]{0pt}{9.7pt}$(\rep{5},0)$ & 5.861 & $0.00150$ \\
\rule[-2pt]{0pt}{9.7pt}$(\rep{5},1)$ & 3.773 & $0.900$ \\
\rule[-2pt]{0pt}{9.7pt}$(\rep{5},1)$ & 5.172 & $0.0296$ \\
\rule[-2pt]{0pt}{9.7pt}$(\rep{5},2)$ & 3.867 & $0.0673$ \\
\rule[-2pt]{0pt}{9.7pt}$(\rep{5},2)$ & 4.754 & $0.480$ \\
\rule[-2pt]{0pt}{9.7pt}$(\rep{5},2)$ & 5.633 & $0.245$ \\
\rule[-2pt]{0pt}{9.7pt}$(\rep{5},3)$ & 4.308 & $0.0161$ \\
\rule[-2pt]{0pt}{9.7pt}$(\rep{5},3)$ & 4.682 & $0.00211$ \\
\rule[-2pt]{0pt}{9.7pt}$(\rep{5},3)$ & 5.732 & $0.377$ \\
\rule[-2pt]{0pt}{9.7pt}$(\rep{5},3)$ & 6.044 & $0.150$ \\
\rule[-2pt]{0pt}{9.7pt}$(\rep{5},4)$ & 5.018 & $0.00114$ \\
\rule[-2pt]{0pt}{9.7pt}$(\rep{5},4)$ & 5.383 & $0.00183$ \\
\rule[-2pt]{0pt}{9.7pt}$(\rep{5},4)$ & 6.036 & $0.00770$ \\
\addlinespace[4pt]
\multicolumn{3}{c}{$AB=st$} \\[2pt]
$(\mathcal R_C,\ell_C)$ & $\Delta_C$ & $\lambda^2$-weight\\\midrule
\rule[-2pt]{0pt}{9.7pt}$(\rep{14},0)$ & 4.000$^{\dagger}$ & ${0.644}^{\ddagger}$ \\
\rule[-2pt]{0pt}{9.7pt}$(\rep{14},0)$ & 5.003 & $0.354$ \\
\rule[-2pt]{0pt}{9.7pt}$(\rep{14},0)$ & 5.875 & $0.377$ \\
\rule[-2pt]{0pt}{9.7pt}$(\rep{14},1)$ & 2.229 & $0.0370$ \\
\rule[-2pt]{0pt}{9.7pt}$(\rep{14},1)$ & 5.529 & $1.27$ \\
\rule[-2pt]{0pt}{9.7pt}$(\rep{14},2)$ & 3.000$^{\dagger}$ & $2.54\times10^{-13}$ \\
\rule[-2pt]{0pt}{9.7pt}$(\rep{14},2)$ & 3.176 & $0.497$ \\
\rule[-2pt]{0pt}{9.7pt}$(\rep{14},2)$ & 3.238 & $0.704$ \\
\rule[-2pt]{0pt}{9.7pt}$(\rep{14},2)$ & 4.577 & $0.0408$ \\
\rule[-2pt]{0pt}{9.7pt}$(\rep{14},2)$ & 5.356 & $0.108$ \\
\rule[-2pt]{0pt}{9.7pt}$(\rep{14},2)$ & 5.568 & $0.547$ \\
\rule[-2pt]{0pt}{9.7pt}$(\rep{14},2)$ & 6.206 & $1.18$ \\
\rule[-2pt]{0pt}{9.7pt}$(\rep{14},4)$ & 5.000$^{\dagger}$ & ${3.92\times10^{-9}}^{\ddagger}$ \\
\rule[-2pt]{0pt}{9.7pt}$(\rep{14},4)$ & 5.028 & $0.0977$ \\
\rule[-2pt]{0pt}{9.7pt}$(\rep{14},4)$ & 5.238 & $0.485$ \\
\bottomrule\end{tabular}\end{minipage}
\hfill%
\begin{minipage}[t]{.244\textwidth}\vspace{0pt}\centering
\begin{tabular}{@{}crr@{}}\toprule
\multicolumn{3}{c}{$AB=t\phi $} \\[2pt]
$(\mathcal R_C,\ell_C)$ & $\Delta_C$ & $\lambda^2$-weight\\\midrule
\rule[-2pt]{0pt}{9.7pt}$(\rep{5},0)$ & 3.000$^{\dagger}$ & $0.0236$ \\
\rule[-2pt]{0pt}{9.7pt}$(\rep{5},0)$ & 3.561 & $0.214$ \\
\rule[-2pt]{0pt}{9.7pt}$(\rep{5},0)$ & 5.861 & $0.00168$ \\
\rule[-2pt]{0pt}{9.7pt}$(\rep{5},1)$ & 3.773 & $0.211$ \\
\rule[-2pt]{0pt}{9.7pt}$(\rep{5},1)$ & 5.172 & $0.0178$ \\
\rule[-2pt]{0pt}{9.7pt}$(\rep{5},2)$ & 3.867 & $0.561$ \\
\rule[-2pt]{0pt}{9.7pt}$(\rep{5},2)$ & 4.754 & $0.0897$ \\
\rule[-2pt]{0pt}{9.7pt}$(\rep{5},2)$ & 5.633 & $0.0176$ \\
\rule[-2pt]{0pt}{9.7pt}$(\rep{5},3)$ & 4.308 & $3.79\times10^{-8}$ \\
\rule[-2pt]{0pt}{9.7pt}$(\rep{5},3)$ & 4.682 & $0.243$ \\
\rule[-2pt]{0pt}{9.7pt}$(\rep{5},3)$ & 5.732 & $0.00194$ \\
\rule[-2pt]{0pt}{9.7pt}$(\rep{5},3)$ & 6.044 & $0.0457$ \\
\rule[-2pt]{0pt}{9.7pt}$(\rep{5},4)$ & 5.018 & $0.00130$ \\
\rule[-2pt]{0pt}{9.7pt}$(\rep{5},4)$ & 5.383 & $0.0245$ \\
\rule[-2pt]{0pt}{9.7pt}$(\rep{5},4)$ & 6.036 & $0.185$ \\
\rule[-2pt]{0pt}{9.7pt}$(\rep{30},0)$ & 2.493 & $2.03$ \\
\rule[-2pt]{0pt}{9.7pt}$(\rep{30},0)$ & 5.395 & $0.0100$ \\
\rule[-2pt]{0pt}{9.7pt}$(\rep{30},1)$ & 5.900 & $0.00673$ \\
\rule[-2pt]{0pt}{9.7pt}$(\rep{30},2)$ & 4.214 & $0.688$ \\
\rule[-2pt]{0pt}{9.7pt}$(\rep{30},2)$ & 4.546 & $0.184$ \\
\rule[-2pt]{0pt}{9.7pt}$(\rep{30},3)$ & 4.956 & $0.159$ \\
\rule[-2pt]{0pt}{9.7pt}$(\rep{30},4)$ & 5.666 & $0.00860$ \\
\rule[-2pt]{0pt}{9.7pt}$(\rep{30},4)$ & 6.238 & $0.264$ \\
\rule[-2pt]{0pt}{9.7pt}$(\rep{35},0)$ & 3.000$^{\dagger}$ & $0.00159$ \\
\rule[-2pt]{0pt}{9.7pt}$(\rep{35},0)$ & 5.190 & $0.00564$ \\
\rule[-2pt]{0pt}{9.7pt}$(\rep{35},1)$ & 3.048 & $1.27$ \\
\rule[-2pt]{0pt}{9.7pt}$(\rep{35},1)$ & 4.697 & $0.0453$ \\
\rule[-2pt]{0pt}{9.7pt}$(\rep{35},2)$ & 3.497 & $0.309$ \\
\rule[-2pt]{0pt}{9.7pt}$(\rep{35},2)$ & 4.342 & $0.166$ \\
\rule[-2pt]{0pt}{9.7pt}$(\rep{35},3)$ & 4.621 & $0.0452$ \\
\rule[-2pt]{0pt}{9.7pt}$(\rep{35},3)$ & 5.131 & $0.396$ \\
\rule[-2pt]{0pt}{9.7pt}$(\rep{35},4)$ & 5.765 & $0.137$ \\
\bottomrule\end{tabular}\end{minipage}
\end{table}
\endgroup

\clearpage
\section{Fuzzy-sphere setup, state identification, and finite-size data}
\label{app:fuzzydetails}
The physical controls are the orbital number $N$, the contact-amplitude ratio $V/U$, and the symmetry sector. The flavour number is fixed at four. Spatial spin $\ell$, internal subgroup spins $s_A,s_B$, exchange $X$, and physical parity $\mathcal P$ are different labels. Eigensolver residuals test a numerical eigenpair; coverage of the $K_\mu$ target spaces and class weights test operator identification; variation with $N$ tests finite-size effects. Passing the first test does not imply passing the other two. The microscopic ratio denoted $g_N=V/U$ is distinct from the continuum sigma-model coupling $g_\sigma$ in Eq.~\eqref{eq:fuzzyNLSM}. The $0.90$ class-weight gate is a squared-projection criterion. Mixed states, unresolved parent families, and missing $K_\mu$ matrix elements are recorded as different outcomes.
\subsection{Lowest-Landau-level conventions and projected interactions}
\label{app:fuzzysetup}
We use an area measure $\dd A=R^2\dd\Omega$ and orthonormal LLL orbitals $\phi_m$, with $m=-q,\ldots,q$:
\begin{equation}
\begin{gathered}
\hat\Psi_\alpha(\Omega)=\sum_{m=-q}^{q}\phi_m(\Omega)c_{m\alpha},
\qquad\alpha=1,\ldots,4,\\
\int_{S^2}\dd A\,\phi_m^*(\Omega)\phi_{m'}(\Omega)=\delta_{mm'}.
\end{gathered}
\label{eq:LLLexpansion}
\end{equation}
The $\phi_m$ are normalized monopole harmonics; this convention includes the radius dependence in the orbitals. Flux quantization gives $R^2=q\ell_B^2$, where $\ell_B$ is the magnetic length. Thus the large-sphere limit at fixed $\ell_B$ is $N_{\rm orb}=2q+1\to\infty$. The size notation is $N=N_{\rm orb}=2q+1$, with $N_e=2N$ at half filling; the flavor number remains four throughout.

One choice of invariant tensor is
\begin{equation}
J_{\rm sp}=\begin{pmatrix}0&\rep{1}_2\\-\rep{1}_2&0\end{pmatrix},
\qquad M^{T}J_{\rm sp}M=J_{\rm sp},\quad M\in\mathrm{Sp}(2).
\end{equation}
At fixed particle number, the projected kinetic energy is constant. We use the contact Hamiltonian of Ref.~\cite{Zhou2024},
\begin{equation}
H_N=\int_{S^2}\dd A\left[U:\!\hat\rho^2\!:
-\frac V2\hat B^\dagger\hat B\right],
\label{eq:fuzzyH}
\end{equation}
where $\hat\rho=\hat\Psi^\dagger\hat\Psi$ and $\hat B=\hat\Psi^T J_{\rm sp}\hat\Psi$. The density interaction preserves SU(4), while the pairing interaction restricts flavor rotations to $M^TJM=J$, giving $\mathrm{Sp}(2)\simeq\mathrm{Spin}(5)$. The fermions are Spin(5) spinors and their neutral bilinears carry SO(5) representations. In particular, $\hat n_i=\hat\Psi^\dagger\Gamma_i\hat\Psi$ transforms as the SO(5) vector whose leading continuum contribution is $\phi_i$.

Hermitian matrices $\Gamma_i$ satisfying $\{\Gamma_i,\Gamma_j\}=2\delta_{ij}\rep{1}_4$ specify the vector bilinears in a compatible spinor basis. The contact Hamiltonian in Eq.~\eqref{eq:fuzzyH} is equivalent, at fixed charge and up to an energy shift, to the un-normal-ordered interaction of Ref.~\cite{Zhou2024}. In the area convention above, the shift is $U K_q N_e$, with $K_q=\sum_m|\phi_m(\Omega)|^2=N_{\rm orb}/(4\pi R^2)$.

For the projected contact interaction define
\begin{equation}
\begin{gathered}
\mathcal I_{1234}=\int_{S^2}\dd A\,
\phi_{m_1}^{*}\phi_{m_2}^{*}\phi_{m_3}\phi_{m_4},\\
\hat B_{mn}=\sum_{\alpha\beta}(J_{\rm sp})_{\alpha\beta}c_{m\alpha}c_{n\beta}.
\end{gathered}
\end{equation}
Then
\begin{align}
H_N={}&U\sum_{m_1\ldots m_4}\mathcal I_{1234}
\sum_{\alpha\beta}c_{m_1\alpha}^\dagger c_{m_2\beta}^\dagger
c_{m_3\beta}c_{m_4\alpha}\nonumber\\
&-\frac V2\sum_{m_1\ldots m_4}\mathcal I_{1234}
\hat B_{m_1m_2}^\dagger\hat B_{m_3m_4}.
\label{eq:projectedH}
\end{align}
Rotational invariance enforces $m_1+m_2=m_3+m_4$. For these contact interactions only the relative-angular-momentum-zero Haldane pseudopotential is nonzero~\cite{Zhou2024}. The same orbital overlap determines the density and pair matrix elements; $U$ and $V$ are their two amplitudes. The orbital integrals convert the area-dimensional contact amplitudes to pseudopotential energies. We set the density pseudopotential amplitude to one, leaving $V/U$ as the parameter varied within the SO(5)-symmetric family.
\paragraph{Finite-size diagonalization and spectral matching.}
\label{app:fuzzynumerics}
Both density and symplectic-pairing interactions retain only the relative-angular-momentum-zero pseudopotential. Its density-channel amplitude is one in the numerical energy units. We use double precision, dense diagonalization for sectors of dimension at most 400, and Lanczos iteration otherwise. The production spectral tolerance is $10^{-10}$; acceptance requires Hamiltonian and Casimir residual norms and the Frobenius orthogonality error at most $10^{-6}$. These are eigensolver checks, not finite-size error estimates. The software versions are Julia 1.11.7, FuzzifiED 2.0.1, and KrylovKit 0.10.4.

The vacuum, current, and stress anchors are the lowest $C_2=0$, $\ell=0$ exchange-even state, the lowest $C_2=3$, $\ell=1$ state in $(s_A,s_B)=(0,1)$, and the lowest $C_2=0$, $\ell=2$ exchange-even state, respectively. They define
\begin{equation}
\alpha_N=\frac{2}{E_J-E_0},\qquad F_N(g)=\alpha_N(E_T-E_0)-3.
\end{equation}
The final production-window matching residual satisfies $|F_N|<10^{-5}$. Couplings are fixed before production; no finite-size extrapolation enters this calibration.

\paragraph{Spectral and target-window coverage.}
The production spectrum is extended until each selected sector is complete or its upper level reaches $\Delta^{\rm eff}\geq6$. Evaluating $Q_K$ additionally requires target windows at $\ell'=\ell-1,\ell,\ell+1$ (only $\ell'=1$ for $\ell=0$). Each target window must extend beyond the required source energy by $10^{-5}$ in energy units, with its edge outside numerical degeneracies. An independent Lanczos calculation in its orthogonal complement checks for missed lower levels or an unresolved edge cluster. Sources above $\Delta^{\rm eff}=6.2$, or lacking any required target window, retain partial or incomplete coverage status. In particular, an $N=12$, $\ell=3$ source does not have complete coverage of the $K_\mu$ target states without its required $\ell'=4$ target.

\paragraph{Fixed-coupling spectra.}
The additional scan uses Lanczos tolerance $10^{-11}$ and requires target Hamiltonian and physical-parity residuals below $10^{-8}$. Numerical energy clusters are resolved by $C_2$ and parity. Every level below the selected target and its complete terminal cluster must be resolved before acceptance. Targets are the lowest states in the specified spin, Casimir, parity, and exchange sector. The scalar singlet search uses $X=+1$; it does not exclude a lower even singlet in an unsearched exchange sector.

The $N=12$ integer-sector calculation includes the required spin-four target windows. Window sensitivity and incomplete parent-family assignments are recorded separately from a missing target.

\subsection{Fuzzy coordinates and the WZW normalization}
\label{app:fuzzyWZW}
The indices $a,b,c=1,2,3$ in the fuzzy-coordinate algebra label spatial coordinates; the five indices of the WZW volume form below label internal target-space components. The $X_a$ are dimensionless, and $RX_a$ have length units. Let $P_q$ project onto the orbital angular-momentum-$q$ LLL. For dimensionless coordinates $x_a$ on the unit sphere, $X_a=P_qx_aP_q$ obey
\begin{equation}
\begin{gathered}
\relax[X_a,X_b]=\frac{i}{q+1}\epsilon_{abc}X_c,\\
\sum_aX_a^2=\frac{q}{q+1}\rep{1}.
\end{gathered}
\end{equation}
Thus spatial rotations act exactly at finite $q$, while the commutative unit sphere is recovered as $q\to\infty$~\cite{ZhuFuzzy2023,Zhou2024}.

At half filling, slow flavor orientations can be represented by a Hermitian matrix $Q$ with $Q^2=\rep{1}_4$ and $\operatorname{Tr}Q=0$, taking values in $\mathrm{U}(4)/[\mathrm{U}(2)\times\mathrm{U}(2)]$. The SO(5)-favoring interaction restricts these orientations to $Q=n_i\Gamma_i$, with $\boldsymbol n^2=1$, identifying the $S^4$ target space. Integrating out amplitude and charge fluctuations in this regime gives~\cite{Ippoliti2018,WangSO52021,Zhou2024}
\begin{equation}
\begin{split}
S_{\rm eff}[\boldsymbol n]
={}&\frac{1}{2g_\sigma}\int\dd^3x\,(\partial_\mu\boldsymbol n)^2\\
&+2\pi i\,\mathcal W[\boldsymbol n]+\cdots .
\end{split}
\label{eq:fuzzyNLSM}
\end{equation}
The topological response of the occupied Landau-level states fixes the WZW level to one. The effective sigma-model coupling $g_\sigma$ is distinct from $g=V/U$; the microscopic ratio changes the stiffness, while the WZW term encodes the intertwined defect quantum numbers of the N\'eel and VBS components.

For the sigma model, the local occupied-subspace projector is $P_{\rm occ}=(\rep{1}_4+Q)/2$. On the SO(5) submanifold, $Q=n_i\Gamma_i$ and $\boldsymbol n^2=1$. On compactified Euclidean spacetime $M_3$, extend $\boldsymbol n$ to $\widetilde{\boldsymbol n}$ on $B_4$ with $\partial B_4=M_3$. A geometric convention is
\begin{equation}
\begin{gathered}
\mathcal W[\boldsymbol n]
=\frac{1}{\Omega_4}\int_{B_4}\widetilde{\boldsymbol n}^{\,*}\omega_4,
\qquad \Omega_4=\frac{8\pi^2}{3},\\
\omega_4=\frac{1}{4!}\epsilon_{abcde}n_a\,
\dd n_b\wedge\dd n_c\wedge\dd n_d\wedge\dd n_e.
\end{gathered}
\end{equation}
The volume form satisfies $\int_{S^4}\omega_4=\Omega_4$. Two extensions differ by an integer winding number, making the Euclidean weight with $S_{\rm WZW}=2\pi i k\mathcal W$ independent of the extension for integer $k$. The Landau-level construction yields $k=1$, with an orientation convention fixing its sign~\cite{Ippoliti2018,WangSO52021}. Changing the microscopic stiffness does not continuously change this quantized level. The derivative expansion in Eq.~\eqref{eq:fuzzyNLSM} assumes a regime in which slow neutral orientations provide the appropriate low-energy variables; it is not a proof of an infrared conformal fixed point.

\subsection{Special conformal generators and primary-space selection}
\label{app:Kprojectors}
The generators $P_\mu$ and $K_\mu$ are dimensionless and use the energy normalization defining $D$. In the conformal limit, $K_\mu$ preserves the internal SO(5) representation, transforms as a spatial vector, and lowers the dimension by one. Its matrix therefore connects source and target blocks with the angular momenta allowed by addition of spin one. All target sectors allowed by $K_\mu$ must enter the norm; projecting $K_\mu$ back into only its source block would discard physical transitions generated by $K_\mu$.

The construction starts from the spin-one angular moment of the cylinder energy density. For an exact CFT stress tensor in unit-radius conventions~\cite{FanGenerators2024,FardelliGenerators2024},
\begin{equation}
\Lambda_\mu=P_\mu+K_\mu
=2\int\dd\Omega\,\hat x_\mu T_{00}(\Omega),
\label{eq:LambdaK}
\end{equation}
The moment $\Lambda_\mu$ is unrelated to the bootstrap derivative order $\Lambda$. Together with $[D,P_\mu]=P_\mu$, $[D,K_\mu]=-K_\mu$ and $P_\mu^\dagger=K_\mu$, this gives $P_\mu-K_\mu=[D,\Lambda_\mu]$. At finite size, derivative contributions to the microscopic density may integrate to zero in $H_N$ but contribute to its spin-one moment. Matrix-element improvement conditions control these terms~\cite{FardelliGenerators2024}; energy ordering provides the practical separation of the matrix elements of $P_\mu$ and $K_\mu$ used in the primary-space construction~\cite{FardelliOPE2026}.

If $\Pi_{\rm in}$ selects retained input states and $\Pi_{\rm out}$ selects the target states used to evaluate the action of $K_\mu$, the computed matrix is
\begin{equation}
Q_K^{\rm cut}=\sum_\mu
\Pi_{\rm in}K_\mu^\dagger\Pi_{\rm out}K_\mu\Pi_{\rm in}.
\end{equation}
Enlarging the target space checks that omitted contributions do not produce spurious small eigenvalues. Near-zero eigenvectors determine a candidate primary space; subsequent Hamiltonian diagonalization within nearby primary groups distinguishes unresolved primary mixing from descendant contamination~\cite{FardelliOPE2026}.

Small eigenvalues must also be compared with shortening limits. For example, in three dimensions the scalar descendant $\Box\phi$ has $Q_K$ eigenvalue $4\Delta_\phi-2$ in the conventions of Ref.~\cite{FardelliOPE2026}, approaching zero near the free-scalar bound. A model-specific selection criterion must therefore include the known parent families and their transitions under $K_\mu$. We examine the low-$Q_K$ spectrum, cutoff and size dependence, and generator calibration together with the parent-family projections.
\paragraph{Generator calibration and reduced matrix elements.}
The spin-one moments $O_1,O_2$ of the density and symplectic-pairing interactions give $\Lambda=c_1O_1+c_2O_2$. We minimize its squared vacuum image subject to
\begin{equation}
\sum_{a=1}^2c_a\langle\partial s\|O_a\|s\rangle=\sqrt{2\Delta_s}.
\end{equation}
The reference $s$ is the lowest excited singlet in the selected scalar block; its spin-one calibration partner is the singlet level nearest $E_s+1/\alpha_N$. This improves and normalizes the generator without imposing an exactly unit descendant gap.

For reduced matrix elements $M_{\ell'\ell}$ of $\Lambda$, the matrices representing $K_\mu$ and the corresponding positive semidefinite operator $Q_K$ are
\begin{equation}
\begin{gathered}
(K_{\ell'\ell})_{ji}=(M_{\ell'\ell})_{ji}\,\rep{1}_{E_j<E_i-\varepsilon_E},\\
Q_K=\sum_{\ell'}\frac{2\ell'+1}{2\ell+1}K_{\ell'\ell}^{\dagger}K_{\ell'\ell}.
\end{gathered}
\end{equation}
The energy guard is $\varepsilon_E=\max[10^{-8},10(r_H^{\rm src}+r_H^{\rm tgt})]$, using the largest Hamiltonian residuals in the two windows. Complete tests using $K_\mu$ require a source below every necessary target frontier. The angular factors implement the reduced-matrix-element convention.

\paragraph{Projection and subsequent classification.}
Production retains $Q_K$ eigenmodes with eigenvalue at most $0.25$ and diagonalizes the Hamiltonian within this subspace. Subsequent classification instead uses $\kappa=w_K/(2\bar\Delta)$, with $Q_K$ eigenvalue $w_K$ and its mode's energy expectation $\bar\Delta$. The prespecified threshold $\kappa_c=0.042169650342858224$ is selected from the pooled $N=6$--10 distribution of $\log_{10}\kappa$; raw eigenlevel classification uses a subspace-weight gate of $0.90$. These thresholds specify different operations. For a raw energy eigenstate, the squared projection weights are
\begin{equation}
w_P=\langle\psi|\Pi_P|\psi\rangle,\qquad
w_{p,n}=\langle\psi|\Pi_{p,n}|\psi\rangle,
\label{eq:weightsmain}
\end{equation}
where $\Pi_P$ projects onto the selected primary subspace and $\Pi_{p,n}$ onto level $n$ of parent family $p$. A primary requires $w_P\geq0.90$ and a single-parent descendant requires $w_{p,n}\geq0.90$. The total descendant weight $w_D$ uses the orthonormal span of all parent images, since individual parent spans need not be mutually orthogonal. Every source also requires all target spins allowed by $K_\mu$, with target frontiers above its energy and outside unresolved degeneracies. Strict primaries satisfy $w_P\geq0.90$ at each of $\kappa_c/3$, $\kappa_c$, and $3\kappa_c$ with the same generator matrices, together with copy and retained-window consistency. Small $Q_K$ expectation values and large projection weights support finite-size candidates; family patterns, shortening limits, coverage, Casimir ambiguity, and threshold sensitivity remain necessary checks.

\paragraph{State identity along the coupling scan.}
Assignments away from $g_*(N)$ are tested through overlaps with the anchor and between neighboring coupling points, whenever the vectors share the same basis. An overlap squared above $1/2$ gives a uniquely dominant overlap with the anchor among the orthonormal eigenstates at that coupling. It does not replace an analysis using the special conformal generators at that coupling or establish one continuum operator throughout the interval. Products of neighboring overlaps depend on the mesh and are not endpoint fidelities. Matching energies across different bases is recorded separately from a vector-overlap test. Same-label level separations are measured in energy units; an upper separation remains unresolved when no higher same-label level appears in the saved window.

The reduced-matrix convention is $\langle\ell'm'|O_{1q}|\ell m\rangle=\langle\ell m;1q|\ell'm'\rangle M_{\ell'\ell}$. Thus the summed norm for fixed source magnetic number contains $(2\ell'+1)/(2\ell+1)|M_{\ell'\ell}|^2$. For a scalar source and a spin-one target this factor is three: the reduced calibration $\sqrt{2\Delta_s}$ gives the full norm $6\Delta_s$ in Eq.~\eqref{eq:raisingnormtest}. The first matrix index labels the output target. The energy-order indicator retains $E_j<E_i-\varepsilon_E$; Hamiltonian residuals $r_H$ and $\varepsilon_E$ are measured in the same numerical energy units.

The retained-window test reduces the top of every source and target window by $0.25$ and $0.50$ in dimension units, holding the remaining parameters fixed. A raw level is stable if its class is unchanged in every reduction where it remains present. A mode also requires best squared vector overlap at least $0.90$ and dimension shift below $0.01$ where its sector survives. Ambiguous copy/primary--descendant overlap is checked before primary or descendant acceptance; insufficient descendant-family coverage is recorded separately from mixed primary--descendant weight. For a normalized projected mode, $\sigma_\Delta^2=\langle u|D_N^2|u\rangle-\bar\Delta^2$ measures energy-basis spread. This quantity differs from uncertainty in the limiting CFT dimension.

\paragraph{Microscopic insertion routes.}
In Eq.~\eqref{eq:extractionmain}, the angular factor is $1$, $\sqrt3$, or $\sqrt{15/2}$ for insertion rank $r=0,1,2$, respectively. The vector density ($\Gamma_2$) uses the contraction $m_1$ and normalization $|Z_\phi|$. The tensor density ($G$ component) uses $m_1+(-1)^r m_2$ and $|Z_t|$. The singlet density combination uses $a_1m_1+a_2m_2-O_{\rm vac}U_{ZX}$ and the signed normalization $Z_{s,\rm raw}$, with $O_{\rm vac}$ the vacuum expectation of the singlet insertion.

Here $U_{ZX}=\langle Z|X\rangle$ in the common retained Hilbert space, so vacuum subtraction remains well defined for nonorthogonal projected states. Each $m_a$ is the contraction $\langle Z|O_Y^{(a,r)}|X\rangle$ of one of the two microscopic operator components. The CG component is $C_{\rm iso}=\langle Z_{\rm int}|\mathcal C_{XY\to Z}|X_{\rm int}\otimes Y_{\rm int}\rangle$, with the isometry normalized by $\mathcal C\mathcal C^\dagger=1_Z$. Each route fixes the internal tensor orientation, spatial projection, and singlet phase. For spinning states, the insertion rank specifies the measured angular contraction; it does not enumerate every possible CFT three-point tensor structure.

\subsection{Checks of the joint symmetry projection}
\label{app:symmetryprojectors}
For orthonormal symmetry-adapted basis columns, $B^\dagger B=\rep{1}$ and $\mathbb P=BB^\dagger$ obey
\begin{equation}
\mathbb P^\dagger=\mathbb P,
\quad\mathbb P^2=\mathbb P,
\quad[H_N,\mathbb P]=0.
\end{equation}
The implemented basis retains internal spins $s_A,s_B\in\{0,1,2\}$ and couples orbital multiplets to definite total angular momentum $\ell$. At $N=5$--11 the production source space includes all six unordered spin pairs with $\ell=0,\ldots,4$; $N=12$ uses $(0,0),(0,1),(0,2)$ with $\ell=0,\ldots,3$. Exchange reduction is exact within the selected blocks. This choice limits representation coverage and does not cover the full fixed-charge Hilbert space. The Casimir normalization is $C_2(\rep{1},\rep{5},\rep{10},\rep{14})=(0,2,3,5)$. Within numerical energy degeneracies we diagonalize $C_2$ and, where needed, the particle--hole map. Higher representations sharing a Casimir value remain unresolved without additional information.

The raw particle--hole eigenvalue $p_{\rm seg}$ includes a rotation and an $N$-dependent vacuum sign. Physical parity is
\begin{equation}
\mathcal P(\psi)=p_{\rm seg}(\psi)\,p_{\rm seg}(0)\,(-1)^\ell,
\label{eq:fuzzyphysicalparity}
\end{equation}
with $|0\rangle$ the lowest singlet in the exchange-even scalar block at the same coupling. We require a particle--hole purity residual below $10^{-6}$ before assigning parity. Repeated SO(5) multiplets in different subgroup blocks are symmetry copies, not independent operators.

Energy gaps are converted to dimensions using the lowest current, while the stress level determines the matching coupling:
\begin{equation}
\alpha_N=\frac{2}{E_J-E_0},\qquad
\Delta_{\mathcal O}^{\rm eff}=\alpha_N(E_{\mathcal O}-E_0),\qquad
\alpha_N(E_T-E_0)=3\quad\text{at }g=g_*(N).
\label{eq:fuzzycalibration}
\end{equation}
The vacuum, current, and stress anchors lie in $(\rep1,0^+)$, $(\rep{10},1^+)$, and $(\rep1,2^+)$, respectively. We use dense diagonalization for small blocks and Lanczos iteration otherwise, checking the Hamiltonian residuals, Casimir labels, and orthogonality of the retained eigenstates. These numerical checks do not estimate finite-size errors. At fixed coupling, each size instead uses its own current gap without imposing the stress condition.
\subsection{Additional conformal and finite-size checks}
\label{app:fuzzysupport}
\subsubsection{Conformal checks}
\label{app:fuzzychecks}
Figure~\ref{fig:fuzzyvalidation} collects three complementary checks. The first follows the vector order parameter $m_i=N^{-1}\sum_m c_m^\dagger\Gamma_i c_m$ across the coupling $g$. The rescaling $N^{1/2}\langle m_i^2\rangle$ uses the scalar unitarity bound only as a plotting reference. Its finite-size ordering locates the regime used for spectroscopy; it does not by itself establish a thermodynamic phase boundary.

The second test compares descendant spacings with integer levels,
\begin{equation}
\epsilon_{\mathcal O,n}=\Delta_{\mathcal O^{(n)}}^{\rm eff}-\Delta_{\mathcal O}^{\rm eff}-n.
\label{eq:descendantresidualappendix}
\end{equation}
The nearly integer spacings coexist with residual finite-size deviations, which need not decrease monotonically: the residual of the first tensor descendant grows from $0.038$ at $N=5$ to $0.081$ at $N=11$, while that of the vector descendant decreases from $0.099$ at $N=8$ to $0.091$ at $N=12$. The third check uses the raising matrix elements. For a normalized primary in three dimensions, the conformal algebra gives
\begin{equation}
\mathcal C_{\mathcal O}=\frac{\sum_\mu\|P_\mu|\mathcal O\rangle\|^2}{6\Delta_{\mathcal O}}=1.
\label{eq:raisingnormtestappendix}
\end{equation}
For $\phi$, $t$, and $T_{\mu\nu}$, the measured ratios at $N=12$ are $1.05837$, $1.02728$, and $1.00770$. Their approach toward unity supplies the matrix-element check quoted in Sec.~\ref{sec:fuzzyvalidation}, independently of the singlet transition used to calibrate the generators. At $N=11$, calibrating the generators with the vector or tensor instead of the singlet preserves the classifications, while tripling $\kappa_c$ changes four primary--descendant assignments; the strict primaries are those stable under this change. Taken together, these observations support conformal organization over the accessible sizes while retaining measurable corrections.

\clearpage
\subsubsection{Additional primary families}
\label{app:publishedcomparison}
Table~\ref{tab:newprimarycontinuations} follows the 18 primaries identified within the published $N=10$ window to larger sizes. Twelve have $N=12$ continuations that pass the strict criteria, six in the mixed-bootstrap channels and six outside them; the others retain their $N=10$ identification without a larger-size partner. The 13 primaries with $5.5<\Delta\leq6.25$ in Table~\ref{tab:newprimarymain} are identified by the same criteria at $N=10$.

The bootstrap entries in Table~\ref{tab:newprimarymain} are the nearest roots of the $\Lambda=23$ solution with the same $SO(5)$ representation, spin, and parity, excluding roots at imposed gaps, with combined OPE-coefficient magnitude $\lambda_{\rm eff}\geq0.01$ as defined in Eq.~\eqref{eq:efmdiagnostics}. Some channels contain several nearby roots; for example, the $N=12$ $(\rep{10},3^+)$ level at $5.521$ lies between roots at $5.250$ and $5.688$, and the $(\rep{14},2^+)$ level at $4.898$ between roots at $4.577$ and $5.356$. The full same-channel root lists and the complete comparison with the published spectrum are retained in the numerical archive.

\Needspace{26\baselineskip}\begingroup\appendixtablestyle\setlength{\LTcapwidth}{\textwidth}
\begin{longtable}{ccrr}
\caption{Size evolution of the 18 primaries of Table~\ref{tab:newprimarymain} within the published $N=10$ window. The couplings are $g_*(10)=0.9437$ and $g_*(12)=1.00924$. A dash means that no $N=12$ continuation passes the strict criteria.}\label{tab:newprimarycontinuations}\\
\toprule \SOrephead & $\ell^{\mathcal P}$ & $\Delta_{10}$ & $\Delta_{12}$\\\midrule\endfirsthead
\toprule \SOrephead & $\ell^{\mathcal P}$ & $\Delta_{10}$ & $\Delta_{12}$\\\midrule\endhead\bottomrule\endfoot
\multicolumn{4}{l}{Families in mixed-bootstrap channels}\\
\rowstrut $\rep5$ & $3^-$ & 5.382 & ---\\
\rowstrut $\rep{10}$ & $1^+$ & 4.515 & 4.505\\
\rowstrut $\rep{10}$ & $3^+$ & 5.497 & 5.521\\
\rowstrut $\rep{14}$ & $2^+$ & 4.887 & 4.898\\
\rowstrut $\rep{14}$ & $2^+$ & 5.376 & ---\\
\rowstrut $\rep{30}$ & $3^-$ & 5.063 & ---\\
\rowstrut $\rep{35}$ & $2^-$ & 4.787 & 4.796\\
\rowstrut $\rep{35}^{\prime}$ & $0^+$ & 4.885 & 4.863\\
\rowstrut $\rep{81}$ & $3^+$ & 5.185 & 5.219\\
\midrule\multicolumn{4}{l}{Families outside mixed-bootstrap channels}\\
\rowstrut $\rep{10}$ & $1^-$ & 5.144 & ---\\
\rowstrut $\rep{10}$ & $3^-$ & 4.418 & 4.424\\
\rowstrut $\rep{14}$ & $2^-$ & 5.299 & 5.303\\
\rowstrut $\rep{14}$ & $3^-$ & 5.500 & ---\\
\rowstrut $\rep{35}$ & $1^+$ & 4.839 & ---\\
\rowstrut $\rep{35}$ & $2^+$ & 4.805 & 4.810\\
\rowstrut $\rep{35}$ & $3^+$ & 5.336 & 5.353\\
\rowstrut $\rep{35}^{\prime}$ & $2^-$ & 5.169 & 5.169\\
\rowstrut $\rep{81}$ & $2^+$ & 4.792 & 4.808\\
\end{longtable}\endgroup
\subsubsection{Singlet calibration and size sequence}
\label{app:singletsequences}
Table~\ref{tab:singletcomparison} lists the tuned sequence shown in Fig.~\ref{fig:singletsize}(a). Its $g_*(N)$ values specify the calibration used for the main spectral and OPE comparisons. Panel (b) of that figure instead holds $g$ fixed and normalizes each size with its own current gap. In each scan, $E_s$ is the lowest nonvacuum state in the searched $(C_2,\ell,\mathcal P,X)=(0,0,+,+)$ sector; the unsearched exchange sector is not included. The complete fixed-coupling scans are retained in the numerical archive. The two protocols remain separate because retuning the microscopic Hamiltonian and increasing the orbital cutoff need not produce the same corrections to scaling.

\begin{figure*}[t]
\centering\includegraphics[width=\textwidth]{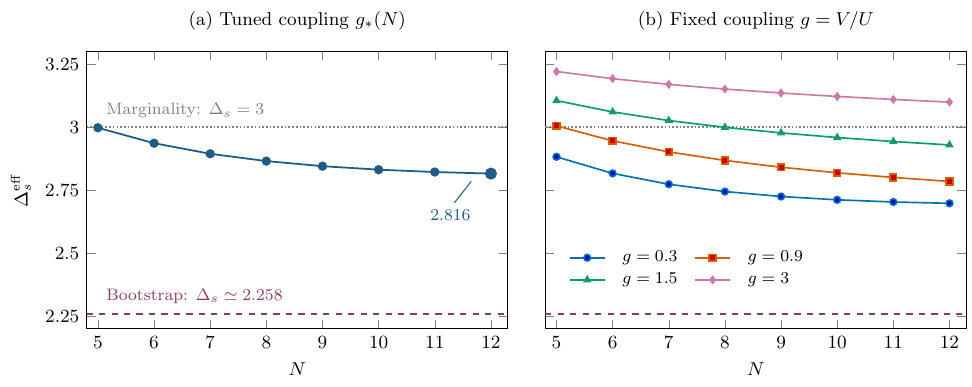}
\caption{Size dependence of the lowest $SO(5)$ singlet. (a) Stress-matched sequence at $g_*(N)$. (b) Representative fixed-coupling sequences. At each point, $\Delta_s^{\rm eff}=2(E_s-E_0)/(E_J-E_0)$ uses the current gap at the same size and coupling. The dotted line marks marginality; the dashed line shows the separately optimized $\Lambda=23$ bootstrap candidate.}
\label{fig:singletsize}
\end{figure*}

\clearpage\subsection{Complete retained \texorpdfstring{$N=12$}{N=12} spectrum}
\label{app:reviewedN12}\label{app:extendeddata}
Table~\ref{tab:N12fullspectrum} collects the $N=12$ energy levels and their primary/descendant assignments that underlie the spectrum analysis in Sec.~\ref{sec:fuzzyspectrum}. It includes all 264 retained raw levels with $\ell\leq3$ and $\Delta\leq6.25$ in the computed symmetry sectors specified in Appendix~\ref{app:symmetryprojectors}. The catalogue contains the identity, 44 nonvacuum primaries, 154 descendants, and 65 unresolved levels. These are finite-size assignments obtained using the criteria of Appendix~\ref{app:Kprojectors}.

\begingroup
\setlength{\LTcapwidth}{\textwidth}
\def\cmpcell#1{{\fontfamily{ptm}\fontsize{8.3}{11.2}\selectfont#1}}
\let\cmphead\cmpcell
\setlength{\tabcolsep}{2.7pt}\renewcommand{\arraystretch}{1.02}
\begin{longtable}{@{}ccrc@{\hspace{26pt}}ccrc@{\hspace{26pt}}ccrc@{\hspace{26pt}}ccrc@{}}
\caption{Complete retained $N=12$ raw spectrum at $g_*(12)=1.0092383$, $\ell\leq3$ and $\Delta\leq6.25$, within the sectors described in Appendix~\ref{app:symmetryprojectors}. The label $\ell^{\mathcal P}$ combines spatial spin and physical parity. Dimensions use $\Delta=2(E-E_0)/(E_J-E_0)$. Of the 264 levels, 45 are labeled P (the identity and 44 nonvacuum primaries), 154 are descendants (D), and 65 remain unassigned (-). Read down each block from left to right. Superscripts $\dagger$ and $\ddagger$ mark the five dashed squares in Fig.~\ref{fig:mixedspectrum}, with stable and unstable retained-window classifications, respectively. The projection weights are given below the table.}
\label{tab:N12fullspectrum}\label{tab:N12currentprimaries}\label{tab:descendants}\label{tab:unresolvedlevelweights}\\
\toprule \cmphead{$\ell^{\mathcal P}$} & \cmphead{\SOrephead} & \cmphead{$\Delta_{12}$} & \cmphead{Class} & \cmphead{$\ell^{\mathcal P}$} & \cmphead{\SOrephead} & \cmphead{$\Delta_{12}$} & \cmphead{Class} & \cmphead{$\ell^{\mathcal P}$} & \cmphead{\SOrephead} & \cmphead{$\Delta_{12}$} & \cmphead{Class} & \cmphead{$\ell^{\mathcal P}$} & \cmphead{\SOrephead} & \cmphead{$\Delta_{12}$} & \cmphead{Class}\\\midrule
\endfirsthead
\multicolumn{16}{c}{TABLE \thetable\ (continued)}\\[4pt]
\toprule \cmphead{$\ell^{\mathcal P}$} & \cmphead{\SOrephead} & \cmphead{$\Delta_{12}$} & \cmphead{Class} & \cmphead{$\ell^{\mathcal P}$} & \cmphead{\SOrephead} & \cmphead{$\Delta_{12}$} & \cmphead{Class} & \cmphead{$\ell^{\mathcal P}$} & \cmphead{\SOrephead} & \cmphead{$\Delta_{12}$} & \cmphead{Class} & \cmphead{$\ell^{\mathcal P}$} & \cmphead{\SOrephead} & \cmphead{$\Delta_{12}$} & \cmphead{Class}\\\midrule
\endhead
\bottomrule
\endfoot
\rule[-2pt]{0pt}{10.4pt}\cmpcell{$0^{+}$} & \cmpcell{$\rep{1}$} & \cmpcell{0.000} & \cmpcell{P} & \cmpcell{$1^{-}$} & \cmpcell{$\rep{5}$} & \cmpcell{5.049} & \cmpcell{D} & \cmpcell{$2^{-}$} & \cmpcell{$\rep{5}$} & \cmpcell{6.122} & \cmpcell{D} & \cmpcell{$1^{+}$} & \cmpcell{$\rep{10}$} & \cmpcell{5.541} & \cmpcell{P} \\
\rule[-2pt]{0pt}{10.4pt}\cmpcell{$0^{+}$} & \cmpcell{$\rep{1}$} & \cmpcell{2.816} & \cmpcell{P} & \cmpcell{$1^{+}$} & \cmpcell{$\rep{5}$} & \cmpcell{6.043} & \cmpcell{D} & \cmpcell{$3^{+}$} & \cmpcell{$\rep{5}$} & \cmpcell{6.225} & \cmpcell{-} & \cmpcell{$1^{-}$} & \cmpcell{$\rep{10}$} & \cmpcell{6.157} & \cmpcell{-} \\
\rule[-2pt]{0pt}{10.4pt}\cmpcell{$1^{+}$} & \cmpcell{$\rep{1}$} & \cmpcell{3.882} & \cmpcell{D} & \cmpcell{$2^{+}$} & \cmpcell{$\rep{5}$} & \cmpcell{3.986} & \cmpcell{P} & \cmpcell{$3^{-}$} & \cmpcell{$\rep{5}$} & \cmpcell{6.231} & \cmpcell{D} & \cmpcell{$3^{+}$} & \cmpcell{$\rep{10}$} & \cmpcell{4.786} & \cmpcell{D} \\
\rule[-2pt]{0pt}{10.4pt}\cmpcell{$0^{+}$} & \cmpcell{$\rep{1}$} & \cmpcell{5.002} & \cmpcell{D} & \cmpcell{$3^{+}$} & \cmpcell{$\rep{5}$} & \cmpcell{4.764} & \cmpcell{D} & \cmpcell{$1^{+}$} & \cmpcell{$\rep{10}$} & \cmpcell{2.000} & \cmpcell{P} & \cmpcell{$2^{-}$} & \cmpcell{$\rep{10}$} & \cmpcell{4.905} & \cmpcell{-} \\
\rule[-2pt]{0pt}{10.4pt}\cmpcell{$2^{+}$} & \cmpcell{$\rep{1}$} & \cmpcell{3.000} & \cmpcell{P} & \cmpcell{$1^{+}$} & \cmpcell{$\rep{5}$} & \cmpcell{5.024} & \cmpcell{D} & \cmpcell{$2^{+}$} & \cmpcell{$\rep{10}$} & \cmpcell{2.940} & \cmpcell{D} & \cmpcell{$2^{-}$} & \cmpcell{$\rep{10}$} & \cmpcell{5.056} & \cmpcell{D} \\
\rule[-2pt]{0pt}{10.4pt}\cmpcell{$3^{+}$} & \cmpcell{$\rep{1}$} & \cmpcell{3.807} & \cmpcell{D} & \cmpcell{$2^{+}$} & \cmpcell{$\rep{5}$} & \cmpcell{5.752} & \cmpcell{D} & \cmpcell{$1^{-}$} & \cmpcell{$\rep{10}$} & \cmpcell{3.058} & \cmpcell{D} & \cmpcell{$3^{+}$} & \cmpcell{$\rep{10}$} & \cmpcell{5.204} & \cmpcell{-} \\
\rule[-2pt]{0pt}{10.4pt}\cmpcell{$2^{-}$} & \cmpcell{$\rep{1}$} & \cmpcell{4.007} & \cmpcell{D} & \cmpcell{$2^{+}$} & \cmpcell{$\rep{5}$} & \cmpcell{5.852} & \cmpcell{D} & \cmpcell{$3^{+}$} & \cmpcell{$\rep{10}$} & \cmpcell{3.683} & \cmpcell{D} & \cmpcell{$3^{+}$} & \cmpcell{$\rep{10}$} & \cmpcell{5.344} & \cmpcell{-} \\
\rule[-2pt]{0pt}{10.4pt}\cmpcell{$3^{-}$} & \cmpcell{$\rep{1}$} & \cmpcell{4.743} & \cmpcell{D} & \cmpcell{$1^{-}$} & \cmpcell{$\rep{5}$} & \cmpcell{5.927} & \cmpcell{D} & \cmpcell{$2^{-}$} & \cmpcell{$\rep{10}$} & \cmpcell{3.939} & \cmpcell{D} & \cmpcell{$3^{+}$} & \cmpcell{$\rep{10}$} & \cmpcell{5.484} & \cmpcell{D} \\
\rule[-2pt]{0pt}{10.4pt}\cmpcell{$0^{-}$} & \cmpcell{$\rep{1}$} & \cmpcell{5.303} & \cmpcell{P} & \cmpcell{$0^{+}$} & \cmpcell{$\rep{5}$} & \cmpcell{6.018} & \cmpcell{D} & \cmpcell{$1^{+}$} & \cmpcell{$\rep{10}$} & \cmpcell{4.074} & \cmpcell{D} & \cmpcell{$2^{-}$} & \cmpcell{$\rep{10}$} & \cmpcell{5.565} & \cmpcell{D} \\
\rule[-2pt]{0pt}{10.4pt}\cmpcell{$1^{-}$} & \cmpcell{$\rep{1}$} & \cmpcell{6.159} & \cmpcell{D} & \cmpcell{$0^{-}$} & \cmpcell{$\rep{5}$} & \cmpcell{4.308} & \cmpcell{P} & \cmpcell{$3^{-}$} & \cmpcell{$\rep{10}$} & \cmpcell{4.609} & \cmpcell{D} & \cmpcell{$3^{-}$} & \cmpcell{$\rep{10}$} & \cmpcell{5.590} & \cmpcell{-} \\
\rule[-2pt]{0pt}{10.4pt}\cmpcell{$0^{+}$} & \cmpcell{$\rep{1}$} & \cmpcell{5.879} & \cmpcell{P} & \cmpcell{$1^{-}$} & \cmpcell{$\rep{5}$} & \cmpcell{5.319} & \cmpcell{D} & \cmpcell{$2^{+}$} & \cmpcell{$\rep{10}$} & \cmpcell{4.884} & \cmpcell{D} & \cmpcell{$3^{-}$} & \cmpcell{$\rep{10}$} & \cmpcell{5.627} & \cmpcell{-} \\
\rule[-2pt]{0pt}{10.4pt}\cmpcell{$2^{+}$} & \cmpcell{$\rep{1}$} & \cmpcell{4.719} & \cmpcell{-} & \cmpcell{$1^{-}$} & \cmpcell{$\rep{5}$} & \cmpcell{4.486} & \cmpcell{P} & \cmpcell{$1^{-}$} & \cmpcell{$\rep{10}$} & \cmpcell{5.049} & \cmpcell{D} & \cmpcell{$3^{-}$} & \cmpcell{$\rep{10}$} & \cmpcell{5.766} & \cmpcell{D} \\
\rule[-2pt]{0pt}{10.4pt}\cmpcell{$2^{+}$} & \cmpcell{$\rep{1}$} & \cmpcell{4.904} & \cmpcell{-} & \cmpcell{$1^{+}$} & \cmpcell{$\rep{5}$} & \cmpcell{5.555} & \cmpcell{D} & \cmpcell{$2^{-}$} & \cmpcell{$\rep{10}$} & \cmpcell{5.782} & \cmpcell{D} & \cmpcell{$2^{+}$} & \cmpcell{$\rep{10}$} & \cmpcell{5.837} & \cmpcell{D} \\
\rule[-2pt]{0pt}{10.4pt}\cmpcell{$2^{+}$} & \cmpcell{$\rep{1}$} & \cmpcell{4.952} & \cmpcell{-} & \cmpcell{$0^{-}$} & \cmpcell{$\rep{5}$} & \cmpcell{5.611} & \cmpcell{D} & \cmpcell{$1^{+}$} & \cmpcell{$\rep{10}$} & \cmpcell{5.949} & \cmpcell{D} & \cmpcell{$2^{+}$} & \cmpcell{$\rep{10}$} & \cmpcell{5.930} & \cmpcell{-} \\
\rule[-2pt]{0pt}{10.4pt}\cmpcell{$3^{+}$} & \cmpcell{$\rep{1}$} & \cmpcell{5.279} & \cmpcell{D} & \cmpcell{$2^{-}$} & \cmpcell{$\rep{5}$} & \cmpcell{5.379} & \cmpcell{-} & \cmpcell{$1^{+}$} & \cmpcell{$\rep{10}$} & \cmpcell{3.159} & \cmpcell{P} & \cmpcell{$2^{+}$} & \cmpcell{$\rep{10}$} & \cmpcell{6.016} & \cmpcell{-} \\
\rule[-2pt]{0pt}{10.4pt}\cmpcell{$3^{+}$} & \cmpcell{$\rep{1}$} & \cmpcell{5.621} & \cmpcell{D} & \cmpcell{$2^{-}$} & \cmpcell{$\rep{5}$} & \cmpcell{5.449} & \cmpcell{-} & \cmpcell{$2^{+}$} & \cmpcell{$\rep{10}$} & \cmpcell{4.079} & \cmpcell{D} & \cmpcell{$1^{-}$} & \cmpcell{$\rep{10}$} & \cmpcell{6.021} & \cmpcell{D} \\
\rule[-2pt]{0pt}{10.4pt}\cmpcell{$2^{-}$} & \cmpcell{$\rep{1}$} & \cmpcell{5.673} & \cmpcell{D} & \cmpcell{$3^{-}$} & \cmpcell{$\rep{5}$} & \cmpcell{5.392} & \cmpcell{P} & \cmpcell{$1^{-}$} & \cmpcell{$\rep{10}$} & \cmpcell{4.196} & \cmpcell{D} & \cmpcell{$2^{+}$} & \cmpcell{$\rep{10}$} & \cmpcell{6.040} & \cmpcell{-} \\
\rule[-2pt]{0pt}{10.4pt}\cmpcell{$3^{+}$} & \cmpcell{$\rep{1}$} & \cmpcell{5.738} & \cmpcell{D} & \cmpcell{$3^{+}$} & \cmpcell{$\rep{5}$} & \cmpcell{5.538} & \cmpcell{D} & \cmpcell{$0^{+}$} & \cmpcell{$\rep{10}$} & \cmpcell{4.292} & \cmpcell{D} & \cmpcell{$3^{-}$} & \cmpcell{$\rep{10}$} & \cmpcell{6.046} & \cmpcell{-} \\
\rule[-2pt]{0pt}{10.4pt}\cmpcell{$1^{+}$} & \cmpcell{$\rep{1}$} & \cmpcell{5.871} & \cmpcell{D} & \cmpcell{$2^{+}$} & \cmpcell{$\rep{5}$} & \cmpcell{5.590} & \cmpcell{P} & \cmpcell{$1^{+}$} & \cmpcell{$\rep{10}$} & \cmpcell{5.183} & \cmpcell{D} & \cmpcell{$1^{+}$} & \cmpcell{$\rep{10}$} & \cmpcell{6.050} & \cmpcell{-} \\
\rule[-2pt]{0pt}{10.4pt}\cmpcell{$2^{-}$} & \cmpcell{$\rep{1}$} & \cmpcell{5.880} & \cmpcell{D} & \cmpcell{$3^{-}$} & \cmpcell{$\rep{5}$} & \cmpcell{4.582} & \cmpcell{-} & \cmpcell{$1^{+}$} & \cmpcell{$\rep{10}$} & \cmpcell{5.247} & \cmpcell{D} & \cmpcell{$1^{+}$} & \cmpcell{$\rep{10}$} & \cmpcell{6.069} & \cmpcell{-} \\
\rule[-2pt]{0pt}{10.4pt}\cmpcell{$1^{+}$} & \cmpcell{$\rep{1}$} & \cmpcell{6.061} & \cmpcell{D} & \cmpcell{$3^{-}$} & \cmpcell{$\rep{5}$} & \cmpcell{4.695} & \cmpcell{-} & \cmpcell{$3^{+}$} & \cmpcell{$\rep{10}$} & \cmpcell{4.222} & \cmpcell{P} & \cmpcell{$3^{-}$} & \cmpcell{$\rep{10}$} & \cmpcell{6.070} & \cmpcell{-} \\
\rule[-2pt]{0pt}{10.4pt}\cmpcell{$3^{-}$} & \cmpcell{$\rep{1}$} & \cmpcell{6.152} & \cmpcell{-} & \cmpcell{$2^{-}$} & \cmpcell{$\rep{5}$} & \cmpcell{4.761} & \cmpcell{D} & \cmpcell{$3^{-}$} & \cmpcell{$\rep{10}$} & \cmpcell{5.120} & \cmpcell{D} & \cmpcell{$3^{+}$} & \cmpcell{$\rep{10}$} & \cmpcell{6.078} & \cmpcell{D} \\
\rule[-2pt]{0pt}{10.4pt}\cmpcell{$2^{+}$} & \cmpcell{$\rep{1}$} & \cmpcell{6.161} & \cmpcell{-} & \cmpcell{$2^{-}$} & \cmpcell{$\rep{5}$} & \cmpcell{4.912} & \cmpcell{D} & \cmpcell{$2^{+}$} & \cmpcell{$\rep{10}$} & \cmpcell{5.483} & \cmpcell{D} & \cmpcell{$1^{-}$} & \cmpcell{$\rep{10}$} & \cmpcell{6.107} & \cmpcell{D} \\
\rule[-2pt]{0pt}{10.4pt}\cmpcell{$0^{-}$} & \cmpcell{$\rep{5}$} & \cmpcell{0.590} & \cmpcell{P} & \cmpcell{$3^{-}$} & \cmpcell{$\rep{5}$} & \cmpcell{5.335} & \cmpcell{D} & \cmpcell{$3^{+}$} & \cmpcell{$\rep{10}$} & \cmpcell{5.940} & \cmpcell{D} & \cmpcell{$3^{-}$} & \cmpcell{$\rep{10}$} & \cmpcell{6.193} & \cmpcell{-} \\
\rule[-2pt]{0pt}{10.4pt}\cmpcell{$1^{-}$} & \cmpcell{$\rep{5}$} & \cmpcell{1.681} & \cmpcell{D} & \cmpcell{$3^{+}$} & \cmpcell{$\rep{5}$} & \cmpcell{5.581} & \cmpcell{-} & \cmpcell{$3^{-}$} & \cmpcell{$\rep{10}$} & \cmpcell{4.424} & \cmpcell{P} & \cmpcell{$0^{+}$} & \cmpcell{$\rep{14}$} & \cmpcell{1.469} & \cmpcell{P} \\
\rule[-2pt]{0pt}{10.4pt}\cmpcell{$2^{-}$} & \cmpcell{$\rep{5}$} & \cmpcell{2.622} & \cmpcell{D} & \cmpcell{$3^{-}$} & \cmpcell{$\rep{5}$} & \cmpcell{5.613} & \cmpcell{D} & \cmpcell{$1^{+}$} & \cmpcell{$\rep{10}$} & \cmpcell{4.505} & \cmpcell{P} & \cmpcell{$1^{+}$} & \cmpcell{$\rep{14}$} & \cmpcell{2.550} & \cmpcell{D} \\
\rule[-2pt]{0pt}{10.4pt}\cmpcell{$0^{-}$} & \cmpcell{$\rep{5}$} & \cmpcell{2.944} & \cmpcell{D} & \cmpcell{$2^{-}$} & \cmpcell{$\rep{5}$} & \cmpcell{5.700} & \cmpcell{D} & \cmpcell{$2^{+}$} & \cmpcell{$\rep{10}$} & \cmpcell{5.413} & \cmpcell{D} & \cmpcell{$2^{+}$} & \cmpcell{$\rep{14}$} & \cmpcell{3.481} & \cmpcell{D} \\
\rule[-2pt]{0pt}{10.4pt}\cmpcell{$3^{-}$} & \cmpcell{$\rep{5}$} & \cmpcell{3.356} & \cmpcell{D} & \cmpcell{$3^{+}$} & \cmpcell{$\rep{5}$} & \cmpcell{5.717} & \cmpcell{-} & \cmpcell{$1^{-}$} & \cmpcell{$\rep{10}$} & \cmpcell{5.541} & \cmpcell{D} & \cmpcell{$0^{+}$} & \cmpcell{$\rep{14}$} & \cmpcell{3.736} & \cmpcell{D} \\
\rule[-2pt]{0pt}{10.4pt}\cmpcell{$1^{-}$} & \cmpcell{$\rep{5}$} & \cmpcell{3.942} & \cmpcell{D} & \cmpcell{$2^{-}$} & \cmpcell{$\rep{5}$} & \cmpcell{5.816} & \cmpcell{D} & \cmpcell{$0^{+}$} & \cmpcell{$\rep{10}$} & \cmpcell{5.656} & \cmpcell{D} & \cmpcell{$2^{+}$} & \cmpcell{$\rep{14}$} & \cmpcell{3.337} & \cmpcell{P} \\
\rule[-2pt]{0pt}{10.4pt}\cmpcell{$0^{-}$} & \cmpcell{$\rep{5}$} & \cmpcell{5.036} & \cmpcell{D} & \cmpcell{$2^{-}$} & \cmpcell{$\rep{5}$} & \cmpcell{5.913} & \cmpcell{D} & \cmpcell{$1^{-}$} & \cmpcell{$\rep{10}$} & \cmpcell{5.120} & \cmpcell{-} & \cmpcell{$2^{-}$} & \cmpcell{$\rep{14}$} & \cmpcell{4.351} & \cmpcell{D} \\
\rule[-2pt]{0pt}{10.4pt}\cmpcell{$1^{-}$} & \cmpcell{$\rep{5}$} & \cmpcell{5.897} & \cmpcell{D} & \cmpcell{$3^{-}$} & \cmpcell{$\rep{5}$} & \cmpcell{6.059} & \cmpcell{-} & \cmpcell{$2^{-}$} & \cmpcell{$\rep{10}$} & \cmpcell{5.959} & \cmpcell{D} & \cmpcell{$3^{-}$} & \cmpcell{$\rep{14}$} & \cmpcell{5.080} & \cmpcell{D} \\
\rule[-2pt]{0pt}{10.4pt}\cmpcell{$2^{-}$} & \cmpcell{$\rep{5}$} & \cmpcell{3.849} & \cmpcell{P} & \cmpcell{$0^{-}$} & \cmpcell{$\rep{5}$} & \cmpcell{6.074} & \cmpcell{-} & \cmpcell{$0^{-}$} & \cmpcell{$\rep{10}$} & \cmpcell{6.142} & \cmpcell{D} & \cmpcell{$1^{-}$} & \cmpcell{$\rep{14}$} & \cmpcell{5.658} & \cmpcell{D} \\
\rule[-2pt]{0pt}{10.4pt}\cmpcell{$2^{+}$} & \cmpcell{$\rep{5}$} & \cmpcell{4.865} & \cmpcell{D} & \cmpcell{$0^{-}$} & \cmpcell{$\rep{5}$} & \cmpcell{6.109} & \cmpcell{-} & \cmpcell{$3^{+}$} & \cmpcell{$\rep{10}$} & \cmpcell{5.521} & \cmpcell{P} & \cmpcell{$2^{-}$} & \cmpcell{$\rep{14}$} & \cmpcell{6.099} & \cmpcell{D} \\
\rule[-2pt]{0pt}{10.4pt}\cmpcell{$0^{+}$} & \cmpcell{$\rep{14}$} & \cmpcell{4.339} & \cmpcell{P} & \cmpcell{$0^{-}$} & \cmpcell{$\rep{30}$} & \cmpcell{5.827} & \cmpcell{P} & \cmpcell{$2^{+}$} & \cmpcell{$\rep{35}$} & \cmpcell{5.694} & \cmpcell{D} & \cmpcell{$0^{+}$} & \cmpcell{$\rep{35}^{\prime}$} & \cmpcell{4.863} & \cmpcell{P} \\
\rule[-2pt]{0pt}{10.4pt}\cmpcell{$1^{+}$} & \cmpcell{$\rep{14}$} & \cmpcell{5.335} & \cmpcell{D} & \cmpcell{$3^{+}$} & \cmpcell{$\rep{30}$} & \cmpcell{6.045} & \cmpcell{D} & \cmpcell{$1^{-}$} & \cmpcell{$\rep{35}$} & \cmpcell{5.798} & \cmpcell{D} & \cmpcell{$1^{+}$} & \cmpcell{$\rep{35}^{\prime}$} & \cmpcell{5.812} & \cmpcell{D} \\
\rule[-2pt]{0pt}{10.4pt}\cmpcell{$2^{+}$} & \cmpcell{$\rep{14}$} & \cmpcell{4.898} & \cmpcell{P} & \cmpcell{$1^{-}$} & \cmpcell{$\rep{30}$} & \cmpcell{6.228} & \cmpcell{-} & \cmpcell{$0^{+}$} & \cmpcell{$\rep{35}$} & \cmpcell{5.846} & \cmpcell{D} & \cmpcell{$2^{-}$} & \cmpcell{$\rep{35}^{\prime}$} & \cmpcell{5.169} & \cmpcell{P} \\
\rule[-2pt]{0pt}{10.4pt}\cmpcell{$2^{-}$} & \cmpcell{$\rep{14}$} & \cmpcell{5.874} & \cmpcell{D} & \cmpcell{$2^{-}$} & \cmpcell{$\rep{30}$} & \cmpcell{4.425$^{\dagger}$} & \cmpcell{-} & \cmpcell{$3^{+}$} & \cmpcell{$\rep{35}$} & \cmpcell{5.353} & \cmpcell{P} & \cmpcell{$3^{-}$} & \cmpcell{$\rep{35}^{\prime}$} & \cmpcell{5.955} & \cmpcell{D} \\
\rule[-2pt]{0pt}{10.4pt}\cmpcell{$2^{-}$} & \cmpcell{$\rep{14}$} & \cmpcell{5.303} & \cmpcell{P} & \cmpcell{$2^{-}$} & \cmpcell{$\rep{30}$} & \cmpcell{4.570$^{\dagger}$} & \cmpcell{-} & \cmpcell{$3^{-}$} & \cmpcell{$\rep{35}$} & \cmpcell{5.778} & \cmpcell{-} & \cmpcell{$2^{+}$} & \cmpcell{$\rep{35}^{\prime}$} & \cmpcell{6.087} & \cmpcell{D} \\
\rule[-2pt]{0pt}{10.4pt}\cmpcell{$3^{-}$} & \cmpcell{$\rep{14}$} & \cmpcell{6.083} & \cmpcell{D} & \cmpcell{$3^{-}$} & \cmpcell{$\rep{30}$} & \cmpcell{5.079} & \cmpcell{-} & \cmpcell{$1^{-}$} & \cmpcell{$\rep{35}$} & \cmpcell{6.010} & \cmpcell{-} & \cmpcell{$0^{+}$} & \cmpcell{$\rep{35}^{\prime}$} & \cmpcell{6.102} & \cmpcell{P} \\
\rule[-2pt]{0pt}{10.4pt}\cmpcell{$2^{+}$} & \cmpcell{$\rep{14}$} & \cmpcell{5.377} & \cmpcell{-} & \cmpcell{$3^{-}$} & \cmpcell{$\rep{30}$} & \cmpcell{5.118} & \cmpcell{-} & \cmpcell{$1^{-}$} & \cmpcell{$\rep{35}$} & \cmpcell{6.196} & \cmpcell{-} & \cmpcell{$2^{+}$} & \cmpcell{$\rep{35}^{\prime}$} & \cmpcell{6.250} & \cmpcell{P} \\
\rule[-2pt]{0pt}{10.4pt}\cmpcell{$3^{-}$} & \cmpcell{$\rep{14}$} & \cmpcell{5.526} & \cmpcell{P} & \cmpcell{$3^{-}$} & \cmpcell{$\rep{30}$} & \cmpcell{5.370} & \cmpcell{D} & \cmpcell{$0^{-}$} & \cmpcell{$\rep{35}$} & \cmpcell{6.140} & \cmpcell{-} & \cmpcell{$1^{-}$} & \cmpcell{$\rep{35}^{\prime}$} & \cmpcell{6.192} & \cmpcell{-} \\
\rule[-2pt]{0pt}{10.4pt}\cmpcell{$0^{+}$} & \cmpcell{$\rep{14}$} & \cmpcell{5.908} & \cmpcell{P} & \cmpcell{$1^{-}$} & \cmpcell{$\rep{30}$} & \cmpcell{5.670} & \cmpcell{D} & \cmpcell{$1^{-}$} & \cmpcell{$\rep{35}$} & \cmpcell{4.450} & \cmpcell{-} & \cmpcell{$0^{+}$} & \cmpcell{$\rep{55}$} & \cmpcell{3.926} & \cmpcell{P} \\
\rule[-2pt]{0pt}{10.4pt}\cmpcell{$3^{+}$} & \cmpcell{$\rep{14}$} & \cmpcell{4.093} & \cmpcell{D} & \cmpcell{$1^{-}$} & \cmpcell{$\rep{30}$} & \cmpcell{5.755} & \cmpcell{D} & \cmpcell{$3^{-}$} & \cmpcell{$\rep{35}$} & \cmpcell{4.707} & \cmpcell{-} & \cmpcell{$1^{+}$} & \cmpcell{$\rep{55}$} & \cmpcell{4.949} & \cmpcell{D} \\
\rule[-2pt]{0pt}{10.4pt}\cmpcell{$3^{+}$} & \cmpcell{$\rep{14}$} & \cmpcell{4.258} & \cmpcell{D} & \cmpcell{$2^{-}$} & \cmpcell{$\rep{30}$} & \cmpcell{6.156} & \cmpcell{-} & \cmpcell{$1^{-}$} & \cmpcell{$\rep{35}$} & \cmpcell{4.720} & \cmpcell{-} & \cmpcell{$0^{+}$} & \cmpcell{$\rep{55}$} & \cmpcell{6.073} & \cmpcell{D} \\
\rule[-2pt]{0pt}{10.4pt}\cmpcell{$1^{+}$} & \cmpcell{$\rep{14}$} & \cmpcell{4.641} & \cmpcell{D} & \cmpcell{$1^{-}$} & \cmpcell{$\rep{35}$} & \cmpcell{3.040} & \cmpcell{P} & \cmpcell{$3^{-}$} & \cmpcell{$\rep{35}$} & \cmpcell{5.035} & \cmpcell{-} & \cmpcell{$2^{+}$} & \cmpcell{$\rep{55}$} & \cmpcell{5.700$^{\ddagger}$} & \cmpcell{-} \\
\rule[-2pt]{0pt}{10.4pt}\cmpcell{$1^{+}$} & \cmpcell{$\rep{14}$} & \cmpcell{4.709} & \cmpcell{D} & \cmpcell{$2^{-}$} & \cmpcell{$\rep{35}$} & \cmpcell{3.981} & \cmpcell{D} & \cmpcell{$3^{+}$} & \cmpcell{$\rep{35}$} & \cmpcell{5.270} & \cmpcell{-} & \cmpcell{$2^{+}$} & \cmpcell{$\rep{55}$} & \cmpcell{5.864$^{\dagger}$} & \cmpcell{-} \\
\rule[-2pt]{0pt}{10.4pt}\cmpcell{$2^{+}$} & \cmpcell{$\rep{14}$} & \cmpcell{5.246} & \cmpcell{D} & \cmpcell{$1^{+}$} & \cmpcell{$\rep{35}$} & \cmpcell{4.114} & \cmpcell{D} & \cmpcell{$2^{-}$} & \cmpcell{$\rep{35}$} & \cmpcell{5.339} & \cmpcell{D} & \cmpcell{$1^{+}$} & \cmpcell{$\rep{81}$} & \cmpcell{4.314} & \cmpcell{P} \\
\rule[-2pt]{0pt}{10.4pt}\cmpcell{$2^{+}$} & \cmpcell{$\rep{14}$} & \cmpcell{5.488} & \cmpcell{D} & \cmpcell{$0^{-}$} & \cmpcell{$\rep{35}$} & \cmpcell{4.218} & \cmpcell{D} & \cmpcell{$1^{+}$} & \cmpcell{$\rep{35}$} & \cmpcell{5.476} & \cmpcell{D} & \cmpcell{$2^{+}$} & \cmpcell{$\rep{81}$} & \cmpcell{5.236} & \cmpcell{D} \\
\rule[-2pt]{0pt}{10.4pt}\cmpcell{$2^{+}$} & \cmpcell{$\rep{14}$} & \cmpcell{5.550} & \cmpcell{D} & \cmpcell{$2^{+}$} & \cmpcell{$\rep{35}$} & \cmpcell{4.986} & \cmpcell{D} & \cmpcell{$3^{+}$} & \cmpcell{$\rep{35}$} & \cmpcell{5.539} & \cmpcell{D} & \cmpcell{$1^{-}$} & \cmpcell{$\rep{81}$} & \cmpcell{5.370} & \cmpcell{D} \\
\rule[-2pt]{0pt}{10.4pt}\cmpcell{$3^{+}$} & \cmpcell{$\rep{14}$} & \cmpcell{5.663} & \cmpcell{-} & \cmpcell{$1^{-}$} & \cmpcell{$\rep{35}$} & \cmpcell{5.134} & \cmpcell{D} & \cmpcell{$0^{-}$} & \cmpcell{$\rep{35}$} & \cmpcell{5.563} & \cmpcell{D} & \cmpcell{$0^{+}$} & \cmpcell{$\rep{81}$} & \cmpcell{5.463} & \cmpcell{D} \\
\rule[-2pt]{0pt}{10.4pt}\cmpcell{$0^{+}$} & \cmpcell{$\rep{14}$} & \cmpcell{5.686} & \cmpcell{D} & \cmpcell{$1^{-}$} & \cmpcell{$\rep{35}$} & \cmpcell{5.188} & \cmpcell{D} & \cmpcell{$2^{-}$} & \cmpcell{$\rep{35}$} & \cmpcell{5.582} & \cmpcell{D} & \cmpcell{$2^{+}$} & \cmpcell{$\rep{81}$} & \cmpcell{4.808} & \cmpcell{P} \\
\rule[-2pt]{0pt}{10.4pt}\cmpcell{$3^{+}$} & \cmpcell{$\rep{14}$} & \cmpcell{5.730} & \cmpcell{-} & \cmpcell{$2^{-}$} & \cmpcell{$\rep{35}$} & \cmpcell{5.918} & \cmpcell{D} & \cmpcell{$3^{+}$} & \cmpcell{$\rep{35}$} & \cmpcell{5.677} & \cmpcell{-} & \cmpcell{$3^{+}$} & \cmpcell{$\rep{81}$} & \cmpcell{5.617} & \cmpcell{D} \\
\rule[-2pt]{0pt}{10.4pt}\cmpcell{$0^{+}$} & \cmpcell{$\rep{14}$} & \cmpcell{5.788} & \cmpcell{D} & \cmpcell{$1^{+}$} & \cmpcell{$\rep{35}$} & \cmpcell{6.100} & \cmpcell{D} & \cmpcell{$1^{+}$} & \cmpcell{$\rep{35}$} & \cmpcell{5.689} & \cmpcell{D} & \cmpcell{$2^{-}$} & \cmpcell{$\rep{81}$} & \cmpcell{5.790} & \cmpcell{D} \\
\rule[-2pt]{0pt}{10.4pt}\cmpcell{$3^{+}$} & \cmpcell{$\rep{14}$} & \cmpcell{5.805} & \cmpcell{D} & \cmpcell{$0^{-}$} & \cmpcell{$\rep{35}$} & \cmpcell{6.223} & \cmpcell{D} & \cmpcell{$3^{+}$} & \cmpcell{$\rep{35}$} & \cmpcell{5.724} & \cmpcell{-} & \cmpcell{$3^{+}$} & \cmpcell{$\rep{81}$} & \cmpcell{5.219} & \cmpcell{P} \\
\rule[-2pt]{0pt}{10.4pt}\cmpcell{$1^{+}$} & \cmpcell{$\rep{14}$} & \cmpcell{6.019} & \cmpcell{D} & \cmpcell{$2^{-}$} & \cmpcell{$\rep{35}$} & \cmpcell{3.567} & \cmpcell{P} & \cmpcell{$0^{-}$} & \cmpcell{$\rep{35}$} & \cmpcell{5.778} & \cmpcell{D} & \cmpcell{$3^{-}$} & \cmpcell{$\rep{81}$} & \cmpcell{6.153} & \cmpcell{D} \\
\rule[-2pt]{0pt}{10.4pt}\cmpcell{$1^{+}$} & \cmpcell{$\rep{14}$} & \cmpcell{6.068$^{\ddagger}$} & \cmpcell{-} & \cmpcell{$3^{-}$} & \cmpcell{$\rep{35}$} & \cmpcell{4.379} & \cmpcell{D} & \cmpcell{$3^{+}$} & \cmpcell{$\rep{35}$} & \cmpcell{5.912} & \cmpcell{D} & \cmpcell{$1^{+}$} & \cmpcell{$\rep{81}$} & \cmpcell{6.172} & \cmpcell{-} \\
\rule[-2pt]{0pt}{10.4pt}\cmpcell{$2^{+}$} & \cmpcell{$\rep{14}$} & \cmpcell{6.088} & \cmpcell{-} & \cmpcell{$2^{+}$} & \cmpcell{$\rep{35}$} & \cmpcell{4.555} & \cmpcell{D} & \cmpcell{$2^{-}$} & \cmpcell{$\rep{35}$} & \cmpcell{5.981} & \cmpcell{D} & \cmpcell{$1^{+}$} & \cmpcell{$\rep{81}$} & \cmpcell{5.882} & \cmpcell{-} \\
\rule[-2pt]{0pt}{10.4pt}\cmpcell{$3^{+}$} & \cmpcell{$\rep{14}$} & \cmpcell{6.117} & \cmpcell{D} & \cmpcell{$2^{-}$} & \cmpcell{$\rep{35}$} & \cmpcell{5.480} & \cmpcell{D} & \cmpcell{$3^{-}$} & \cmpcell{$\rep{35}$} & \cmpcell{5.985} & \cmpcell{D} & \cmpcell{$3^{+}$} & \cmpcell{$\rep{81}$} & \cmpcell{5.923} & \cmpcell{-} \\
\rule[-2pt]{0pt}{10.4pt}\cmpcell{$3^{+}$} & \cmpcell{$\rep{14}$} & \cmpcell{6.136} & \cmpcell{D} & \cmpcell{$2^{-}$} & \cmpcell{$\rep{35}$} & \cmpcell{4.796} & \cmpcell{P} & \cmpcell{$3^{-}$} & \cmpcell{$\rep{35}$} & \cmpcell{6.075} & \cmpcell{D} & \cmpcell{$1^{+}$} & \cmpcell{$\rep{81}$} & \cmpcell{6.008} & \cmpcell{-} \\
\rule[-2pt]{0pt}{10.4pt}\cmpcell{$3^{+}$} & \cmpcell{$\rep{14}$} & \cmpcell{6.219} & \cmpcell{D} & \cmpcell{$3^{-}$} & \cmpcell{$\rep{35}$} & \cmpcell{5.602} & \cmpcell{D} & \cmpcell{$3^{-}$} & \cmpcell{$\rep{35}$} & \cmpcell{6.165} & \cmpcell{D} & \cmpcell{$2^{-}$} & \cmpcell{$\rep{81}$} & \cmpcell{6.087} & \cmpcell{-} \\
\rule[-2pt]{0pt}{10.4pt}\cmpcell{$2^{+}$} & \cmpcell{$\rep{14}$} & \cmpcell{6.238} & \cmpcell{D} & \cmpcell{$2^{+}$} & \cmpcell{$\rep{35}$} & \cmpcell{5.790} & \cmpcell{D} & \cmpcell{$2^{+}$} & \cmpcell{$\rep{35}$} & \cmpcell{6.225} & \cmpcell{-} & \cmpcell{$3^{+}$} & \cmpcell{$\rep{81}$} & \cmpcell{6.223} & \cmpcell{-} \\
\rule[-2pt]{0pt}{10.4pt}\cmpcell{$0^{-}$} & \cmpcell{$\rep{30}$} & \cmpcell{2.592} & \cmpcell{P} & \cmpcell{$1^{-}$} & \cmpcell{$\rep{35}$} & \cmpcell{5.935} & \cmpcell{D} & \cmpcell{$2^{-}$} & \cmpcell{$\rep{35}$} & \cmpcell{6.231} & \cmpcell{D} & \cmpcell{$2^{-}$} & \cmpcell{$\rep{81}$} & \cmpcell{6.231} & \cmpcell{-} \\
\rule[-2pt]{0pt}{10.4pt}\cmpcell{$1^{-}$} & \cmpcell{$\rep{30}$} & \cmpcell{3.649} & \cmpcell{D} & \cmpcell{$2^{+}$} & \cmpcell{$\rep{35}$} & \cmpcell{4.810} & \cmpcell{P} & \cmpcell{$2^{+}$} & \cmpcell{$\rep{35}^{\prime}$} & \cmpcell{4.714} & \cmpcell{P} & \cmpcell{$2^{+}$} & \cmpcell{$\rep{81}$} & \cmpcell{6.237} & \cmpcell{-} \\
\rule[-2pt]{0pt}{10.4pt}\cmpcell{$0^{-}$} & \cmpcell{$\rep{30}$} & \cmpcell{4.807} & \cmpcell{D} & \cmpcell{$2^{-}$} & \cmpcell{$\rep{35}$} & \cmpcell{5.765} & \cmpcell{D} & \cmpcell{$3^{+}$} & \cmpcell{$\rep{35}^{\prime}$} & \cmpcell{5.527} & \cmpcell{D} & \cmpcell{$2^{-}$} & \cmpcell{$\rep{105}$} & \cmpcell{6.140} & \cmpcell{P} \\
\rule[-2pt]{0pt}{10.4pt}\cmpcell{$2^{+}$} & \cmpcell{$\rep{30}$} & \cmpcell{5.432} & \cmpcell{D} & \cmpcell{$1^{+}$} & \cmpcell{$\rep{35}$} & \cmpcell{5.939} & \cmpcell{D} & \cmpcell{$2^{-}$} & \cmpcell{$\rep{35}^{\prime}$} & \cmpcell{5.723} & \cmpcell{D} & \cmpcell{$1^{-}$} & \cmpcell{$\rep{154}$} & \cmpcell{5.792} & \cmpcell{P} \\
\rule[-2pt]{0pt}{10.4pt}\cmpcell{$3^{+}$} & \cmpcell{$\rep{30}$} & \cmpcell{6.112} & \cmpcell{D} & \cmpcell{$1^{+}$} & \cmpcell{$\rep{35}$} & \cmpcell{4.822} & \cmpcell{-} & \cmpcell{$1^{+}$} & \cmpcell{$\rep{35}^{\prime}$} & \cmpcell{5.894} & \cmpcell{D} & \cmpcell{$0^{-}$} & \cmpcell{$\rep{91}$} & \cmpcell{5.449} & \cmpcell{P} \\
\end{longtable}
\endgroup
\noindent\textit{Projection weights of the marked levels.}
At nominal $\kappa_c$, the pairs $(w_P,w_D)$ in percent are
$6.068$ ($89.60,\ 10.05$), $4.425$ ($78.43,\ 21.56$), $4.570$ ($21.57,\ 78.43$), $5.700$ ($63.34,\ 36.66$), and $5.864$ ($36.66,\ 63.34$).
The weights come from separate projectors onto the candidate primary and total descendant spans. All five raw levels fail the $0.90$ primary gate. Their symmetry copies agree, and changing $\kappa_c$ to $\kappa_c/3$ or $3\kappa_c$ changes $w_P$ by less than $0.02$ percentage points. The two $\ddagger$ marks denote failures of the independent retained-window check.

The table lists raw energy eigenstates; a primary mode spread over several eigenstates does not supply an additional table entry. Levels without a resolved individual assignment retain a dash, as illustrated by the projection weights in the notes to Table~\ref{tab:unresolvedlevelweights}. The Casimir values $C_2=15$ and $20$ each admit two representations, $\rep{84}=[0,6]$ or $\rep{154}=[3,2]$ and $\rep{91}=[5,0]$ or $\rep{231}=[1,6]$. The symmetry copies select one of them: at $N=5$--11 the $C_2=15$ levels occur in the $(s_A,s_B)$ blocks $(0,1)$, $(1,1)$, $(1,2)$, and $(2,2)$, as required by the branching of $\rep{154}$, whereas $\rep{84}$ would occur only in $(1,2)$; the $C_2=20$ level occurs in $(0,0)$, $(1,1)$, and $(2,2)$, as required for $\rep{91}$ rather than $\rep{231}$. The $N=12$ levels, computed only in the $(0,0)$, $(0,1)$, and $(0,2)$ blocks, continue these families. The full-precision energies, projection weights, and family diagnostics are preserved in the accompanying numerical archive.
\section{OPE extraction and normalization}
\label{app:numericalconventions}
This appendix specifies how the fuzzy-sphere matrix elements and the bootstrap amplitudes are converted to the common OPE convention of Table~\ref{tab:ope}. The complete contraction catalogue, insertion routes, and finite-size sequences are retained in the numerical data accompanying this work.
\subsection{Matrix-element extraction}
\label{app:opeextractiondetails}
The state--operator correspondence relates a CFT three-point coefficient to a matrix element on the cylinder. A microscopic insertion with the required quantum numbers has an expansion $\widehat{\mathcal O}_{\rm micro}=Z_{\mathcal O}\mathcal O+\cdots$, where the omitted operators and descendants produce finite-size corrections. Vacuum-to-state matrix elements fix the leading overlap $Z_{\mathcal O}$. Primary-state identification and matching of the inserted observable are separate steps: the $K$-generator analysis controls the former, while size dependence and comparisons between insertion routes test the latter.

For input state $X$, microscopic insertion $Y$ of angular rank $r$, and output state $Z$, our reduced matrix elements give
\begin{equation}
\lambda_{\rm iso}(X\otimes Y\to Z)=
\sqrt{\frac{(2r+1)!!}{r!}}\,
\frac{m_{ZX}^{(Y)}}{Z_Y C_{\rm iso}}.
\label{eq:extractionappendix}
\end{equation}
Here $C_{\rm iso}$ is the relevant component of the internal $SO(5)$ Clebsch--Gordan isometry, normalized by $\mathcal C\mathcal C^\dagger=\rep1_Z$, and $m_{ZX}^{(Y)}$ is the spatially projected microscopic matrix element. The connected singlet contraction subtracts $\langle0|O_s|0\rangle\langle Z|X\rangle$. The factors for insertion ranks $r=0,1,2$ are $1,\sqrt3,\sqrt{15/2}$. For spinning external states, the measured angular projection must be specified; one such contraction need not determine all independent three-point tensor structures.

Different insertions can estimate the same continuum coefficient with different finite-size corrections; we keep these routes separate rather than average them. The signed scalar ratios in Table~\ref{tab:ope} use $s$ insertion for $\phi\phi s$, $stt$, and $sss$, and $t$ insertion for $ttt$ and the common $\phi\phi t$ denominator. The current and stress entries use $\phi$ insertion. All entries are converted from $\lambda_{\rm iso}$ to $\lambda$ with the factors of Table~\ref{tab:conversion}.

\paragraph{Insertion orderings.}
By the state--operator correspondence, $\lambda_{\phi\phi\mathcal O}$ can be obtained by inserting $\phi$ between the states $\phi$ and $\mathcal O$, or by inserting $\mathcal O$ between two $\phi$ states. The two orderings coincide in the CFT. Their microscopic insertions contain different subleading operators and descendants, and hence have different finite-size corrections.

For $\phi$ insertion we use the rank-zero harmonic $n_{00}$ of the vector density $n^a=\hat\Psi^\dagger\Gamma^a\hat\Psi$ defined in Appendix~\ref{app:fuzzysetup}, with $\Gamma^2=I\otimes\tau^z$. Choose the vector phase so that $Z_\phi=\langle\phi|n_{00}|0\rangle>0$, and fix the singlet phase by $\langle\phi|n_{00}|s\rangle>0$. In the convention of Table~\ref{tab:ope}, the scalar coefficients are
\begin{equation}
\begin{aligned}
\lambda^{(\phi)}_{\phi\phi s}&=\frac{\langle\phi|n_{00}|s\rangle}{Z_\phi},\\
\lambda^{(\phi)}_{\phi\phi t}&=\frac{\sqrt5}{2}\,
\frac{|\langle\phi|n_{00}|t_G\rangle|}{Z_\phi}.
\end{aligned}
\label{eq:phi_insertion_routes}
\end{equation}
Here $t_G$ has internal polarization $E_G=\operatorname{diag}(-1,4,-1,-1,-1)/\sqrt{20}$, and the corresponding local tensor component is $Q_G=(-Q_{11}+4Q_{22}-Q_{33}-Q_{44}-Q_{55})/\sqrt{20}$. Thus $\sqrt5/2=1/(E_G)_{22}$ removes the fixed-component Clebsch factor, as in the vector-insertion prescription of Ref.~\cite{Zhou2024}.

For $s$ insertion, let $\mathcal B_\rho$ and $\mathcal B_B$ denote the projected local density-square and symplectic-pair-square observables built from $\hat\rho$ and $\hat B$ in Eq.~\eqref{eq:fuzzyH}, with the same contact prescription. We form $O_s=a_1\mathcal B_\rho+a_2\mathcal B_B$ with microscopic coefficients $a_1,a_2$. Their relative weight suppresses the stress overlap, and their overall magnitude normalizes the singlet overlap:
\begin{equation}
\langle T_{\mu\nu}|O_s^{(2)}|0\rangle=0,\qquad
Z_{s,\rm raw}=\langle s|O_s^{(0)}|0\rangle,\quad |Z_{s,\rm raw}|=1.
\label{eq:singlet_insertion_calibration}
\end{equation}
The superscripts specify the angular ranks of the harmonics of the same local insertion. In particular, the stress constraint uses rank two. The singlet phase has already been fixed by the vector matrix element, so $Z_{s,\rm raw}$ is retained with its sign. Equivalently, the rescaled insertion $\widetilde O_s=O_s/Z_{s,\rm raw}$ has unit positive vacuum overlap. For the tensor insertion set $Z_t=|\langle t_G|Q_G|0\rangle|$, with the phase of $t_G$ aligned to this overlap. Then
\begin{equation}
\begin{aligned}
\lambda^{(s)}_{\phi\phi s}&=
\frac{\langle\phi|O_s^{(0)}|\phi\rangle-\langle0|O_s^{(0)}|0\rangle}{Z_{s,\rm raw}},\\
\lambda^{(t)}_{\phi\phi t}&=\frac{\sqrt5}{2}\,
\frac{\langle\phi|Q_G|\phi\rangle}{Z_t}.
\end{aligned}
\label{eq:scalar_insertion_routes}
\end{equation}
The first expression subtracts the disconnected vacuum term. For general projected states, that subtraction is $\langle0|O_s|0\rangle\langle Z|X\rangle$, retaining their overlap.

Along the tuned $N=5$--$12$ sequence, the $\phi$-insertion estimates for both scalar coefficients increase with $N$, whereas the $s$- and $t$-insertion estimates decrease. The ratio $\lambda^{(s)}_{\phi\phi s}/\lambda^{(t)}_{\phi\phi t}$ changes from $0.2057$ to $0.2033$, about $1.1\%$, and the $\phi$-insertion ratio from $0.1279$ to $0.1681$. Separate $a+b/N$ fits of the numerator and denominator over $N=6$--$12$ or $8$--$12$ give ratios of fitted intercepts near $0.20$ for both orderings.

In the mixed-crossing normalization $\beta$ these ratios are larger by $(25/14)^{1/4}$; for example, the $N=12$ value $0.2033$ corresponds to $\beta_{\phi\phi s}/\beta_{\phi\phi t}=0.2350$. The $\phi\phi J$ and $\phi\phi T_{\mu\nu}$ coefficients use $\phi$ insertion with the rank-one and rank-two harmonics of the same vector density and the angular factors of Eq.~\eqref{eq:extractionappendix}.

\paragraph{Size extrapolation of the current and stress couplings.} Linear fits $a+b/N$ of the $N=5$--12 sequences give intercepts that increase as the fit window moves to larger $N$: $0.776$, $0.781$, and $0.786$ for $|\lambda_{\phi\phi J}|$ over $N=6$--12, $8$--12, and $10$--12, and $0.355$, $0.363$, and $0.369$ for $|\lambda_{\phi\phi T}|$. Table~\ref{tab:ope} quotes the $N=8$--12 fit to three decimals.
\subsection{OPE conventions for each channel}
\label{app:opeunit}
The external scalars have unit two-point functions in the tensor convention of Eq.~\eqref{eq:norm}. For the current and stress tensor we use unit two-point structures
\begin{align}
\langle\widehat J_{ij}(x,z)\widehat J_{kl}(0,w)\rangle
&=\frac{A_{ij,kl}\,z\cdot I(x)\cdot w}{|x|^{2\Delta_J}},\\
\langle\widehat T(x,z)\widehat T(0,w)\rangle
&=\frac{[z\cdot I(x)\cdot w]^2}{|x|^{2\Delta_T}},
\end{align}
where $A_{ij,kl}=\tfrac12(\delta_{ik}\delta_{jl}-\delta_{il}\delta_{jk})$, $I_{\mu\nu}(x)=\delta_{\mu\nu}-2x_\mu x_\nu/x^2$, and $z,w$ are null polarizations. Restoring physical current and stress normalizations multiplies the hatted operators by $\sqrt{C_J}$ and $\sqrt{C_T}$. The three-point structures use the conformal vector without an additional factor, and the convolved blocks incorporate the spatial factor $\ell!/(2\ell-1)!!$, so no spin-dependent factor enters the conversions below.

Three normalizations of a three-point coefficient for external scalars $A,B$ and an exchanged primary $\mathcal O$ in representation $R$ appear in this work.
\begin{enumerate}
\item The isometric coefficient $\lambda_{\rm iso}(A\otimes B\to R)$ multiplies the internal Clebsch--Gordan tensor $\mathcal C$ of $A\otimes B\to R$ normalized by $\mathcal C\mathcal C^\dagger=\rep1_R$. The fuzzy-sphere extraction, Eq.~\eqref{eq:extractionappendix}, yields this quantity.
\item The mixed-crossing coefficient $\beta_{AB\mathcal O}$ is the amplitude in the crossing equations~\eqref{eq:crossing} solved by the bootstrap,
\begin{equation}
\beta_{AB\mathcal O}=G_{AB\to R}\,\lambda_{\rm iso}(A\otimes B\to R),\qquad
G_{AB\to R}=\left(\frac{d_R}{d_A d_B}\right)^{1/4}.
\label{eq:appendixOPEconversion}
\end{equation}
The crossing equations identify the repeated appearances of one coupling in different channels, such as $\beta_{\phi\phi t}$ in $\phi\phi\to t$ and in $t\phi\to\phi$.
\item The $v$-singlet coefficient $\lambda$ is used for every OPE coefficient reported in this paper.
\end{enumerate}

The $v$-singlet convention is defined for the couplings of two vectors in Ref.~\cite{LiShen2026}, Table~II and Appendix~A,
\begin{equation}
\begin{aligned}
\lambda_{\phi\phi s}&=\beta_{\phi\phi s}=\lambda_s^{\rm iso}/\sqrt5,\\
\lambda_{\phi\phi t}&=(25/14)^{1/4}\beta_{\phi\phi t}=\lambda_t^{\rm iso},\\
\lambda_{\phi\phi J}&=(5/8)^{1/4}\beta_{\phi\phi J}=\lambda_J^{\rm iso}/\sqrt2,\\
\lambda_{\phi\phi T}&=\beta_{\phi\phi T}=\lambda_T^{\rm iso}/\sqrt5 .
\end{aligned}
\label{eq:coreOPEconversion}
\end{equation}
Starting instead from the raw amplitudes of the vector-correlator crossing equations, the corresponding factors are $1/\sqrt5$, $1/\sqrt{14}$, $1/\sqrt{20}$, and $1/\sqrt5$. In terms of invariant tensors, $\lambda_{\phi\phi s}$ and $\lambda_{\phi\phi T}$ multiply $\delta_{ij}$, and $\lambda_{\phi\phi t}$ multiplies $\Pi_{ij,kl}$ of Eq.~\eqref{eq:norm}, which is itself an isometry from $\rep{14}$ into $\rep5\otimes\rep5$. For the current, $\lambda_{\phi\phi J}$ is the coefficient of a fixed component $J_{[ab]}$ normalized by $A_{ij,kl}$, smaller than the isometric coefficient by $\sqrt2$.

Only $\phi\phi$ exchanges occur in the vector-correlator crossing equations, so Ref.~\cite{LiShen2026} does not fix the other channels. We extend the convention by the rule that reproduces Eq.~\eqref{eq:coreOPEconversion}:
\begin{enumerate}
\item[(a)] If $A$, $B$, or $R$ is a singlet, the invariant tensor is the two-point tensor of the remaining pair ($\delta_{ij}$, $\Pi_{ij,kl}$, or $1$), and $\lambda=\beta$.
\item[(b)] If $R=\rep{10}$, the exchanged antisymmetric operator uses the fixed-component normalization of the current, $\lambda=\lambda_{\rm iso}/\sqrt2=(d_Ad_B/40)^{1/4}\beta$.
\item[(c)] Otherwise the invariant tensor is the isometric Clebsch--Gordan tensor for the exchanged representation, $\lambda=\lambda_{\rm iso}=(d_Ad_B/d_R)^{1/4}\beta$.
\end{enumerate}
The five couplings among the external scalars are defined through $\phi\phi\to s$, $\phi\phi\to t$, $ss\to s$, $st\to t$, and $tt\to t$. Rule (a) fixes $\lambda_{sss}=\beta_{sss}$ and $\lambda_{stt}=\beta_{stt}$ uniquely; $tt\to s$ gives the same $\lambda_{stt}$. For $\lambda_{ttt}$, the cubic invariant of $\rep{14}$ is unique up to normalization, and rule (c) chooses the same isometric normalization as $\lambda_{\phi\phi t}$, $\lambda_{ttt}=\lambda_{\rm iso}(t\otimes t\to t)=14^{1/4}\beta_{ttt}$. A trace normalization $d_{abc}=\mathrm{tr}(e_ae_be_c)$ in an orthonormal basis $e_a$ of traceless symmetric matrices would instead give $\sqrt{20/27}\,\lambda_{ttt}$, since $\sum_{ab}d_{abc}d_{abd}=\tfrac{27}{20}\delta_{cd}$. Each three-point coefficient thus has a single value, and $\lambda/\beta$ is a fixed constant for each channel. The bootstrap spectrum, the scanned ratios $r_{XYZ}$ of Eq.~\eqref{eq:operatioscan}, and the relative differences between bootstrap and fuzzy-sphere values are independent of this choice.

Table~\ref{tab:conversion} lists $G=\beta/\lambda_{\rm iso}$ and $\kappa=\lambda/\beta$ for every mixed channel. A bootstrap amplitude is converted by $\lambda=\kappa\beta$; a fuzzy-sphere coefficient by $\lambda=\kappa G\lambda_{\rm iso}$, which equals $\lambda_{\rm iso}$ under rule (c). For the scalar ratios of Table~\ref{tab:ope} this gives
\begin{equation}
\frac{\lambda_{\phi\phi s}}{\lambda_{\phi\phi t}}=\left(\frac{14}{25}\right)^{1/4}r_{\phi\phi s},\quad
\frac{\lambda_{ttt}}{\lambda_{\phi\phi t}}=\left(\frac{196}{25}\right)^{1/4}r_{ttt},\quad
\frac{\lambda_{stt}}{\lambda_{\phi\phi t}}=\left(\frac{14}{25}\right)^{1/4}r_{stt},
\label{eq:ratioconversion}
\end{equation}
and likewise for $sss$. The combined OPE-coefficient magnitude $\lambda_{\rm eff}$ of Fig.~\ref{fig:mixedspectrum} is defined in Eq.~\eqref{eq:efmdiagnostics} using the crossing normalization, so each reported coefficient enters as $\lambda_{i,a}/\kappa_i$.

An operator sign change reverses every OPE coefficient containing that operator an odd number of times. Signed comparisons therefore require consistent phase choices across routes. Where a fuzzy-sphere route determines only a modulus, we compare $|\lambda|$ and infer no relative sign. The same convention is applied at every size.

The rank-two extraction factor follows from the angular integral. A normalized tensor polarization is $Q=\mathrm{diag}(-1,-1,2)/\sqrt6$. Its angular matrix element is $\lambda_{\phi\phi T}(3\cos^2\theta-1)/\sqrt6$. Here $M_{20}$ is the stress-target matrix element of the $Y_{20}$ harmonic of the vector insertion, and $M_{00}$ is the vacuum normalization matrix element of its $Y_{00}$ harmonic, with $\int|Y_{\ell m}|^2\dd\Omega=1$. Direct angular integration gives
\begin{equation}
 r_{20}=\frac{|M_{20}|}{|M_{00}|}=\sqrt{\frac{2}{15}}|\lambda_{\phi\phi T}|,\qquad
 |\lambda_{\phi\phi T}|=\sqrt{\frac{15}{2}}r_{20},
\end{equation}
consistent with $|\lambda_{\phi\phi T}|=|\lambda_T^{\rm iso}|/\sqrt5$. At $N=12$ it yields $0.273$. A coefficient defined with half the canonical stress normalization, $\sqrt{15/8}\,r_{20}$, is converted once by the factor two.

\Needspace{14\baselineskip}\begingroup\appendixtablestyle\setlength{\LTcapwidth}{\textwidth}
\setlength{\tabcolsep}{4pt}
\begin{longtable}{cccc@{\hspace{22pt}}cccc}
\caption{Conversion factors for every mixed channel: $G=\beta/\lambda_{\rm iso}$ of Eq.~\eqref{eq:appendixOPEconversion} and $\kappa=\lambda/\beta$ of rules (a)--(c). $A,B$ are the external pair and $R$ is the exchanged representation; names in parentheses identify leading exchanges. The spin and parity restrictions are given in Table~\ref{tab:mixedselection}. For the external couplings, $\lambda_{\phi\phi t}$ is defined through $\phi\phi\to t$; the $t\phi\to\rep5$ factor applies to the other exchanged vectors.}\label{tab:conversion}\\
\toprule $AB$ & \SOrephead & $G$ & $\kappa$ & $AB$ & \SOrephead & $G$ & $\kappa$\\\midrule
\endfirsthead
\multicolumn{8}{c}{TABLE \thetable\ (continued)}\\
\toprule $AB$ & \SOrephead & $G$ & $\kappa$ & $AB$ & \SOrephead & $G$ & $\kappa$\\\midrule
\endhead\bottomrule\endfoot
\rowstrut $\phi \phi $ & $\rep{1}\;(s,T_{\mu\nu})$ & $1/\sqrt5$ & $1$ & $ss$ & $\rep{1}\;(s,T_{\mu\nu})$ & $1$ & $1$ \\
\rowstrut $\phi \phi $ & $\rep{10}\;(J_\mu)$ & $(2/5)^{1/4}$ & $(5/8)^{1/4}$ & $st$ & $\rep{14}\;(t)$ & $1$ & $1$ \\
\rowstrut $\phi \phi $ & $\rep{14}\;(t)$ & $14^{1/4}/\sqrt5$ & $(25/14)^{1/4}$ & $tt$ & $\rep{1}\;(s,T_{\mu\nu})$ & $1/\sqrt{14}$ & $1$ \\
\rowstrut $s\phi $ & $\rep{5}\;(\phi)$ & $1$ & $1$ & $tt$ & $\rep{10}\;(J_\mu)$ & $(5/98)^{1/4}$ & $(49/10)^{1/4}$ \\
\rowstrut $t\phi $ & $\rep{5}$ & $14^{-1/4}$ & $14^{1/4}$ & $tt$ & $\rep{14}\;(t)$ & $14^{-1/4}$ & $14^{1/4}$ \\
\rowstrut $t\phi $ & $\rep{30}\;(\mathcal M_{6\pi})$ & $(3/7)^{1/4}$ & $(7/3)^{1/4}$ & $tt$ & $\rep{35^{\prime}}$ & $(5/28)^{1/4}$ & $(28/5)^{1/4}$ \\
\rowstrut $t\phi $ & $\rep{35}$ & $2^{-1/4}$ & $2^{1/4}$ & $tt$ & $\rep{55}\;(\mathcal M_{8\pi})$ & $(55/196)^{1/4}$ & $(196/55)^{1/4}$ \\
\rowstrut  &  &  &  & $tt$ & $\rep{81}$ & $3/\sqrt{14}$ & $\sqrt{14}/3$ \\
\end{longtable}\endgroup
\clearpage\twocolumngrid


%
\end{document}